\documentclass[preprint,12pt, sort&compress]{elsarticle}
\usepackage{fullpage}
\usepackage{color}
\usepackage{subcaption}
\usepackage{multirow}
\usepackage{amsmath}
\usepackage{amssymb}
\usepackage{hyperref}
\usepackage{bm}
\usepackage{pict2e}
\usepackage{epstopdf}
\usepackage{epsfig}
\usepackage{url}

\journal{Journal of Computational Physics}

\newcommand{\rz}{\mathbb{R}}

\newcommand{\cz}{\mathbb{C}}

\newcommand{\nz}{\mathbb{N}}

\newcommand{\bfx}{{\bf x}}
\newcommand{\bfy}{{\bf y}}

\newcommand{\bfX}{{\bf X}}
\newcommand{\bfY}{{\bf Y}}

\newcommand{\beq}{\begin{equation}}
\newcommand{\eeq}{\end{equation}}
\newcommand{\beqs}{\begin{eqnarray}}
\newcommand{\eeqs}{\end{eqnarray}}
\newcommand{\beql}{\begin{equation} \label}
\newcommand{\half}{\frac{1}{2}}

\newcommand{\calB}{{\cal B}}

\newcommand{\calE}{{\cal E}}

\newcommand{\calI}{{\cal I}}
\newcommand{\calJ}{{\cal J}}

\newcommand{\calL}{{\cal L}}

\newcommand{\calN}{{\cal N}}
\newcommand{\calO}{{\cal O}}
\newcommand{\calP}{{\cal P}}

\newcommand{\calR}{{\cal R}}

\newcommand{\calW}{{\cal W}}

\newcommand{\calY}{{\cal Y}}

\newcommand{\thrbyto}{\frac{3}{2}}

\newcommand{\abs}[1]{\lvert#1\rvert}

\newcommand{\innprod}[3]{\langle#1,#2\rangle_{#3}}
\newcommand{\biginnprod}[3]{\mathbf{\Big\langle}#1,#2\mathbf{\Big\rangle}_{#3}}

\newcommand{\Lpspc}[3]{\textsf{L}^{#1}_{#2}(#3)}

\newcommand{\pd}[2]{\frac{\partial#1}{\partial#2}}

\let\oldFootnote\footnote
\newcommand\nextToken\relax

\renewcommand\footnote[1]{%
    \oldFootnote{#1}\futurelet\nextToken\isFootnote}

\newcommand\isFootnote{%
    \ifx\footnote\nextToken\textsuperscript{,}\fi}

\begin{document}

\begin{frontmatter}

\title{A spectral scheme for Kohn-Sham density functional theory of clusters}

\author[aem]{Amartya S. Banerjee}
\ead{asb@aem.umn.edu}
\author[aem]{Ryan S. ̃Elliott\corref{cor1}}
\ead{elliott@aem.umn.edu}
\author[aem]{Richard D. ̃James}
\ead{james@aem.umn.edu}

\cortext[cor1]{Corresponding author}

\address[aem]{Department of Aerospace Engineering and Mechanics,
University of Minnesota, Minneapolis, MN 55455, U.S.A.}

\begin{abstract}
Starting from the observation that one of the most successful methods for solving the Kohn-Sham equations for periodic systems -- the plane-wave method -- is a spectral method based on eigenfunction expansion, we formulate a spectral method designed towards solving the Kohn-Sham equations for clusters. This allows for efficient calculation of the electronic structure of clusters (and molecules) with high accuracy and systematic convergence properties without the need for any artificial periodicity. The basis functions in this method form a complete orthonormal set and are expressible in terms of spherical harmonics and spherical Bessel functions. Computation of the occupied eigenstates of the discretized Kohn-Sham Hamiltonian is carried out using a combination of preconditioned block eigensolvers and Chebyshev polynomial filter accelerated subspace iterations. Several algorithmic and computational aspects of the method, including computation of the electrostatics terms and parallelization are discussed. We have implemented these methods and algorithms into an efficient and reliable package called ClusterES (Cluster Electronic Structure). A variety of benchmark calculations employing local and non-local pseudopotentials are carried out using our package and the results are compared to the literature. Convergence properties of the basis set are discussed through numerical examples. Computations involving large systems that contain thousands of electrons are demonstrated to highlight the efficacy of our methodology. The use of our method to study clusters with arbitrary point group symmetries is briefly discussed.
\end{abstract}

\end{frontmatter}

% Introduction
\section{Introduction}
\label{sec:introduction}
Over the past few decades, quantum mechanical calculations based on Kohn-Sham Density Functional Theory (KS-DFT) have provided important insights into a variety of material systems \citep{Martin_ES, LeBris_ReviewBook, Saad_Chelikowsky_Shontz_review}. One of the most widely used and successful methods for numerical solution of the equations of Kohn-Sham theory is the pseudopotential plane-wave method \citep{Kresse_abinitio_iterative, Kresse_metal_semiconductor, Hutter_abinitio_MD, Teter_Payne_Allan_2, Barnett_Landman}, currently available in a number of software packages \citep{Kresse_abinitio_iterative, Gonze_ABINIT_1, CASTEP_1, Quantum_Espresso_1}. 

The advantages of plane-waves include the fact that they are orthonormal and therefore result in simple discretized expressions. Also, they form a complete basis, thus allowing for systematic convergence with increasing basis set size, governed by a single parameter, the energy cutoff. The global nature of the plane wave basis also results in minimum user interference in terms of basis set choice. Being a Fourier basis, plane-waves allow for spectral convergence leading to highly accurate numerical solutions \citep{Cances_planewave_numerical_analysis}.  Further, independence of the basis functions on atomic positions results in the absence of (the otherwise difficult to compute) Pulay forces \citep{CASTEP_1}. On the downside, while the plane-wave method is ideally suited for the study of periodic systems such as crystals, its application to non-periodic systems such as molecules and clusters is more limited due to the need for introducing artificial periodicity in the form of the supercell method \citep{Martin_ES, Hutter_abinitio_MD, Rappe_planewaves_molecules}. In addition, while studying such systems, plane-wave method codes only take advantage of symmetry groups which are {compatible} with translational symmetry (such as some of the crystallographic point groups).

Alternatives to the plane-wave approach include the use of atom centered basis functions such as Gaussians and atomic orbitals \citep{LCAO_3, LCAO_famous, SIESTA_1}, as well as real space discretization approaches such as finite differences and finite elements \citep{Chelikowsky_Saad_1, Octopus_1, Pask_FEM_review, Gavini_Kohn_Sham, Gavini_higher_order}. 
Atom centered basis functions generally require fewer basis functions per atom compared to plane-waves but these basis sets are usually incomplete and they suffer from basis set superposition errors \cite{LeBris_ReviewBook}. Thus, they have issues with systematic convergence. {Finite element methods, in contrast, have systematic convergence properties but the quality of the solution as well as the efficiency of the method is heavily dependent on the quality of the mesh as well as the type of element used for the calculation \citep{Gavini_higher_order}. {With the use of higher order finite elements in pseudopotential calculations, simple uniform meshes usually suffice, but mesh-coarsening is generally required for obtaining high efficiency in the vacuum region while dealing with isolated systems \citep{Gavini_higher_order}.} {Readers may refer to \citep{Saad_Chelikowsky_Shontz_review} for a more general review of various basis sets and numerical methods that are in common use today for solution of the Kohn-Sham problem.}

From the above discussion, it is quite clear that it would be highly desirable to have methods which are very similar to the plane-wave method but that are designed for systems which are non-periodic. Accordingly, in this work, we develop a scheme that is in many respects an exact analog of the plane-wave method but one which is designed with isolated systems such as clusters and molecules in mind. {Ab initio} studies of clusters, including various fullerenes and nanostructures, have received and continue to receive a lot of attention in different contexts \citep{Optical_magnetic_Boron_fullerene, Chelikowsky_silicon_nanostructures, PARSEC, Parallel_Chebyshev, Abinito_Fullerenes_Science, Small_Metal_Clusters, B80_abinitio, Gold_atomic_electronic_structure, Super_Atoms_1, Super_Atoms_2}. The methodology developed in this work therefore is likely to be useful for carrying out first principles studies of such systems in a consistent, systematic and efficient manner.

In order to formulate the appropriate basis functions for our method, we first make the observation that plane-waves are eigenfunctions of the periodic Laplacian. Using eigenfunctions of the Laplacian as basis functions leads to numerous advantages, including the fact that the kinetic energy operator is diagonalized in such cases. Accordingly, our method uses eigenfunctions of the Dirichlet Laplacian (i.e., the Laplacian operator with Dirichlet boundary conditions) in a spherical domain as the basis set. Our basis functions are expressible as the product of spherical harmonics with spherical Bessel functions.\footnote{For the purpose of clarity, we emphasize that our basis functions are centered on the origin; i.e., we are dealing with a molecule centered basis set as opposed to an atom centered basis set.} Let us remark that the use of a spherical (or near spherical) domain for the study of cluster systems has been used earlier in finite difference and finite element methods \cite{PARSEC, Gavini_Kohn_Sham}. To the best of our knowledge however, this is the first work to make systematic use of Laplacian eigenfunction expansions in non-periodic domains for use in electronic structure calculations.

Spherical basis functions have been used in earlier {work} to compute electronic properties of small metallic clusters \citep{electronic_sodium_magnesium_clusters, spherical_averaged_jellium} as well as that of $\textrm{C}_{60}$ \citep{C60_in_spherical_basis, Broglia_original_paper}. These basis functions have the distinct advantage that for many cluster systems, the Kohn-Sham eigenstates (molecular orbitals) and their symmetry properties are relatively easy to interpret using the quantum numbers associated with the basis functions\footnote{This allows systems such as super atoms \citep{Super_Atoms_1, Super_Atoms_2} to be studied conveniently.} themselves \citep{solid_state_finite}. As explained in \citep{Broglia_original_paper} the choice of spherical basis functions is usually motivated by the fact that the systems under study are nearly spherical. We show in this work however, that such a constraint on the system under study is unnecessary\footnote{This is owing to the fact that we have a complete orthonormal basis.} and that a wide variety of cluster systems including ones which are far from being spherical can be studied efficiently with our method. In contrast to our use of spherical Bessel functions, the radial part of the spherical basis functions used in {these aforementioned studies} has typically been obtained by solving a one dimensional radial eigenvalue problem.

In order to avoid computational complexity, many of the aforementioned {studies} use a simplified treatment of the electron-nucleus interaction in the form of simple-jellium or pseudo-jellium models \citep{spherical_averaged_jellium}. The use of these simplified models however, can often lead to inaccuracies, even while studying simple metallic clusters \citep{Review_metal_clusters}. In our view, one of the main reasons behind the computational difficulties encountered by these workers is due to the formulation of their methods in which convolution sums are carried out in reciprocal space by means of coupling coefficients (e.g. \cite{spherical_averaged_jellium} and \cite{solid_state_finite}). This makes certain operations such as computation of the electron density from the wavefunctions unmanageable beyond relatively small system sizes, unless approximations are used. In addition, these studies also rely on setting up of the full Hamiltonian matrix and then performing diagonalization of this matrix using direct methods, at each self-consistent field iteration cycle. This is quite unlike the approach employed by modern plane-wave codes where a dual representation of various quantities is employed for efficiency purposes and the Fast Fourier Transform (FFT) is used to switch between real and reciprocal space \citep{Hutter_abinitio_MD}. In addition, instead of direct diagonalization methods, most plane-wave codes employ matrix free iterative diagonalization methods to compute the occupied eigenspace of the Hamiltonian \citep{Kresse_abinitio_iterative, Teter_Payne_Allan_2}. We adopt similar strategies in this work and show that this leads to a method where accurate ground state electronic structure calculations for cluster systems containing many hundreds of electrons can be done routinely using our code. In particular, employing widely used, accurate {ab initio} norm conserving pseudopotentials for modelling the electron-nucleus interaction, without resorting to any form of spherical averaging of the potentials \citep{spherical_averaged_jellium}, poses no difficulty in our method.

As mentioned earlier, one of the key aspects of the plane-wave method is the use of three dimensional FFTs to switch between quantities expressed in real and reciprocal space. Analogously, we require efficient transforms to switch between quantities expressed on an appropriate grid used to discretize our spherical domain and the expansion coefficients of that quantity when expanded using our basis set (i.e., reciprocal space). Our strategy for obtaining efficient transforms is to use separation of variables into radial and angular parts and handling each of these parts through efficient transforms individually. Specifically, the radial part is computed through Gauss-Jacobi quadrature \citep{Gauss_Jacobi_Quad} while the angular part is handled using high performance Spherical Harmonics Transforms (SHTs)\citep{shtns}.

Another key requirement for carrying out accurate Kohn-Sham calculations is the ability to accurately evaluate the electrostatics terms. We accomplish this task here by developing an expansion of the Green's function of the associated Poisson problem in terms of our basis functions. This is followed by computing the convolution of the Green's function with the electronic charge. This is somewhat similar in spirit to the Green's function based methods developed in the context of plane-wave codes \citep{Hutter_abinitio_MD, Hockney_1970, Eastwood_Brownrig, Reciprocal_Hockney} {and free-boundary problems \citep{genovese2006efficient}}. The calculation of the Green's function (in terms of its expansion) can be done ahead of time and does not have to be repeated. As explained later, this means that during the SCF (Self-Consistent Field) cycle, our method allows for the computation of the Hartree potential from the electron density at the cost of a single forward and inverse transform pair. Also, the use of the Green's function ensures that the appropriate decay of the electrostatic potentials is correctly handled, without having to use large computational domains or non-trivial boundary conditions.

Computation of the occupied eigenspace of the discretized Kohn-Sham eigenvalue problem is the most computationally demanding step in a typical self consistent field calculation. Accordingly, a number of strategies have been devised over the years for an efficient solution of this problem through iterative diagonalization methods \citep{Teter_Payne_Allan_2, Kresse_abinitio_iterative, various_eigensolvers, Hutter_abinitio_MD}. We have adopted the 
Locally Optimal Block Preconditioned Conjugate Gradient (LOBPCG) algorithm \citep{LOBPCG_1} for this purpose in our code. This robust method has been implemented with success in other state-of-the-art Kohn-Sham codes \citep{ABINIT_LOBPCG, Octopus_LOBPCG}. With the aid of a simple diagonal preconditioner (described later) we have found it to work well for a variety of systems. For relatively large system sizes however, especially while running under distributed memory environments, LOBPCG-like methods suffer from the need to repeatedly orthogonalize the computed eigenstates. For dealing with such situations, we have adopted a highly efficient Chebyshev polynomial filtered subspace iteration algorithm \citep{Serial_Chebyshev, Parallel_Chebyshev} which avoids explicit diagonalization and minimizes orthonormalizations. For these large scale calculations therefore, LOBPCG is used only in the first SCF step (to generate a good guess for the occupied eigenspace) while Chebyshev polynomial filtering is used exclusively on subsequent SCF steps.

Spectral methods like the plane-wave method and the method presented here\footnote{{More correctly, both the plane-wave method and the method presented here should be referred to as pseudospectral since they rely on interpolatory transforms.}} are susceptible to suffer from scalability issues while running under distributed memory environments, since the global nature of the basis functions involved tends to induce communication between the processing elements. To ameliorate this difficulty, we have adopted a two-level parallelization scheme over electronic states as well as physical space, much in the spirit of some large scale plane-wave codes \citep{Gygi_2D_parallel}. This effectively reduces global communications to an order of the square root of the number of processes (instead of the total number of processes) and it has resulted in speed-critical portions of our code scaling well up to 512 processing units.

These strategies and methods combine to give an unusually efficient and accurate method, as seen in the examples presented in Section \ref{sec:examples}. We employ norm conserving {ab initio} pseudopotentials for most of our calculations. We first study the convergence properties of our basis set and demonstrate faster than polynomial rates of convergence (i.e., spectral convergence). Then, starting from light atoms, small molecules and clusters (metallic and non metallic), we visit various examples involving organic molecules, fullerenes and large face centered cubic (FCC) aluminum clusters. The largest example system considered here contains 1688 aluminum atoms (over 5000 electrons). Timing comparisons reveal that our method outperforms competing finite element and plane-wave method codes in benchmark calculations involving aluminum clusters. {Also, comparison against well converged plane-wave method results allow us to show that extremely high accuracies in ground state energy calculations (of the order of $10^{-6}$ Hartrees per atom) are routinely achievable by our code.} By visiting an example that involves the calculation of electrostatic multipole moments, we demonstrate that systematic convergence properties of our basis set allow for easy and accurate calculations of important physical properties.

The rest of the paper is organized as follows. Section \ref{sec:formulation} describes the formulation of our method while Section \ref{sec:implementation_details} describes various implementation aspects. Section \ref{sec:examples} presents the example problems solved using our method and compares our results with the literature to assess the efficacy of our method. Section \ref{sec:conclusions} concludes the paper with a summary and comments on future directions.
\section{Formulation}
\label{sec:formulation}
We describe the KS-DFT energy functional and the associated Kohn-Sham equations in this section. {We outline the key aspects of discretization of the Kohn-Sham equations using our spectral basis and also describe our approach to computation of the various terms that appear in these equations.} The atomic unit system with $m_e=1,e=1,\hbar=1,\frac{1}{4\pi\epsilon_0}=1$, is chosen for the rest of the work, unless otherwise mentioned.
\subsection{The Kohn-Sham eigenvalue problem}
\label{subsec:KS_eigenvalue}
We consider a system consisting of $N_e$ electrons moving in the fields produced by $M$ nuclei. We assume that the nuclei have charges $(z_1, \ldots, z_M)$ and that they are clamped to the positions $(\bfx_1,\ldots,\bfx_M) \in \rz^{3M}$. For the sake of simplicity, we consider a system in which spin polarization effects are absent and we consider $N_e$ to be even. The extension of the present work to study spin-polarized systems is straight-forward.\footnote{Some of our example calculations presented in section \ref{sec:examples} do use spin-polarization.}

In line with the Born-Oppenheimer approximation \citep{LeBris_ReviewBook}, the Kohn-Sham model \citep{KohnSham_DFT,LeBris_ReviewBook} computes the total energy of this system (denoted here as $E^{\text{KS}}_{N_e, M}$) at absolute zero, by splitting it into an electronic part (denoted here as $\calE^{\text{KS}}_{N_e, M}$) and a nucleus-nucleus interaction part:
\begin{align}
E^{\text{KS}}_{N_e, M} = \calE^{\text{KS}}_{N_e} + \sum_{1\leq k < l \leq M}
\frac{z_kz_l}{\abs{\bfx_k-\bfx_l}}\;.
\label{KS_total_Energy}
\end{align}
The electronic part of the energy in the Kohn-Sham model is computed in terms of orbitals; i.e., an $N_e / 2$ tuple of complex valued scalar fields $\{\phi_i\}_{i = 1}^{N_e / 2}$, as follows:\footnote{In the equations that follow, $\mathsf{L}^2(\rz^3)$ is used to denote the usual space of square integrable functions on $\rz^3$ while $\mathsf{H}^1(\rz^3)$ denotes the Sobolev space of functions in $\mathsf{L}^2(\rz^3)$ whose first order weak derivatives also lie in $\mathsf{L}^2(\rz^3)$.}
\begin{align}
\label{KS_model}
\calE^{\text{KS}}_{N_e} &=\,\displaystyle \inf_{\substack{{\phi_i \in \mathsf{H}^1(\rz^3),}\\{\innprod{\phi_i}{\phi_j}{\mathsf{L}^2(\rz^3)}=\delta_{ij}}}}
{\biggl\{\half \sum_{i=1}^{N_e/2} \int_{\rz^3}\!f_i\abs{\nabla \phi_i}^2
+\int_{\rz^3}\!\rho V_{\text{nu}} + \half\int_{\rz^3}\int_{\rz^3}\!\frac{\rho(\bfx)\rho(\bfy)}{\abs{\bfx-\bfy}}\;d\bfx\,d\bfy}
+E_{\text{xc}}(\rho)\biggr\}\\
\label{density_expression}
&\textrm{with the electron density,}\quad\rho(\bfx)=2 \sum_{i=1}^{N_e / 2} f_i\abs{\phi_i(\bfx)}^2\;.
\end{align}
The scalars $f_i$ denote orbital occupations and have values $0 \leq f_i \leq 1$. {These values need to be specified apriori or they can be computed as part of the solution (using thermalization for instance, see Section \ref{subsubsec:mixing_smearing})}. The factor of two in eq.~\ref{density_expression} above, is due to the assumption of dealing with a spin-unpolarized system, as a consequence of which, each orbital is doubly occupied. The orthonormalization constraint on the orbitals implies that the electron density $\rho$ satisfies the normalization condition:
\begin{align}
\label{density_normalized}
\int_{\rz^3}\!\rho = N_e\;.
\end{align} 
The first term in eq.~\ref{KS_model} (involving the gradient of the orbitals) models the kinetic energy of the electrons. The second term models the interaction of the nuclei with the electrons. The nuclear potential appearing in that term is given as:
\begin{align}
\label{Vnu_original}
V_{\text{nu}}(\bfx) = - \sum_{k=1}^{M}\frac{z_k}{\abs{\bfx-\bfx_k}}\;.
\end{align}
{In practice, the Coulombic singularities present in eq.~\ref{Vnu_original} cause problems with efficient numerical solution of the equations and  spectral methods (like the plane-wave method and the present one) are particularly affected due to appearance of Gibbs phenomenon \citep{Folland_Real}. The pseudopotential approximation\footnote{{In particular, ab initio pseudopotentials provide a well defined recipe of carrying out this smoothening procedure such that the energy and length scales of the problem are dictated by the chemically relevant electronic states. See \citep{Troullier_Martins_pseudo} for more details.}} allows these issues to be mitigated by smoothening out these singularities \citep{LeBris_ReviewBook, Martin_ES, Troullier_Martins_pseudo}.} The bulk of the present work is devoted to pseudopotential calculations.

{The third term in eq.~\ref{KS_model} represents the mutual electrostatic repulsion of the electrons (Hartree energy). Finally, the term $E_{\text{xc}}(\rho)$ is the exchange correlation energy. We adopt here the widely used Local Density Approximation (LDA) \citep{Parr_Yang, KohnSham_DFT} of this term by using the parametrization presented in \citep{perdew_zunger, ceperley_alder}. An extension of our method to density gradient corrected functionals \citep{GGA_made_simple_perdew} poses no particular difficulty.}

The Euler-Lagrange equations of the minimization problem (eq.~\ref{KS_model}) are the celebrated Kohn-Sham equations \citep{KohnSham_DFT}, {which, with the definition of the electron density introduced in eq.~\ref{density_expression}} can be written as follows:
\begin{align}
\label{KS_equations}
K(\rho)\,\phi_i&=\lambda_i\,\phi_i\;;\;\innprod{\phi_i}{\phi_j}{\mathsf{L}^2(\rz^3)}=\delta_{ij}\;,\\
\label{KS_explain1}
\text{with}\quad
K(\rho)&=-\half \Delta + V_{\text{nu}} + 
\int_{\rz^3}\!\frac{\rho(\bfy)}{\abs{\bfx-\bfy}}\;d\bfy+V_{\text{xc}}(\rho)\;,\\
\text{where}\quad
V_{\text{xc}}(\rho)&=\pd{E_{\text{xc}}(\rho)}{\rho}\;.
\label{KS_explain2}
\end{align}
The $\lambda_i$ that appear in eq.~\ref{KS_equations} are the Lagrange multipliers of the orthonormality constraints on the orbitals. They are taken to be the lowest $N_e / 2$ eigenvalues of the Kohn-Sham operator $K(\rho)$. 

The usual method of solution of the Kohn-Sham equations is by a Self-Consistent Field (SCF) approach \citep{KohnSham_DFT,
Martin_ES, LeBris_ReviewBook}. In practice, a variety of mixing schemes are employed to accelerate convergence of the SCF iterations \citep{Martin_ES, Dederichs_Zeller_SCF}.
\subsection{Problem set-up and discretization}
\label{subsec:discretization}
Let $\calB_R$ denote the sphere of radius $R$ centered at the origin. For the purpose of this work, we will restrict the physical domain to $\calB_R$ and the cluster / molecular system of interest will be embedded within this spherical region. We will apply Dirichlet boundary conditions to the electron density on the surface of the sphere in accordance with the well-known spatial exponential decay of the electron density \citep{wavefunc_decay1, wavefunc_decay2}. The relation between the electron density and the wavefunctions (eq.~\ref{density_expression}) automatically implies that the Dirichlet boundary conditions apply to the wavefunctions\footnote{Application of Dirichlet boundary conditions to the Kohn-Sham wavefunctions has been considered earlier in the context of real-space methods \citep{Gavini_Kohn_Sham, Chelikowsky_Saad_1, Chelikowsky_Saad_2}} as well. {We do not enforce specific boundary conditions on the various potential terms. These are applied implicitly based on the method of computation: in case of the Hartree term for instance, our method of calculation automatically ensures that the right kind of decay of that term is obtained (Section \ref{subsubsec:hartree_potential}).}
\subsubsection{Basis set}
\label{subsubsec:basis_set}
The particular choice of a spherical domain allows for the Laplacian eigenfunctions in this domain to be represented analytically in spherical coordinates.\footnote{In our notation for spherical coordinates, we denote $r\in [0,R]$ as the radial coordinate, $\vartheta \in [0,\pi]$ as the polar angle and $\varphi\in[0,2\pi]$ as the azimuthal angle. The Cartesian coordinates $(x,y,z)$ are obtained as $x = r\sin{\vartheta}\cos\varphi,\;y = r\sin{\vartheta}\sin\varphi,\; z = r\cos{\vartheta}$.} Specifically, we consider the $\Lpspc{2}{}{\calB_R}$ orthonormal eigenfunctions of the Laplacian operator in the spherical domain and we impose Dirichlet boundary conditions on the surface of the domain. In this setup, a simple separation of variables calculation shows that the eigenfunctions of the Laplacian which are regular at the origin are expressible in terms of spherical Bessel functions of the first kind and spherical harmonics \citep[see e.g.][for details of this calculation]{My_MS_Thesis}. Letting $(l,m,n) \in \Gamma_{\infty}$ with:
\begin{align}
{\Gamma}_{\infty} = \bigg\{(l,m,n):l\in\{0,1,\ldots\},m\in\{-l,\ldots,l\},n\in\{0,1,\ldots\}\bigg\}\;,
\end{align}
the Laplacian eigenfunctions take the form:
\begin{align}
\label{eigfunction_1}
{F}_{l,m,n}(r,\vartheta,\varphi) &= \calR_{l,n}(r)\;\calY^{m}_l(\vartheta,\varphi)\;, \\
\intertext{with the radial part being the spherical Bessel functions of the first kind:}
\label{spherical_bessel_1}
\calR_{l,n}(r) &= \displaystyle\frac{1}{R J_{l+\thrbyto}\bigl(b^{n}_{l+\half}\bigr)}
\sqrt{\frac{2}{r}}\displaystyle
J_{l+\half}\biggl(\displaystyle\frac{b^{n}_{l+\half}}{R}r\biggr)\;,\\
\intertext{and the angular part being the spherical harmonics:}
\calY^{m}_l(\vartheta,\varphi) &=
c_{l,m}\,\calP_l^m (\cos{\vartheta}) \, e^{i m \varphi }\;,
\label{spherical_harmonics_1}
\text{with}\;c_{l,m} = \sqrt{\frac{(2l+1)}{4\pi}\frac{(l-m)!}{(l+m)!}}\,\;.
\end{align}
In eq.~\ref{spherical_bessel_1}, $\displaystyle J_{l+\half}(\cdot)$ denotes the (ordinary) Bessel function of the first kind of order $(l+\half)$, while $\displaystyle b^{n}_{l+\half}$ denotes its $(n + 1)^{\text{th}}$ root. Thus, $\calR_{l,n}(r)$ attains a value of zero $(n + 1)$ times in the interval $[0,R]$. In eq.~\ref{spherical_harmonics_1}, $\calP_l^m(\cdot)$ denotes the associated Legendre polynomial of degree $l$ and order $m$.
The eigenvalue associated with the eigenfunction ${F}_{l,m,n}$ is given by:
\begin{align}
\label{laplacian_eigenvalue}
{\Lambda}_{l,m,n} = \biggl(\frac{b^{n}_{l+\half}}{R}\biggr)^2\;.
\end{align}
Since the Laplacian is a self-adjoint operator with a compact resolvent, the infinite collection of eigenfunctions $\calE_{{\Gamma}_{\infty}} = \big\{{F}_{l,m,n}: {(l,m,n) \in {\Gamma_{\infty}}}\big\}$ form an orthonormal basis of $\Lpspc{2}{}{\calB_R}$ \citep{Evans_PDE, Kato}. Further, elliptic-regularity results \citep{Evans_PDE} imply that each basis function ${F}_{l,m,n}$ is smooth. We now choose a finite subset of $\calE_{\tilde{\Gamma}}$ as our basis set.

We fix $\calL,\calN \in \nz$ (henceforth referred to as the angular and radial cutoff, respectively), and form $\Gamma \subset {\Gamma_{\infty}} $ by restricting\footnote{For each $l$, $m$ is allowed to vary in $\{-l,\ldots,l\}$ as before.}\footnote{{For the purpose of this article, we have used a uniform basis set in which the maximum value of $n = (\calN -1)$ for \emph{every} $l$. We are aware however, that a non-uniform basis set in which the maximum value of $n$ is allowed to vary with $l$ allows more flexibility and can lead to a number of desirable effects. For instance, this setting allows (as in plane-wave codes), the introduction of a kinetic energy cut-off in which the basis set only includes spherical waves whose kinetic energies lie below a specified threshold. This results in a basis set which is optimal in the sense of the Sobolev $\mathsf{H}^1$ norm. Also, we are aware that a uniform basis set can sometimes face a loss of approximation power for large radial distances, although we have not encountered any serious issues from this effect in this work. A non-uniform basis set however, can be made to correct for this issue automatically. More details of this methodology is the scope of future work.}} $l\in\{0,\ldots,\calL-1\}$ and $n\in\{0,\ldots,\calN-1\}$. Given any function $f \in \Lpspc{2}{}{\calB_R}$, for the purpose of numerical discretization, we approximate it using the functions in $\calE_{{\Gamma}} = \big\{{F}_{l,m,n}: {(l,m,n) \in {\Gamma}}\big\}$ as:
\begin{align}
\label{basis_approx}
f =\!\sum_{(l,m,n) \in {\Gamma}}\!\hat{f}_{l,m,n}\; {F}_{l,m,n}\;.
\end{align}
We may observe that the span of the functions in $\calE_{\tilde{\Gamma}}$ form a linear subspace of $\Lpspc{2}{}{\calB_R}$ of dimension $\mathit{{d}} = \calL^2 \calN$. The expansion coefficients can be obtained from orthonormality of the basis functions by:
\begin{align}
\label{flmn_def}
\hat{f}_{l,m,n} = \innprod{f}{{F}_{l,m,n}}{\Lpspc{2}{}{\calB_R}} =  \int_{\calB_R}\!f\;\overline{{F}_{l,m,n}}\;,
\end{align}
and the collection of expansion coefficients $\big\{\hat{f}_{l,m,n}:(l,m,n) \in {\Gamma} \big\}$ will often be interpreted interchangeably with vectors in $\cz^\mathit{d}$.
If the function $f$ is real valued, as it is for example, in case of the electron density, the expansion coefficients obey the additional condition\footnote{If the Condon-Shortley phase \citep{CS_phase_encyclopedia}  is included, this becomes $\hat{f}_{l,-m,n} = (-1)^m \overline{\hat{f}_{l,m,n}}$. We do not make use of the Condon-Shortley phase in this work.} $\hat{f}_{l,-m,n} = \overline{\hat{f}_{l,m,n}}$.

\subsubsection{Basis transforms}
\label{subsubsec:basis_transform}
In order to perform the quadratures required for evaluation of the expansion coefficients via eq.~\ref{flmn_def} we introduce
a discretization of the domain $B \subset \calB_R$. Akin to the terminology used in the plane-wave literature, we will often refer to the representation of a given function in terms of its expansion coefficients as the \textit{reciprocal space representation} while the representation of the same function on the grid points in $B$ will be referred to as the \textit{real space representation}. The operations that allow us to  switch between these two representations will be referred to as \textit{basis transforms}. 

The specific choice of the grid points is made as follows. Let $N_r, N_{\vartheta}$ and $N_{\varphi}$ denote the number of discretization points in the radial, polar and azimuthal directions respectively. These quantities are dependent on the radial and angular cutoffs and are chosen in accordance with constraints of the sampling theorem. We discretize the unit sphere by choosing $N_{\vartheta}$ Gauss quadrature points in $\cos(\vartheta)$ over the interval $[-1,1]$ and $N_{\varphi}$ equally spaced points in $\varphi$ over the interval $[0,2\pi]$. In the radial direction, we choose $N_r$ Gauss-Jacobi quadrature nodes \citep{Gauss_Jacobi_Quad} associated with the quadrature weight of $r^2$ over the interval $[0,R]$. The set $B$ is now taken to be a Cartesian product of the radial quadrature points and the unit sphere discretization points. This allows a separation of variables in the angular and radial directions while carrying out the basis transforms, thereby reducing computational complexity.

Given the real space representation $\tilde{f}:B\to\cz$, we obtain the reciprocal space representation by first computing  spherical harmonic transforms holding the radial variable fixed:
\begin{align}
\nonumber
A(r;\,l,m) &= \int_{0}^{2\pi}\int_{0}^{\pi}f(r;\vartheta, \varphi)\;\overline{\calY^{m}_l(\vartheta,\varphi)}\,\sin(\vartheta)\;d\vartheta\,d\varphi\;,\\
\label{Ar_lm}
&= c_{l,m}\int_{-1}^{1}\calP_l^m(t)\bigg[\int_{0}^{2\pi}\!f(r;\,\cos^{-1}(t), \varphi)\,e^{-i m \varphi }\,d\varphi\bigg]\,dt\,,
\end{align}
and then performing radial quadratures using the quadrature nodes $\{r_{k_r}\}_{k_r=1}^{N_r}$ and corresponding weights $\{w_{k_r}\}_{k_r=1}^{N_r}$:
\begin{align}
\label{Arlm_to_coeff}
\hat{f}_{l,m,n} = \int_{0}^{R}\!A(r;\,l,m)\,\calR_{l,n}(r)\,r^2\,dr  \approx \sum_{k_r = 1}^{N_r}w_{k_r}A(r_{k_r};\,l,m)\,\calR_{l,n}(r_{k_r})\;.
\end{align}
The spherical harmonic transform as expressed in eq.~\ref{Ar_lm} itself consists of two steps: first holding $\vartheta$ fixed, the Fast Fourier Transform (FFT)\citep{FFT_Cooley_Tukey} is used to evaluate the inner integral involving $\varphi$. Subsequently, a quadrature in $t = \cos(\vartheta)$ is carried out on the result to evaluate the outer integral. 

Similarly, given the reciprocal space representation $\hat{f}:\Gamma \to \cz$,  the inverse transform can be carried out by first computing:
\begin{align}
\label{eq:G_lmr}
{G}(l,m;r_{k_r}) = \sum_{n = 0}^{\calN-1}\hat{f}_{l,m,n}\,\calR_{l,n}(r_{k_r})\,,
\end{align}
while holding $l$ and $m$ fixed and then, for each radial grid node $r_{k_r}$, performing inverse spherical harmonics transforms (using inverse FFTs and dot products as in \citep{shtns}).

The basis transforms as described above, have a time complexity of $O(\calL^3\calN + \calL^2\calN^2)$ in terms of the angular and radial cutoffs.\footnote{A naive implementation of the transforms, that is, one that does not employ the separation of variables structure, would have a time complexity of $O(\calL^4\calN^2)$ in terms of the angular and radial cutoffs.}\footnote{Using more sophisticated techniques for carrying out the associated Legendre polynomial transforms \citep{Mohlenkamp_SHT,Driscoll_Healey_SHT}, this can be reduced to $O(\calL^2(\log{\calL})^2\calN + \calL^2\calN^2)$.} As far as practical implementation is concerned, the use of Gauss quadrature points as well as various numerical and implementation level optimizations \citep[see][for example]{shtns} can be used to ensure that the prefactor for this asymptotic estimate is rather low. This allowed us to carry out basis transforms routinely and efficiently even with basis sets containing millions of basis functions in our code.
\subsubsection{Set up of matrix eigenvalue problem}
\label{subsubsec:matrix_eigprob}
Within the self-consistent field iterations, the governing equations (eq.~\ref{KS_equations} - \ref{KS_explain2}) posed in the spherical domain take the form of the following linearized eigenvalue problem with an effective potential $V^{\text{eff}}$:
\begin{align}
\label{lin_eigen}
\Bigl(-\half \Delta + V^{\text{eff}}\Bigr)\phi_i =& \;\lambda_i\,\phi_i\;\text{for} \;i=1,\ldots,N_e/2\quad,\\
\phi_i =& \;0\;\text{on}\;\partial \calB_R\;,
\label{eq:lin_eigen_BC}
\end{align}
and the effective potential at a point $\bfx \in \calB_R$ is given as:
\begin{align}
V^{\text{eff}}(\bfx) = V_{\text{xc}}(\rho(\bfx)) + 
\int_{\calB_R}\!\frac{\rho(\bfy)}{\abs{\bfx-\bfy}}\,d\bfy + V_{\text{nu}}(\bfx)\quad.
\end{align}
We choose to ignore any non-local contributions to the ionic pseudopotentials at this point. The specific treatment of these non-local terms is discussed later in section \ref{susubbsec:pseudo_pot_terms}.

To discretize eq.~\ref{lin_eigen} we set: 
\begin{align}
\label{phi_expansion}
\displaystyle\phi_i =& \sum_{(l,m,n)\in \Gamma}\!\hat{\phi}^{i}_{l,m,n}\;F_{l,m,n}\;,
\intertext{noting that this ensures that the Dirichlet boundary conditions on the wavefunctions are satisfied automatically. This gives us:}
\half\sum_{\Gamma}\!\hat{\phi}^{i}_{l,m,n}\;\Lambda_{l,m,n}\;F_{l,m,n}\,+& \;V^{\text{eff}}\sum_{\Gamma}\hat{\phi}^{i}_{l,m,n}\;
F_{l,m,n}
= \;\lambda_i\;\sum_{\Gamma}\hat{\phi}^{i}_{l,m,n}\;F_{l,m,n}\;.
\label{temp_gov_eqn_in_basis_set}
\end{align}
Now, if the expansion coefficients of $V^{\text{eff}}$ are known as $\{\hat{V}^{\text{eff}}_{\tilde{l},\tilde{m},\tilde{n}}: (\tilde{l},\tilde{m},\tilde{n}) \in \Gamma\}$, we may substitute the expansion of $V^{\text{eff}}$ into eq.~\ref{temp_gov_eqn_in_basis_set} 
to get:
\begin{align}
\nonumber
\displaystyle\half\sum_{\Gamma}\!\hat{\phi}^{i}_{l,m,n}\;\Lambda_{l,m,n}\;F_{l,m,n}
&+\sum_{\Gamma}\sum_{\Gamma}\,\hat{\phi}^{i}_{l,m,n}\,\hat{V}^{\text{eff}}_{\tilde{l},\tilde{m},\tilde{n}}\;
F_{\tilde{l},\tilde{m},\tilde{n}}\;
F_{l,m,n}\\
&=\lambda_i \sum_{\Gamma}\;\hat{\phi}^{i}_{l,m,n}F_{l,m,n}\;.
\label{gov_eqn_in_basis_set_with_potential}
\end{align}
We now take the inner product of this equation with $F_{l',m',n'}$ and use orthonormality of the basis functions to obtain the following system of linear equations for 
$\hat{\phi}^{i}_{l',m',n'}$, with $\,(l',m',n')\in \Gamma$ :
\begin{align}
\displaystyle\half\Lambda_{l',m',n'}\,\hat{\phi}^{i}_{l',m',n'}+&\sum_{\Gamma}\sum_{\Gamma}
\calW^{(l',m',n')}_{(l,m,n)\,,\,(\tilde{l},\tilde{m},\tilde{n})}\,\hat{V}^{\text{eff}}_{\tilde{l},\tilde{m},\tilde{n}}\,\hat{\phi}^{i}_{l,m,n}
=\lambda_i\,\hat{\phi}^{i}_{l',m',n'}\;,
\label{discretized_eqn_with_W_coefficient}
\end{align}
where $\displaystyle \calW^{(l',m',n')}_{(l,m,n)\,,\,(\tilde{l},\tilde{m},\tilde{n})}$ denote the coupling coefficients of the basis set, i.e.,
\begin{align}
\calW^{(l',m',n')}_{(l,m,n)\,,\,(\tilde{l},\tilde{m},\tilde{n})} =\biginnprod{F_{\tilde{l},\tilde{m},\tilde{n}}\,F_{l,m,n}\,}{F_{l',m',n'}}{\Lpspc{2}{}{\calB_R}}\;.
\label{what_is_W}
\end{align}
It is possible to express these coupling coefficients in terms of Wigner {3-j} symbols \citep{Messiah_3j_symbol} and the integral of the product of three spherical basis functions taken together \citep{My_MS_Thesis}. Such an expression allows us to see that the coupling coefficients are non-zero only when $\abs{l-\tilde{l}}\leq l'\leq l+\tilde{l}$, $m + m' + \tilde{m} = 0$ and $l+l'+\tilde{l}$ is odd.

To recognize the finite dimensional linear eigenvalue problem expressed in eq.~\ref{discretized_eqn_with_W_coefficient}, we may introduce an indexing map $\calI:\Gamma\to\{1,2,\ldots,\mathit{d}\}$ and let $\calJ$ denote its inverse.\footnote{A simple indexing map is, for instance, 
$(l,m,n)\mapsto (l^2 + l + m) * \calN + (n + 1)$.} We rewrite eq.~\ref{discretized_eqn_with_W_coefficient} using the map $\calJ$ to obtain a matrix problem of the form:
\begin{align}
\label{hphi_lambdaphi}
{\mathbf{H}\,\mathbf{X}} = {\mathbf{X}\,\mathfrak{D}}\;,
\end{align}
where $\mathbf{H} \in \cz^{\mathit{d} \times \mathit{d}}, \mathbf{X}\in \cz^{\mathit{d} \times (N_e/2)}$ and $\mathfrak{D} \in \rz^{(N_e/2) \times (N_e/2)}$. {We note that the matrix $\mathbf{H}$ is Hermitian, while the matrix $\mathbf{X}$ is unitary in the sense that $\mathbf{X}^{\dagger}\,\mathbf{X}$ is an identity matrix of dimension $(N_e/2)\times (N_e/2)$.} Denoting $\delta_{\alpha,\beta}$ as the Kronecker delta, we see that  matrices $\mathbf{H}, \mathbf{X}$ and $\mathfrak{D}$ have entries of the following form (in terms of the indexing map):
\begin{align}
\label{Hij_def}
\mathbf{H}_{\alpha,\beta} &= \half\delta_{\alpha,\beta}\, \Lambda_{\calJ(\alpha)}+\sum_{(\tilde{l},\tilde{m},\tilde{n})\in \Gamma}\!\hat{V}^{\text{eff}}_{\tilde{l},\tilde{m},\tilde{n}}\;\calW^{\calJ(\alpha)}_{\calJ(\beta)\,,\,(\tilde{l},\tilde{m},\tilde{n})}\,,\\
\mathbf{X}_{\alpha,\beta} &= \hat{\phi}^{\beta}_{\calJ(\alpha)}\;\text{and}\;
\mathfrak{D}_{\alpha,\beta} = \delta_{\alpha,\beta}\,\lambda_{\calJ(\beta)}\,,
\end{align}
with $\alpha, \beta$ varying within relevant matrix dimensions.

As we mentioned earlier in this article, setting up of the matrix eigenvalue problem followed by direct diagonalization are both expensive operations, although this approach seems to have been adopted by earlier work involving spherical basis functions \citep[see e.g.][]{solid_state_finite}. From eq.~\ref{Hij_def}, for instance, we can see that the matrix $\mathbf{H}$ is dense and therefore, the asymptotic computational complexity of the matrix setup is of cubic order in the total number of basis functions. Direct diagonalization of the Hamiltonian, even by the most efficient algorithms available today \citep[see e.g.][]{MRRR}, will have the same cubic computational complexity in the number of basis functions due to the necessity of reducing the matrix to tridiagonal form. In addition, the memory storage requirement of the full Hamiltonian matrix scales as the square of the number of basis functions and therefore this  becomes an additional constraint while trying to deal with even moderate sized systems.
\subsubsection{Set up of matrix-vector products}
\label{subsubsec:matrix_vector_prods}
To avoid the computational difficulties associated with direct diagonalization, we choose to employ matrix-free iterative methods for computing the occupied eigenspace of the Hamiltonian matrix \citep[see][for a detailed discussion of this class of methods]{Saad_large_eigenvalue_book}. As the name suggests, these methods do not need access to the individual matrix entries but only require matrix vector products to be specified. To see how the product of a given vector with the Hamiltonian matrix may be calculated efficiently, without explicit involvement of the coupling coefficients, we proceed as follows:

Let $\mathbf{Y} \in \cz^{\mathit{d}}$ be a given vector and let $\mathbf{Z} \in \cz^{d}$ be the result of the matrix vector product, that is,  $\mathbf{Z} = {\mathbf{H}\,\mathbf{Y}}$. In terms of components we have :
\begin{align}
\nonumber
\mathbf{Z}_{\alpha} &= \sum_{\beta = 1}^{\mathit{d}} \mathbf{H}_{\alpha,\beta}\,\mathbf{Y}_{\beta}\, =  \sum_{\beta = 1}^{\mathit{d}}\bigg(\half\delta_{\alpha,\beta}\, \Lambda_{\calJ(\alpha)}+\sum_{(\tilde{l},\tilde{m},\tilde{n})\in \Gamma}\!\hat{V}^{\text{eff}}_{\tilde{l},\tilde{m},\tilde{n}}\;\calW^{\calJ(\alpha)}_{\calJ(\beta)\,,\,(\tilde{l},\tilde{m},\tilde{n})}\bigg)\mathbf{Y}_{\beta}\\
&= \half\Lambda_{\calJ(\alpha)}\mathbf{Y}_{\alpha} + \sum_{\beta = 1}^{\mathit{d}}\!\bigg(\sum_{(\tilde{l},\tilde{m},\tilde{n})\in \Gamma}\!\hat{V}^{\text{eff}}_{\tilde{l},\tilde{m},\tilde{n}}\;\calW^{\calJ(\alpha)}_{\calJ(\beta)\,,\,(\tilde{l},\tilde{m},\tilde{n})}\bigg)\mathbf{Y}_{\beta}\,.
\intertext{The second term, by making use of eq.~\ref{what_is_W} and the linearity of the inner product, can be written as:}
\nonumber
&=\sum_{\beta = 1}^{\mathit{d}}\sum_{(\tilde{l},\tilde{m},\tilde{n})\in \Gamma}\!\hat{V}^{\text{eff}}_{\tilde{l},\tilde{m},\tilde{n}}\,\mathbf{Y}_{\beta}\,\biginnprod{F_{\tilde{l},\tilde{m},\tilde{n}}\,F_{\calJ(\beta)}\,}{F_{\calJ(\alpha)}}{\Lpspc{2}{}{\calB_R}} \\\nonumber
&= \biginnprod{\bigg(\sum_{(\tilde{l},\tilde{m},\tilde{n})\in \Gamma}\!\hat{V}^{\text{eff}}_{\tilde{l},\tilde{m},\tilde{n}}F_{\tilde{l},\tilde{m},\tilde{n}}\bigg)\,\bigg(\sum_{\beta = 1}^{\mathit{d}}\mathbf{Y}_{\beta} F_{\calJ(\beta)}\bigg)\,}{F_{\calJ(\alpha)}}{\Lpspc{2}{}{\calB_R}}\,.
\intertext{We recognize the first term in parentheses as the expansion of the effective potential in our basis set, and therefore, the final expression for $\mathbf{Z}_{\alpha}$ becomes:}
\label{matvec_final}
\mathbf{Z}_{\alpha} &= \half\Lambda_{\calJ(\alpha)}\mathbf{Y}_{\alpha} + \biginnprod{{V}^{\text{eff}}\,\bigg(\sum_{\beta = 1}^{\mathit{d}}\mathbf{Y}_{\beta} F_{\calJ(\beta)}\bigg)\,}{F_{\calJ(\alpha)}}{\Lpspc{2}{}{\calB_R}}\,.
\end{align}
The above equation suggests that the computation of the Hamiltonian times vector product is to be carried out in two stages and the results from these stages should be summed. First, the action of the kinetic energy operator is to be carried out in reciprocal space because of the diagonal structure of that operator in that space. In the second stage, the action of the operator expressing the action of the effective potential is to be computed. This operator however, is diagonal in real space. Thus, given the vector $\mathbf{Y}$, we imagine its components $\{\mathbf{Y}_{\alpha}\}_{\alpha = 1}^{\mathit{d}}$ to represent expansion coefficients and we perform an inverse basis transform to obtain a function $Y$ defined on the gridpoints in $B$. We then perform a pointwise multiplication of $Y$ with the effective potential (also defined over $B$) and we finally compute a forward basis transform of the product $({V}^{\text{eff}}\cdot Y)$ to obtain the result of the second stage. 

The principal computational cost of the process described above arises from a pair of basis transforms and therefore the associated time and space complexities are $O(\calL^3\calN + \calL^2\calN^2)$ and $O(\calL^2\calN)$ respectively. In contrast, a direct matrix vector product, once the Hamiltonian matrix has been set up, would involve $O(\calL^4\calN^2)$ complexity both in memory and speed.

Note that the discussion above does not take into account the role played by non-local pseudopotentials. When such non-local terms are present, the Hamiltonian as described above, has an additional projection operator term that acts on the given vector $\mathbf{Y}$. The action of this additional operator on the given vector can be directly computed as a dense linear algebra operation.
\subsubsection{Computation of the electron density}
\label{subsubsec:electronic_density}
Using the expansion of the wavefunctions (eq.~\ref{phi_expansion}) as well as the expression in eq.~\ref{density_expression}, we see that the electron density admits expansion coefficients of the form:
\begin{align}
\label{tau_lmn_expression}
\tau_{l',m',n'}=
2\sum_{j=1}^{N_e/2}\sum_{\Gamma}\sum_{\Gamma}\calW^{(l',m',n')}_{(l,m,n)\,,\,
(\tilde{l},\tilde{m},\tilde{n})}\,\hat{\phi}^{i}_{\tilde{l},\tilde{m},\tilde{n}}\,
\overline{\hat{\phi}^{i}_{l,-m,n}}\;.
\end{align}
Two comments are in order at this stage. First, since the coupling coefficients are non-zero only when $\abs{l-\tilde{l}}\leq l'\leq l+\tilde{l}$ and we have $0 \leq l,\tilde{l} \leq (\calL - 1)$, we see that $\tau_{l',m',n'}$ may have non-zero values for all $l'$ satisfying $0 \leq l' \leq 2(\calL - 1)$. Thus, due to the quadratic non-linearity in eq.~\ref{density_expression}, the electron density needs to be represented using a basis set that is larger than the one used to represent the wavefunctions. A similar situation also arises in the context of the plane-wave method, where sometimes, compared to the wavefunctions, the electron density expansion employs a larger energy cutoff \citep{Large_scale_plane_wave}. Often however, plane wave codes allow the so called \textit{dualing approximation} to be engaged, as a result of which, the electron density is expanded using the same basis set as the wavefunctions \citep{Large_scale_plane_wave}. In the same vein, our implementation allows a larger basis set (as well as a correspondingly denser real space grid) for the electron density to be employed based on user choice.

The second comment is that, while eq.~\ref{tau_lmn_expression} illustrates the basis requirements while computing the electron density, a direct application of this expression to compute the electron density expansion coefficients is prohibitively expensive. The time complexity of the operations involved in that expression, in terms of the basis set size and the number of electrons involved, is $O(N_e\mathit{d}^3)$. Instead, starting from the expansion coefficients of the wavefunctions, we may compute the real space representations of the wavefunctions using inverse basis transforms. We may then use eq.~\ref{density_expression} to compute the electron density at the grid points in $B$ and finally apply a forward basis transform to obtain the required expansion coefficients of the density. This method  results in the reduced time complexity of $O(N_e(\calL^3\calN + \calL^2\calN^2))$ and in practice, it turns out to be much more efficient.

The methods for computation of the electron density (as described above) and matrix vector products (described earlier in section \ref{subsubsec:matrix_vector_prods}) are both based on the general idea of evaluating convolution sums through efficient transforms \citep{Spectral_Methods_book}. While this technique seems to have been used quite commonly in fluid dynamics simulations both for spherical and periodic domains \citep{orszag_1, orszag_2}, it's application to Kohn-Sham density functional theory seems to have been only in the context of the plane-wave method \citep{Hutter_abinitio_MD} and spherical basis function based methods seem to have ignored it \citep[see e.g.][]{solid_state_finite,spherical_averaged_jellium}. However, the two different methods of evaluation of the convolution sums (i.e., direct application of eq.~\ref{tau_lmn_expression} vs. the transform method described above) can lead to very large differences in computation times. To illustrate this point, we carried out computation of the electron density coefficients from randomly generated families of single electronic states using both methods. While using the direct method, we used  hash functions \citep{rasch_wu_hash} for expedited computation  of the coupling coefficients\footnote{Storage of all the coupling coefficients becomes memory intensive quite quickly.} that appear in eq.~\ref{tau_lmn_expression}. For both the direct method as well as the transform method, we varied the angular and radial cutoffs independently in order to directly observe the computational scalings of the density calculation routine.\footnote{For both routines, we investigated the angular cutoff range $10 \leq \calL \leq 60$ and the radial cutoff range  $10 \leq \calN \leq 60$.} Figure~\ref{fig:density_scaling} shows that the transform routine has far better scaling behavior than the direct routine for both discretization parameters. In practice, the computation run times for both these routines can differ by many orders of magnitude\footnote{Similar conclusions can be drawn about the matrix-vector product computation routines described earlier in section \ref{subsubsec:matrix_vector_prods}.} as Table~\ref{table:direct_vs_transform} shows. The angular and radial cutoffs that were chosen for the comparison in that table are very typical for obtaining acceptable levels of convergence in the total energies of small sized cluster systems containing a few light metallic atoms.
\begin{figure}[ht]
\centering
\includegraphics[scale=0.28]{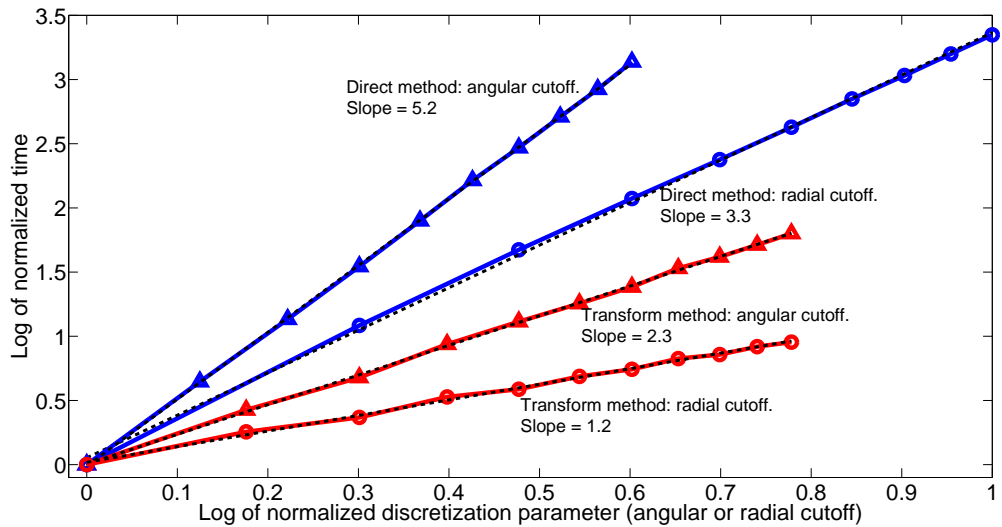}
\caption{Observed scaling of the density computation routine using the direct method (blue curves) and the transform method (red curves) with increasing discretization parameter. Both the angular cutoff variation (curves with triangles) as well as the radial cutoff variation (curves with circles) are shown. Slopes of the fitted lines (black) indicate scaling behavior.}
\label{fig:density_scaling}
\end{figure}

\begin{table}[ht]
\begin{center}
\begin{tabular}{ | c | c | c |}
\hline
Transform method & Direct method: & Direct method:\\
 & without hash functions &  with hash functions\\\hline\hline
$1.00$ & $2.18 \times 10^6$ & $1.71 \times 10^6$\\\hline
\end{tabular}
    \caption{Relative timings of the density computation routines for $\calL = 30, \calN = 30$ normalized using the transform method timing.}
\label{table:direct_vs_transform}
\end{center}  
\end{table}
\subsection{Computation of the potentials}
\label{subsec:potential_terms}
We describe the computation of the various potential terms in this section with particular attention to the Hartree and pseudopotential terms. The exchange correlation terms (within the Local Density Approximation) are evaluated directly using the real space representation of the electron density and deserve no further comments. 
\subsubsection{Computation of the Hartree potential}
\label{subsubsec:hartree_potential}
The Hartree potential at a point $\bfx \in \calB_R$ is given as:
\begin{align}
\label{Hartree_potential}
V_H(\bfx) = \int_{\calB_R} \!\frac{\rho(\bfy)}{\abs{\bfx-\bfy}}\;d\bfy\;.
\end{align}
One of the most popular approaches to solving this equation is by employing Poisson solvers \citep{Gavini_Kohn_Sham, BigDFT, Chelikowsky_Saad_1}. Our approach to the computation of $V_H$ is to directly deal with eq.~\ref{Hartree_potential} by  exploiting the so called Laplace expansion \citep{Jackson_ElectroDyn} of the Green's kernel:
\begin{align}
\frac{1}{\abs{\bfx-\bfy}}=\displaystyle\sum_{l=0}^{\infty}\frac{4\pi}{2l+1}\sum_{m=-l}^{m=l}
\frac{r^l_{<}}{r^{l+1}_{>}}\,\overline{\calY_{l}^{m}(\vartheta_{\bfx},\varphi_{\bfx})}\,
\calY_{l}^{m}(\vartheta_{\bfy},\varphi_{\bfy})\;.
\label{Laplace_expansion}
\end{align}
In the equation above, $r_{<}=\min{(r_{\bfx},r_{\bfy})},r_{>}=\max{(r_{\bfx},r_{\bfy})}$ and
$(r_{\bfx},\vartheta_{\bfx},\varphi_{\bfx})$\\ and $(r_{\bfy},\vartheta_{\bfy},\varphi_{\bfy})$ denote $\bfx$
and $\bfy$ in spherical coordinates respectively. For a typical point $\bfy \in \calB_R$, the electron density $\rho$ is available through a basis expansion  as:
\begin{align}
\label{rho_expansion_spherical}
\rho(r_{\bfy},\vartheta_{\bfy},\varphi_{\bfy})&=\displaystyle\sum_{\hat{\Gamma}}\tau_{\hat{l},\hat{m},\hat{n}}\,
\calR_{\hat{l},\hat{n}}(r_{\bfy})\,
\calY_{\hat{l}}^{\hat{m}}(\vartheta_{\bfy},\varphi_{\bfy})\;,
\end{align}
with $\hat{\Gamma}$ denoting the same basis set as $\Gamma$, or a larger one, depending on whether the dualing approximation has been used or not. Now, if $d\breve{\bfy}$ denotes the volume element in the sphere $\calB_R$, that is, $d\breve{\bfy}=r_{\bfy}^2\,\sin\vartheta_{\bfy}\,dr_{\bfy}d\vartheta_{\bfy}
d\varphi_{\bfy}$, then substituting eqs. \ref{Laplace_expansion} and  \ref{rho_expansion_spherical} in eq.~\ref{Hartree_potential} and using orthonormality of the spherical harmonics, we get:
\begin{align}
\nonumber
{V}_H(r_{\bfx},\vartheta_{\bfx},\varphi_{\bfx})&=\sum_{l=0}^{\infty}\frac{4\pi}{2l+1}\sum_{m=-l}^{m=l}
\sum_{\hat{\Gamma}} \tau_{\hat{l},\hat{m},\hat{n}}
\calY_{l}^{m}(\vartheta_{\bfx},\varphi_{\bfx})\\\nonumber
&\quad\times\;\int_{\calB_R}\!
\frac{r^l_{<}}{r^{l+1}_{>}}\,\overline{\calY_{l}^{m}(\vartheta_{\bfy},\varphi_{\bfy})}
\calR_{\hat{l},\hat{n}}(r_{\bfy})\,
\calY_{\hat{l}}^{\hat{m}}(\vartheta_{\bfy},\varphi_{\bfy})\,d\breve{\bfy}\,,\\\nonumber
&=\sum_{l=0}^{\infty}\frac{4\pi}{2l+1}\sum_{m=-l}^{m=l}
\sum_{\hat{\Gamma}} \tau_{\hat{l},\hat{m},\hat{n}}\,
\calY_{l}^{m}(\vartheta_{\bfx},\varphi_{\bfx})\,\delta_{l,\hat{l}}\,\delta_{m,\hat{m}}\\\nonumber
&\quad\quad\quad\times\;
\int_{r_{\bfy}=0}^{r_{\bfy}=R}\!\frac{r^{\hat{l}}_{<}}{r^{\hat{l}+1}_{>}}\,\calR_{\hat{l},\hat{n}}(r_{\bfy})\,
r_{\bfy}^2\,dr_{\bfy}\,,
\intertext{which we may rewrite as,}
\label{hartree_calc1}
&:=\displaystyle 
\sum_{\hat{\Gamma}}\frac{4\pi}{2\hat{l}+1}\,\tau_{\hat{l},\hat{m},\hat{n}}\,
\calY_{\hat{l}}^{\hat{m}}(\vartheta_{\bfx},\varphi_{\bfx})\,\mathfrak{Z}_{\hat{l},\hat{n}}(r_{\bfx})\;.
\end{align}
This suggests that computing the Hartree potential from the electron density expansion coefficients is very much like performing an inverse basis transform. The key difference is that the functions $\mathfrak{Z}_{\hat{l},\hat{n}}(r)$ need to be used, instead of the usual radial basis functions $\calR_{{l},{n}}(r)$, while carrying out the radial part of the calculation. If the $\mathfrak{Z}_{\hat{l},\hat{n}}(r)$ functions are pre-computed and stored, the method described here turns out to be extremely efficient: in our implementation, the entire calculation of obtaining the real space representation of ${V}_H$, starting from the real space representation of $\rho$, consumes less than $0.03$\% of the total time of a typical SCF cycle.\footnote{{This happens to be true even for the largest example systems considered later in this paper. From the discussion above, it is clear that the performance and scaling of the electrostatics terms will depend entirely on the efficiency and scalability of the basis transforms themselves. Various aspects of performance and scaling of the basis transforms are discussed throughout this paper. We do mention however, that in order to obtain better scalability of the electrostatics computation routines, we adopted a two level scheme that uses a process grid (the same grid discussed in Section \ref{subsec:two_level_scheme}) to obtain parallelization in the radial variable and in the radial basis function number (while using eqs.~\ref{hartree_calc1} and  \ref{rho_expansion_spherical}), thus effectively reducing communication loads.}}

The functions $\mathfrak{Z}_{\hat{l},\hat{n}}(r_{\bfx})$ may be written as follows:
\begin{align}
\nonumber
\mathfrak{Z}_{\hat{l},\hat{n}}(r_{\bfx}) =&\;\frac{1}{RJ_{\hat{l}+\thrbyto}(b^{\hat{n}}_{\hat{l}+\half})}\,
\int_{r_{\bfy}=0}^{r_{\bfy}=R}
\!\frac{r^{\hat{l}}_{<}}{r^{\hat{l}+1}_{>}}\,\sqrt{\frac{2}{r_{\bfy}}}\,
J_{\hat{l}+\half}\bigg(\frac{b^{\hat{n}}_{\hat{l}+\half}}{R}r_{\bfy}\bigg)\,r_{\bfy}^2\,dr_{\bfy}\,,
\\\nonumber
:=&\;\sqrt{2R}\,\,\tilde{\mathfrak{Z}}_{\hat{l},\hat{n}}(s)\,,\text{with}\,
s = r_{\bfx}/R\;\text{and}\; s \in [0,1]\;,
\\\nonumber
\tilde{\mathfrak{Z}}_{\hat{l},\hat{n}}(s)=&\;\frac{1}{J_{\hat{l}+\thrbyto}(b^{\hat{n}}_{\hat{l}+\half})}\,\bigg[s^{\hat{l}+1}
\int_{0}^{s}r_{1}^{\hat{l}+\thrbyto}
J_{\hat{l}+\half}\bigg({b^{\hat{n}}_{\hat{l}+\half}}r_{1}\bigg)\,dr_{1}\\
&\quad\quad\quad\quad\quad\quad
+s^{\hat{l}}
\int_{s}^{1}\frac{1}{r_{1}^{\hat{l}-\half}}
J_{\hat{l}+\half}\bigg({b^{\hat{n}}_{\hat{l}+\half}}r_{1}\bigg)\,dr_{1}
\bigg]\,,
\label{zeta_definition}
\end{align}
and $r_1$ simply denotes an integration variable. The integrals in eq.~\ref{zeta_definition} can be carried out numerically using Gauss quadrature. In our implementation, we have computed $\tilde{\mathfrak{Z}}_{\hat{l},\hat{n}}(s)$ accurately for a large number of values of $\hat{l}$ and $\hat{n}$ over a fine grid of values over $[0,1]$ and stored the results. The values of $\tilde{\mathfrak{Z}}_{\hat{l},\hat{n}}(s)$ at other values of $s \in [0,1]$ are computed using cubic spline interpolation as and when required. During an actual simulation, this procedure is used to quickly set up the functions $\mathfrak{Z}_{\hat{l},\hat{n}}(r_{\bfx})$ at the different radial grid points before the first SCF step.
\subsubsection{Computation of the pseudopotential terms}
\label{susubbsec:pseudo_pot_terms}
Modern pseudopotentials usually consist of local and non-local terms \citep{Martin_ES}. We now look at how each of these terms can be evaluated within our framework. 

The total local pseudopotential at a point $\bfx \in \calB_R$ is a combination of terms of the form:
\begin{align}
\label{local_pseudo_def}
{V}_{\text{nu}}(\bfx) = \sum_{j = 1}^{M} {V}^{j}_{\text{nu}}(\abs{\bfx - \bfx_j})\;,
\end{align}
where each of the functions ${V}^{j}_{\text{nu}}$ is reasonably smooth.\footnote{These functions are actually in $\mathsf{C}^{\infty}(\rz)$ for the pseudopotentials considered in this work. They have a somewhat lower regularity for the popular Troullier-Martins pseudopotentials \citep{Troullier_Martins_pseudo}.} By observing that ${V}_{\text{nu}}$ consists of radially symmetric terms which are centered at the atoms while the basis functions are centered at the origin, it is possible to make use of L\"{o}wdin transformations \citep{Lowdin_1956_quantum} to directly arrive at the expansion coefficients of local pseudopotential terms \citep{solid_state_finite,Broglia_original_paper}.
Our method for dealing with the local pseudopotential however, is to directly evaluate eq.~\ref{local_pseudo_def} at the gridpoints in $B$. This is because the local pseudopotential enters the Kohn-Sham calculation through the total effective potential and as described earlier, the total effective potential is required in real space representation during the computation of matrix vector products. The reciprocal space representation of the local pseudopotential can be evaluated by carrying out forward basis transforms, if required.

Non-local pseudopotentials are used in electronic structure methods in order to account for the effect of the inert core electrons on the chemically active valence electrons, without directly introducing these core states into the calculation  \citep{Martin_ES,LeBris_ReviewBook}. From a computational point of view, the inclusion of a non-local pseudopotential means that a projection operator needs to be added to the Hamiltonian while performing matrix vector products or while computing the total energies / forces. In general, this projection operator can be written as the sum of atom centered rank one operators. By definition, the action of a rank one operator $\calO = p_1\otimes p_2$ on a function $f \in \Lpspc{2}{}{\calB_R}$ is simply given as $\calO\,f = \innprod{p_2}{f}{\Lpspc{2}{}{\calB_R}}\;p_1$. {In our implementation, we first evaluate each of the projector functions $p_1$ and $p_2$ on the underlying real space grid, following which, we compute and store their the expansion coefficients by means of basis transforms (ahead of the first SCF step). From these expansion coefficients, the action of the projector $p_1\otimes p_2$ on an electronic state can be carried out in reciprocal space as the action of a rank one operator as described above.} The collective action of all the atom centered projectors on all the electronic states can be conveniently described through a pair of matrix-matrix multiplications. 

Instead of this reciprocal space formulation of the non-local pseudopotential terms, it is possible to carry out this calculation more efficiently in real space by making use of the fact that the functions $p_1$ and $p_2$ are usually short ranged (in real space). Some additional care is required so as to ensure that aliasing errors are avoided in this approach \citep{king_smith_non_local_pseudo} and therefore, we intend to explore this methodology in future work.
\section{Implementation}
\label{sec:implementation_details}
We outline various implementation related issues and solution strategies in this section. In particular, we discuss methods of obtaining the occupied eigenspace of the Hamiltonian as well as parallelization aspects of some of the key routines and procedures employed in our method.
\subsection{Diagonalization using LOBPCG}
\label{subsec:LOBPCG}
As remarked earlier, efficient eigensolvers for iterative diagonalization of the Hamiltonian matrix are necessary for dealing with large systems. Perhaps the most commonly used diagonalization method in {ab initio} calculations is the band-by-band conjugate gradient algorithm for direct minimization of the total energy \citep{Teter_Payne_Allan_1, Teter_Payne_Allan_2}, later modified to fit the iterative diagonalization framework\citep{Kleinman_Bylander_band_by_band_CG}. In this work, we have adopted the Locally Optimal Block Preconditioned Conjugate Gradient (LOBPCG) method \citep{LOBPCG_1}. The LOBPCG algorithm is much better supported theoretically \citep{LOBPCG_support}, it has been shown to outperform the traditional preconditioned conjugate gradient method \citep{ABINIT_LOBPCG} and it has found applications in numerous electronic structure methods  \citep{Meza_Yang_DCM, E_Lin_LOBPCG_F, Octopus_LOBPCG, ABINIT_LOBPCG} due to its robustness. 

When dealing with relatively small sized example systems (approximately a couple of hundred electrons), we have used the LOBPCG method exclusively to carry out diagonalization in all SCF steps. For some of the larger example systems described later, we used the LOBPCG method only in the first SCF step so as to generate a good guess for the Chebyshev polynomial filtered subspace iteration algorithm (described later) that was used in the subsequent SCF steps.
\subsubsection{Implementation details}
\label{subsubsec:lobpcg_implementation_details}
Our implementation of the LOBPCG method follows the algorithmic steps outlined in \cite{LOBPCG_3}. This allowed us to take advantage of \textit{soft locking} whenever some eigenvectors converged faster than others,\footnote{This can have a considerable impact on speeding up SCF iterations -- in many example systems, we found that diagonalization via LOBPCG in first SCF step is about 1.5 -- 2 times slower than in SCF steps which are close to attaining convergence. The eigenvectors in the latter are already close to their converged values and therefore soft locking allows for faster progression of LOBPCG.} as well as \textit{hard locking} which allowed us to carry out deflation against fixed orthonormal constraints. The latter proves to be particularly useful in calculations on large systems since the total number of electronic states may be too numerous to fit all required eigenstates into one LOBPCG block. This is primarily due to the large computational demands of the Raleigh-Ritz step used by LOBPCG.

A second detail is related to the use of Cholesky factorization for orthonormalization (of the residual vectors and conjugate directions) in the LOBPCG implementation. This technique is more computationally efficient but also known to be less reliable than the traditional approach involving QR factorization \citep{LOBPCG_3}. Computation of the Cholesky decomposition of matrices that are poorly scaled  are often required by the LOBPCG method and so, it is crucial to either use a Cholesky decomposition that is numerically invariant with respect to matrix scaling, or to scale the columns of such matrices\footnote{We are grateful to Andrew Knyazev (Mitsubishi Electric Research Laboratories) for his consistent support and suggestions during our implementation of LOBPCG, and in particular, for pointing out the stability issues related to Cholesky factorization.} before performing the factorization \citep{Knyazev_email}. In addition, our experience has been that numerical noise or round off errors (arising from the transform based matrix-vector product computations, for instance) can sometimes cause the Cholesky factorization or the Raleigh-Ritz procedure to become unstable. In these situations, we have always found it useful to restart the LOBPCG iterations (discarding the computed conjugate direction and residual vectors) by using the most recently computed block of eigenvectors as the initial guess of a fresh set of iterations. This simple strategy seems to result in a much more robust implementation and it does not introduce any computational bottlenecks.  

\subsubsection{Use of the Teter-Payne-Allan preconditioner}
\label{subsubsec:lobpcg_Teter preconditioner}
The need for a good preconditioner for use with the LOBPCG method been emphasized in \citep{LOBPCG_3, LOBPCG_support}. A majority of the generic preconditioners that have been developed over the years, are aimed towards sparse systems. These are unsuitable for our purposes because our method is matrix free, and moreover, the underlying matrix involved is dense. Fortunately, the presence of the Laplacian operator in the Kohn-Sham eigenvalue problem suggests a viable preconditioner \citep{Teter_Payne_Allan_1, Teter_Payne_Allan_2}. These authors introduced a preconditioner within the context of the plane-wave method that has the particular advantage of being diagonal in reciprocal space (and it is therefore, inexpensive to apply). The formal similarities of our spectral method with the plane-wave method allowed us to directly adopt the diagonal preconditioner introduced by these authors. 

Specifically, we used a preconditioning matrix $\textbf{T}_{p_{\alpha,\beta}}$ of the form:
\begin{align}
\textbf{T}_{p_{\alpha,\beta}} = \frac{27 + 18g + 12g^2 + 8g^3}{27 + 18g + 12g^2 + 8g^3 + 16g^4}\;\delta_{\alpha,\beta}\,,
\end{align}
where $g$ is the ratio of the Laplacian eigenvalue to the kinetic energy of the residual vector on which the preconditioner is being applied, i.e., denoting the residual vector as $\bfY \in \cz^{\mathit{d}}$,
\begin{align}
\displaystyle g =\Lambda_{\calJ(\alpha)} \bigg /\bigg( \half\sum_{\alpha = 1}^{\mathit{d}}\Lambda_{\calJ(\alpha)}\abs{\bfY_\alpha}^2\bigg )\;.
\end{align}
As $g$ approaches zero, the preconditioner elements approach one with zero derivatives upto third order and so, $\textbf{T}_{p}$ leaves the low energy states unchanged. On the other hand, above $g = 1$, $\textbf{T}_{p}$ asymptotically approaches the inverse Laplacian thus suitably damping out the high kinetic energy states that are responsible for ill-conditioning.

As can be seen from Figure~\ref{fig:c_teter_prec}, this simple and inexpensive preconditioner makes a marked difference in the rate of convergence of the residuals in LOBPCG. The particular system used for the demonstration was an 18 atom Barium cluster for which 4000 basis functions were used and only the linear part of the Kohn-Sham equations was solved. This preconditioner was therefore adopted in all further calculations wherever the LOBPCG solver was employed.
\begin{figure}[ht]
\centering
\includegraphics[scale=0.28]{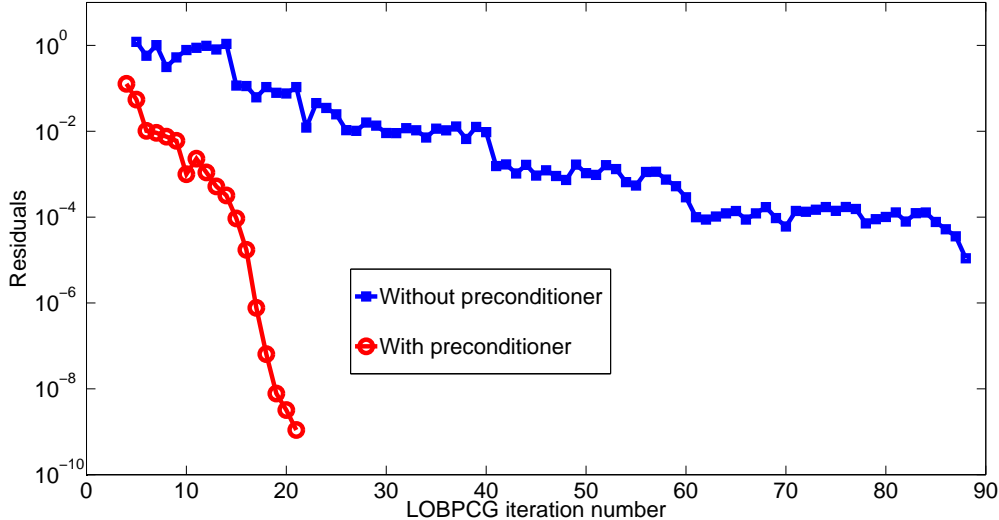}
\caption{Effect of the diagonal preconditioner on LOBPCG iterations. The average residual for the first few eigenvectors has been plotted against the iteration number.}
\label{fig:c_teter_prec}
\end{figure}
\subsubsection{Parallelization scheme and scaling}
\label{subsubsec:lobpcg_parallel_elemental}
Our strategy for parallelization of the LOBPCG method is to carry out relevant  linear algebra operations using a distributed memory dense linear algebra library. For this purpose, we have adopted the state of the art numerical library\footnote{We are grateful to Jack Poulson (Georgia Tech.), the lead author of the Elemental package for his suggestions and help with the package.} Elemental \citep{Elemental_Poulson}. This library has been designed to be a more scalable and easier to interface successor of the ScaLAPACK \citep{ScaLAPACK_1, ScaLAPACK_2} and PLAPACK \citep{PLAPACK_1, PLAPACK_2} libraries that have already found widespread use in other electronic structure codes. Elemental uses an element-wise block-cyclic distribution of matrices over a two-dimensional grid of processors.\footnote{See Figure~\ref{fig:data_redistrib} to see an example of how the data is distributed among processors.} The Message Passing Interface (MPI) is used for interprocess communication while linear algebra operations that are local to each process are carried out by making calls to (serial) BLAS and LAPACK libraries.\footnote{To ensure maximum use of hardware resources, our code was linked to machine optimized BLAS and LAPACK libraries.}

The dimensions of the process grid that underlies the linear algebra operations in Elemental can have an impact on the resulting parallel efficiency of the LOBPCG routine. We have used square process grids in most cases. In some cases however, we found that the use of rectangular process grids, in which the height of the grid was longer than the width of the grid seemed to result in better performance.

In order to judge the parallel efficiency of our LOBPCG implementation, we studied the weak scaling of our routine in the following manner.  We generated random dense hermitian matrices of various sizes and computed the first few hundred eigenstates. We used between 16 and 512 c.p.u cores, the matrix size was increased in proportion to the number of cores used\footnote{The computational platform details are described in a later section.} and the number of states computed was held constant. We measured the average time per LOBPCG step and the results from this study are plotted in Figure~\ref{fig:lobpcg_scal_weak}. Keeping in mind that one of the most  computationally expensive steps in the setting of dense matrices is due to parallel matrix vector multiplications in which the problem size grows quadratically\footnote{This is unlike the weak scaling studies presented in \citep{LOBPCG_3} in which sparse matrices were used, resulting in linear growth of problem size with increasing matrix dimension.} with increasing matrix dimension \citep{poulson_parallel_matmul}, we have also plotted in Figure~\ref{fig:lobpcg_scal_weak}, the computational complexity adjusted parallel efficiency. As is evident from the figure, the adjusted parallel efficiency remains above 90\% up to 256 c.p.u. cores and drops to a little less than 80\% at 512 c.p.u. cores.\footnote{{Here as well as in Section \ref{subsubsec:scaling_perf_and_PG}, we have focussed on weak scalability results. This is because we were primarily interested in ensuring that our code is able to handle even large materials systems within a reasonably constant wall-clock time. Our underlying assumption of course, was that more computational resources would to be allocated to the code, when necessary, in order to meet the wall-clock time requirements. We do mention however that the strong scaling of our LOBPCG routine is reasonably good, although it is not as encouraging as its weak scaling. In a test involving the computation of a few hundred eigenstates of a randomly generated hermitian matrix of dimension $40,960 \times 40,960$, the strong parallel efficiency was about $60$ \% for 128 MPI processes and it dropped to about $40$\% for 256 MPI processes. More details may be found in \citep{My_PhD_Thesis}.}}
\begin{figure}[ht]
\centering
\includegraphics[scale=0.28]{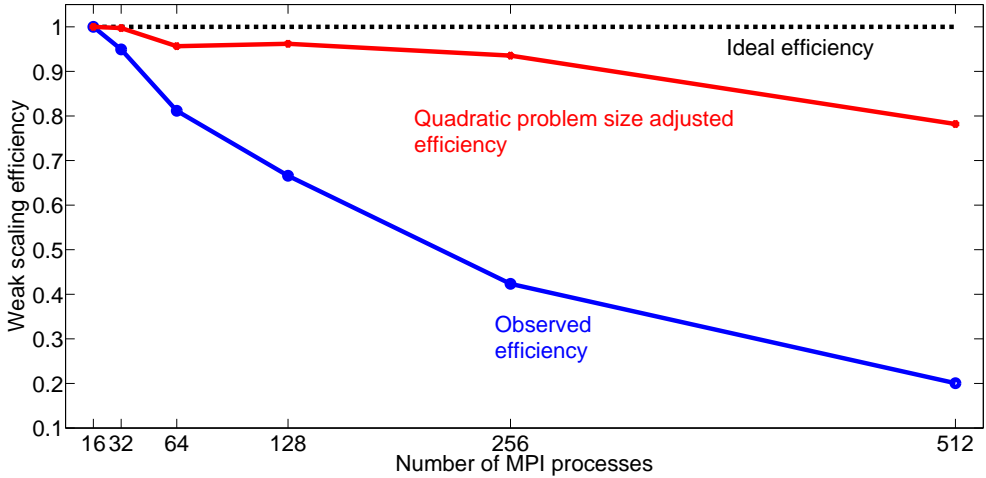}
\caption{Weak parallel scaling behavior of the LOBPCG implementation measured by time taken per LOBPCG step vs. the number of MPI processes employed, while keeping the problem size per process constant.}
\label{fig:lobpcg_scal_weak}
\end{figure}
\subsection{Chebyshev polynomial filter accelerated subspace iterations}
\label{subsec:Chebyshev_polynomial_filter}
Subspace iteration algorithms constitute a generalization of the classical power iterations approach to computation of eigenpairs \citep{Saad_large_eigenvalue_book, Online_Template_Eigenvalue_Problem_Book}. These methods allow the computation of multi-dimensional invariant subspaces rather than one eigenvector at a time.  Since the electron density or the total Kohn-Sham energy do not depend explicitly on the eigenvectors of the Hamiltonian, but only on the occupied subspace, subspace iterations have often been used for electronic structure calculations \citep{stephan1998improved, bekas2005computing, baroni1992towards}. 

The Chebyshev polynomial filtered SCF iteration technique for computing the occupied eigenspace of the Kohn-Sham operator was introduced  in \cite{Serial_Chebyshev, Parallel_Chebyshev}, and was originally presented in the context of the finite difference method. However,  this method has {enjoyed success} in conjunction with finite elements as well \citep{Gavini_higher_order}. The method can be thought of as a form of non-linear subspace iteration which takes advantage of the fact that eigenvectors of the Hamiltonian do not need to be computed accurately at every SCF step since the Hamiltonians involved are approximate as well. This allows one to exploit the non-linear nature of the problem in the sense that the technique removes emphasis on the accurate solution of the intermediate linearized Kohn-Sham eigenvalue problems. By means of spectral mapping, the method employs the exponential growth property of the Chebyshev polynomials outside the region $[-1,1]$ to damp out the unwanted part of the spectrum of the Hamiltonian thus accelerating the subspace iterations. 

Although the Chebyshev polynomial filtered SCF iteration technique does not seem to have been adopted in the plane-wave method literature so far, it was apparent to us that as long as one has access to efficient matrix-vector product routines, the technique is likely to yield large savings compared to traditional diagonalization methods. This is primarily due to the fact that  
orthonormalization and other linear algebra operation costs that accompany traditional diagonalization methods are minimal in this method.
\subsubsection{Implementation details}
\label{subsusec:cheby_implementation}
The Chebyshev polynomial filtered SCF iteration technique is presently the work horse of most medium to large sized computations carried out using the ClusterES package. In our implementation of this method, we first {obtain a guess for the initial electron density by linearly superposing precomputed atomic electron densities. Next, having computed the potentials, we use the LOBPCG method (on a collection of randomly generated vectors used as an initial guess\footnote{{This appears to be a fairly common practice in the literature - see for example \citep{Teter_Payne_Allan_1, zhou_2014_chebyshev}.}}) to obtain a good eigenbasis of the Hamiltonian for the first SCF step.} This is used to serve as a good guess for the occupied subspace of the Hamiltonian at self-consistency.\footnote{Typically, a few extra states (about 10 -- 20) are included from the LOBPCG calculation so that the Raleigh-Ritz step (used in the Chebyshev filtering method) and finite-temperature Fermi-Dirac smearing (used for metallic systems) can be employed.} The Chebyshev polynomial filtered subspace iterations begin after this first SCF step and they attempt to adaptively improve the initial guess subspace by polynomial filtering.\footnote{{We recently became aware of techniques which can omit the first diagonalization step in lieu of filtering \citep{zhou_2014_chebyshev}. We intend to adopt this methodology into our code in the near future.}}

In the original presentation of the Chebyshev filtering method, the authors used the DGKS algorithm \citep{DGKS} for orthonormalization of the basis vectors of the occupied subspace. In the spirit of the LOBPCG method as well as the RMM-DIIS method \citep{Kresse_abinitio_iterative}, we have used the faster (but less stable) Cholesky factorization method instead. This helped speed up the orthonormalization calculation (by a factor of 2--3 in most cases) and we have not witnessed any problematic side effects. 

As described in \cite{Serial_Chebyshev, Parallel_Chebyshev}, the bulk of the Chebyshev filtering algorithm consists of evaluating the polynomial filter using matrix vector products. The additional linear algebra operations involved are in the form of scaling and shifting, orthonormalization and the Raleigh-Ritz step. Therefore, as in our implementation of the LOBPCG method, we used the Elemental package and its underlying process grid structure for carrying out these dense linear algebra operations in parallel. 

Table~\ref{table:cheby_vs_lobpcg} shows the performance gains obtained by our use of the Chebyshev filtered subspace iteration method when compared to LOBPCG. For the two examples presented in that table, each typical Chebyshev filtered SCF step is about 10 -- 20 times faster than each typical LOBPCG based SCF step while the total number of SCF steps used by both methods to reach similar levels of convergence is about the same. Thus, there is an enormous amount of savings in the total computation time for the examples presented. It seems likely that for larger material systems, the savings are even greater.
\begin{table}[ht]
\begin{center}
\resizebox{13.5cm}{!}{
\begin{tabular}{|c|c|c|c|c|c|}
\hline
\multirow{2}{*}{} 
               &   No. of Basis &  No. of &  No. of   &  No. of & Ratio of LOBPCG \\
      Material & functions &   electronic &  LOBPCG   & Chebyshev & step time to\\ 
      System   &  used   &  states used & SCF steps & SCF steps & Chebyshev step time\\\hline\hline
172 atom  &    &   &  &  & \\ 
Aluminum &   512000   &  280 & 22 & 23 & 20.2\\ 
FCC cluster &     &   & &  & \\  
\hline
$\text{C}_{60}$ &     &   & &  & \\
Buckyball & 343000  & 136  & 11 & 13  & 12.3  \\
\hline
\end{tabular}
}
\caption{Performance of Chebyshev Filtered SCF iterations compared against LOBPCG based SCF iterations.}
\label{table:cheby_vs_lobpcg}
\end{center}  
\end{table}
\subsection{Parallelization of Matrix vector products and electron density computation : Two level scheme}
\label{subsec:two_level_scheme}
For systems containing up to a few thousand electronic states, the Hamiltonian matrix times vector computation routine is one of the main computationally intensive steps in the LOBPCG method and it is the principle one in the Chebyshev filtering method. These methods typically require the  product of the Hamiltonian with a block of vectors to be computed. Due to our use of the two dimensional process grid for carrying out dense linear algebra operations (see sections \ref{subsubsec:lobpcg_parallel_elemental}, \ref{subsusec:cheby_implementation}), the block of vectors that needs to be multiplied with the Hamiltonian already appears distributed over the process grid. Specifically, the states are distributed over the process grid columns and each state is further distributed over process grid rows. In this situation, it is natural to parallelize the matrix vector product over the different Kohn-Sham states (i.e., band/state parallelization, as it is often called in the plane-wave literature) since this involves no communication between the processors that store the different states.  

However, we  observe that the inverse basis transform requires access to all the expansion coefficients that constitute an entire state (eq.~\ref{eq:G_lmr}) while the forward basis transform requires access to function values at all the grid points (eqs.~\ref{Ar_lm}, \ref{Arlm_to_coeff}). Since the process grid for the linear algebra operations distributes each state over the process grid rows, this requires the basis transforms to induce communication within process grid columns. The data redistribution over the process grid that is required for these basis transforms is shown schematically in Figure~\ref{fig:data_redistrib}.

A crucial detail is that the forward and inverse spherical harmonics transforms (which constitute the bulk of the operations involved within the basis transforms) can be performed independently over the various radial grid points. Thus, we may adopt a second level of parallelization by distributing real space quantities  over different values of the radial gridpoints since this will ensure that the basis transform routines are mostly communication free. Figure~\ref{fig:big_figure} shows a schematic outline of the individual steps of the matrix vector computation procedure over the process grid.
\begin{figure}[ht]
\centering
\begin{subfigure}[b]{\textwidth}
\centering
\begin{tabular}{|c|c|}
\hline
0 & 2 \\\hline
1 & 3 \\\hline
\end{tabular}
\caption{Numbering of processes in a $2\times2$ process grid. }
\label{fig:proc_grid}
\end{subfigure}
\begin{subfigure}[b]{\textwidth}
\footnotesize
\centering
$\left(\begin{array}{ccccc}
  0 & 2 & 0 & 2\\
  1 & 3 & 1 & 3\\
  0 & 2 & 0 & 2\\
  1 & 3 & 1 & 3\\
  0 & 2 & 0 & 2\\
  1 & 3 & 1 & 3
\end{array}\right)$ 
\begin{picture}(60,60)
\thicklines
\put(0,15){\vector(1,0){60}}
\put(0,20){Fwd. Trans.}
\put(60,6){\vector(-1,0){60}}
\put(8,-5){Inv. Trans.}
\end{picture}
$\left(\begin{array}{cccc}
  \{0,1\} & \{2,3\} & \{0,1\} & \{2,3\} \\
  \{0,1\} & \{2,3\} & \{0,1\} & \{2,3\} \\
  \{0,1\} & \{2,3\} & \{0,1\} & \{2,3\} \\
  \{0,1\} & \{2,3\} & \{0,1\} & \{2,3\} \\
  \{0,1\} & \{2,3\} & \{0,1\} & \{2,3\} \\
  \{0,1\} & \{2,3\} & \{0,1\} & \{2,3\} 
\end{array}\right)$

\caption{Individual matrix entry ownership (by processes) during basis transforms.}
\label{fig:data_ownership}
\end{subfigure}               
\caption{Example of data redistribution during forward and inverse transforms. A process grid of dimension $2\times2$ has been used for storing 4 electronic states and the number of expansion coefficients for each state (i.e., basis set size) is 6. Inter process communication occurs only along process grid columns during the transforms.}
\label{fig:data_redistrib}
\end{figure}

The two level parallelization scheme described above is in the spirit of similar schemes adopted by some large scale plane-wave codes \citep{Large_scale_plane_wave, Gygi_2D_parallel}. Due to this scheme, the only communication involved during the matrix vector product calculations is over individual process grid columns: one time during the broadcast of the different portions of a particular state while the inverse basis transform occurs and a second time during computation of the radial quadratures while the forward basis transform occurs. The important point however, is that the communication load gets reduced from the total number of processors involved, to roughly the square root of the total number of processors (if a square process grid is in use). Also, due to the distribution of various real space quantities over the radial grid points, the memory overhead is reduced as well.

The computation of the electron density from the Kohn-Sham states can also be made to follow this two level parallelization scheme. In this case, while computing the electron density in real space (from the Kohn-Sham states in reciprocal space), the  communication involved is once along individual process grid columns during computation of the inverse basis transform and a second time along individual process grid rows while summing the results from the different Kohn-Sham states according to eq.~\ref{density_expression}. Once again, this means that the communication load scales roughly as the square root of the number of processors. 
%%%%%%%%%%%%%% Start of Big Figure
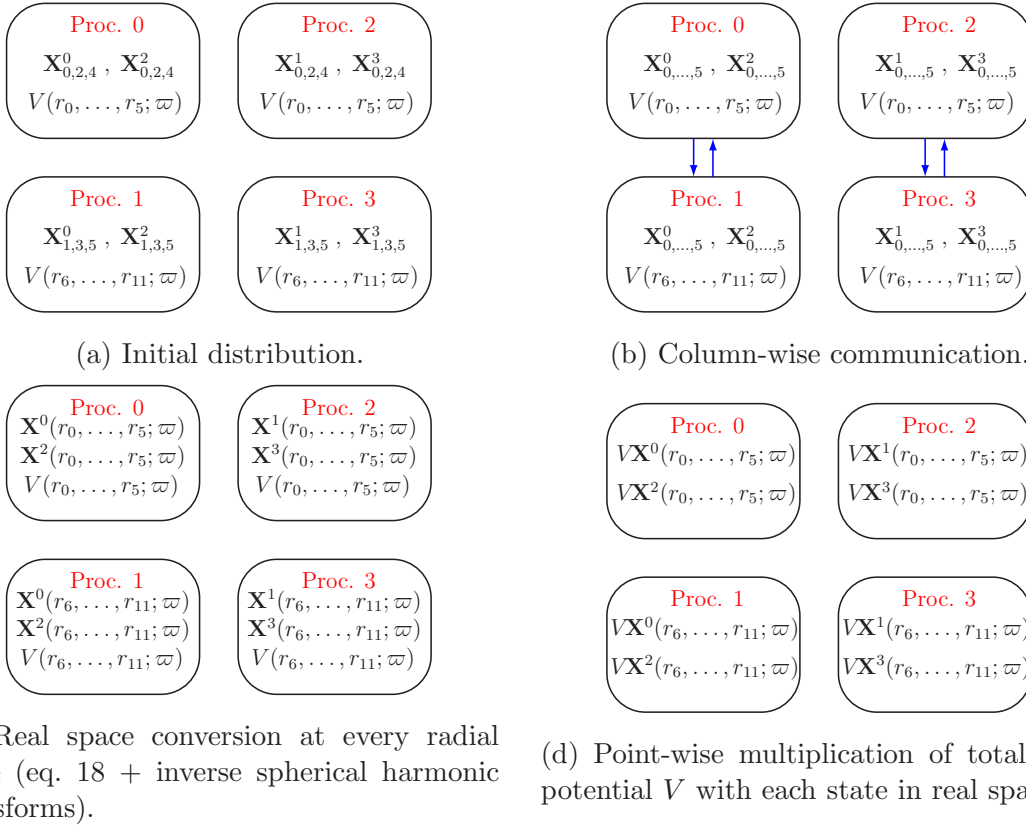
\begin{figure}[ht]
\centering
\begin{subfigure}[h]{0.45\textwidth}
\centering
\scalebox{0.73}{
\begin{picture}(235,170)
\thicklines
\multiput(55, 130)(120, 0 ){2}{\oval(100,70)}
\multiput(55, 40)(120, 0 ){2}{\oval(100,70)}
% Process numbers
\put(38, 150){\color{red}{Proc. 0}}
\put(38, 60){\color{red}{Proc. 1}}
\put(158, 150){\color{red}{Proc. 2}}
\put(158, 60){\color{red}{Proc. 3}}

% Proc 0
\put(24, 130){{$\bfX^0_{0,2,4}\;,\;\bfX^2_{0,2,4}$}}
\put(16, 110){{$V(r_0,\ldots,r_{5};\varpi)$}}

% Proc 1
\put(24, 40){{$\bfX^0_{1,3,5}\;,\;\bfX^2_{1,3,5}$}}
\put(14, 20){{$V(r_6,\ldots,r_{11};\varpi)$}}

% Proc 2
\put(144, 130){{$\bfX^1_{0,2,4}\;,\;\bfX^3_{0,2,4}$}}
\put(136, 110){{$V(r_0,\ldots,r_5;\varpi)$}}

% Proc 3
\put(144, 40){{$\bfX^1_{1,3,5}\;,\;\bfX^3_{1,3,5}$}}
\put(134, 20){{$V(r_6,\ldots,r_{11};\varpi)$}}
\end{picture}}
\caption{Initial distribution.}
\label{fig1:big_fig_1}
\end{subfigure}      
\quad         
\begin{subfigure}[h]{0.45\textwidth}
\centering
\scalebox{0.73}{
\begin{picture}(235,170)
\thicklines
\multiput(55, 130)(120, 0 ){2}{\oval(100,70)}
\multiput(55, 40)(120, 0 ){2}{\oval(100,70)}

% Process numbers
\put(38, 150){\color{red}{Proc. 0}}
\put(38, 60){\color{red}{Proc. 1}}
\put(158, 150){\color{red}{Proc. 2}}
\put(158, 60){\color{red}{Proc. 3}}

% Proc 0
\put(24, 130){{$\bfX^0_{0,\ldots,5}\;,\;\bfX^2_{0,\ldots,5}$}}
\put(16, 110){{$V(r_0,\ldots,r_{5};\varpi)$}}

% Proc 1
\put(24, 40){{$\bfX^0_{0,\ldots,5}\;,\;\bfX^2_{0,\ldots,5}$}}
\put(14, 20){{$V(r_6,\ldots,r_{11};\varpi)$}}

% Proc 2
\put(144, 130){{$\bfX^1_{0,\ldots,5}\;,\;\bfX^3_{0,\ldots,5}$}}
\put(136, 110){{$V(r_0,\ldots,r_5;\varpi)$}}

% Proc 3
\put(144, 40){{$\bfX^1_{0,\ldots,5}\;,\;\bfX^3_{0,\ldots,5}$}}
\put(136, 20){{$V(r_6,\ldots,r_{11};\varpi)$}}

% Vectors
\put(50,95){\color{blue}\vector(0,-1){20}}
\put(60,75){\color{blue}\vector(0,1){20}}

\put(170,95){\color{blue}\vector(0,-1){20}}
\put(180,75){\color{blue}\vector(0,1){20}}
\end{picture}}
\caption{Column-wise communication.}
\label{fig:big_fig_2}
\end{subfigure}
\begin{subfigure}[h]{0.45\textwidth}
\centering
\scalebox{0.73}{
\begin{picture}(235,170)
\thicklines
\multiput(55, 130)(120, 0 ){2}{\oval(100,70)}
\multiput(55, 40)(120, 0 ){2}{\oval(100,70)}
% Process numbers
\put(38, 150){\color{red}{Proc. 0}}
\put(38, 60){\color{red}{Proc. 1}}
\put(158, 150){\color{red}{Proc. 2}}
\put(158, 60){\color{red}{Proc. 3}}

% Proc 0
\put(12, 140){{$\bfX^0(r_0,\ldots,r_5;\varpi)$}}
\put(12, 125){{$\bfX^2(r_0,\ldots,r_5;\varpi)$}}
\put(14, 110){{$V(r_0,\ldots,r_{5};\varpi)$}}

% Proc 1
\put(10, 50){{$\bfX^0(r_6,\ldots,r_{11};\varpi)$}}
\put(10, 35){{$\bfX^2(r_6,\ldots,r_{11};\varpi)$}}
\put(12, 20){{$V(r_6,\ldots,r_{11};\varpi)$}}

% Proc 2
\put(132, 140){{$\bfX^1(r_0,\ldots,r_5;\varpi)$}}
\put(132, 125){{$\bfX^3(r_0,\ldots,r_5;\varpi)$}}
\put(134, 110){{$V(r_0,\ldots,r_{5};\varpi)$}}

% Proc 3
\put(130, 50){{$\bfX^1(r_6,\ldots,r_{11};\varpi)$}}
\put(130, 35){{$\bfX^3(r_6,\ldots,r_{11};\varpi)$}}
\put(132, 20){{$V(r_6,\ldots,r_{11};\varpi)$}}
\end{picture}}
\caption{Real space conversion at every radial node (eq.~\ref{eq:G_lmr} + inverse spherical harmonic transforms).}
\label{fig1:big_fig_3}
\end{subfigure}      
\quad         
\begin{subfigure}[h]{0.45\textwidth}
\centering
\scalebox{0.73}{
\begin{picture}(235,170)
\thicklines
\multiput(55, 130)(120, 0 ){2}{\oval(100,70)}
\multiput(55, 40)(120, 0 ){2}{\oval(100,70)}
% Process numbers
\put(38, 150){\color{red}{Proc. 0}}
\put(38, 60){\color{red}{Proc. 1}}
\put(158, 150){\color{red}{Proc. 2}}
\put(158, 60){\color{red}{Proc. 3}}

% Proc 0
\put(10, 135){{$V\!\bfX^0(r_0,\ldots,r_5;\varpi)$}}
\put(10, 115){{$V\!\bfX^2(r_0,\ldots,r_5;\varpi)$}}

% Proc 1
\put(7, 45){{$V\!\bfX^0(r_6,\ldots,r_{11};\varpi)$}}
\put(7, 25){{$V\!\bfX^2(r_6,\ldots,r_{11};\varpi)$}}

% Proc 2
\put(130, 135){{$V\!\bfX^1(r_0,\ldots,r_5;\varpi)$}}
\put(130, 115){{$V\!\bfX^3(r_0,\ldots,r_5;\varpi)$}}

% Proc 3
\put(127, 45){{$V\!\bfX^1(r_6,\ldots,r_{11};\varpi)$}}
\put(127, 25){{$V\!\bfX^3(r_6,\ldots,r_{11};\varpi)$}}
\end{picture}}
\caption{Point-wise multiplication of total local potential $V$ with each state in real space.}
\label{fig:big_fig_4}
\end{subfigure}
\caption{Schematic of the steps involved in computing the Hamiltonian times block of vectors product using the $2\times2$ process grid. Only the local part of the total potential is considered here. The block of vectors $\bfX$ has 4 states (denoted with superscripts) with 6 expansion coefficients (labelled using subscripts) used for each state. The real space grid has 12 points $r_0,\ldots,r_{11}$ in the radial direction. The angular grid is left unspecified here and denoted as $\varpi = (\vartheta,\varphi)$. Real space quantities are shared along process grid columns by distributing the radial nodes.}
\label{fig:big_figure}
\end{figure}
%%%%%%%%%% End of Figure part 1
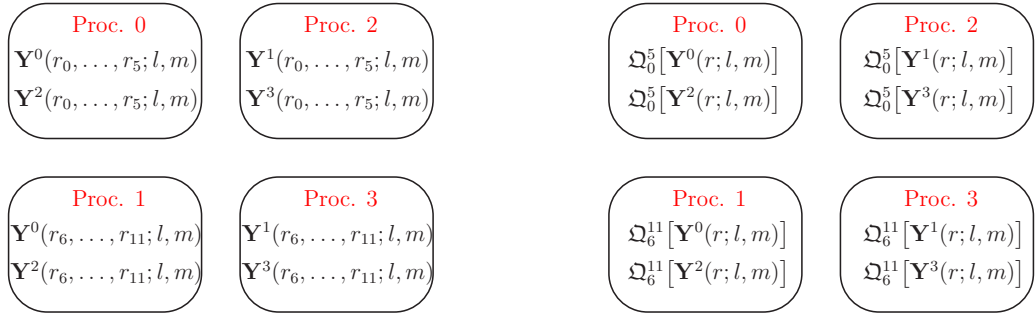
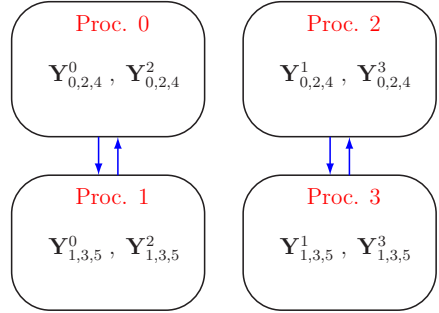
\begin{figure}[ht]
\ContinuedFloat 
\centering
\begin{subfigure}[h]{0.45\textwidth}
\centering
\scalebox{0.73}{
\begin{picture}(235,170)
\thicklines
\multiput(55, 130)(120, 0 ){2}{\oval(100,70)}
\multiput(55, 40)(120, 0 ){2}{\oval(100,70)}
% Process numbers
\put(38, 150){\color{red}{Proc. 0}}
\put(38, 60){\color{red}{Proc. 1}}
\put(158, 150){\color{red}{Proc. 2}}
\put(158, 60){\color{red}{Proc. 3}}

% Proc 0
\put(8, 132){{$\bfY^{0}(r_0,\ldots,r_5;l,m)$}}
\put(8, 112){{$\bfY^{2}(r_0,\ldots,r_5;l,m)$}}

% Proc 1
\put(6, 42){{$\bfY^{0}(r_6,\ldots,r_{11};l,m)$}}
\put(6, 22){{$\bfY^{2}(r_6,\ldots,r_{11};l,m)$}}

% Proc 2
\put(128, 132){{$\bfY^{1}(r_0,\ldots,r_5;l,m)$}}
\put(128, 112){{$\bfY^{3}(r_0,\ldots,r_5;l,m)$}}

% Proc 3
\put(126, 42){{$\bfY^{1}(r_6,\ldots,r_{11};l,m)$}}
\put(126, 22){{$\bfY^{3}(r_6,\ldots,r_{11};l,m)$}}
\end{picture}}
\caption{Spherical harmonic transforms at every radial node (eq.~\ref{Ar_lm}).}
\label{fig1:big_fig_5}
\end{subfigure}      
\quad
  \begin{subfigure}[h]{0.45\textwidth}
\centering
\scalebox{0.73}{
\begin{picture}(235,170)
\thicklines
\multiput(55, 130)(120, 0 ){2}{\oval(100,70)}
\multiput(55, 40)(120, 0 ){2}{\oval(100,70)}
% Process numbers
\put(38, 150){\color{red}{Proc. 0}}
\put(38, 60){\color{red}{Proc. 1}}
\put(158, 150){\color{red}{Proc. 2}}
\put(158, 60){\color{red}{Proc. 3}}

% Proc 0
\put(15, 132){{$\mathfrak{Q}_{0}^{5}\big[\bfY^{0}(r;l,m)\big]$}}
\put(15, 112){{$\mathfrak{Q}_{0}^{5}\big[\bfY^{2}(r;l,m)\big]$}}

% Proc 1
\put(15, 42){{$\mathfrak{Q}_{6}^{11}\big[\bfY^{0}(r;l,m)\big]$}}
\put(15, 22){{$\mathfrak{Q}_{6}^{11}\big[\bfY^{2}(r;l,m)\big]$}}

% Proc 2
\put(137, 132){{$\mathfrak{Q}_{0}^{5}\big[\bfY^{1}(r;l,m)\big]$}}
\put(137, 112){{$\mathfrak{Q}_{0}^{5}\big[\bfY^{3}(r;l,m)\big]$}}

% Proc 3
\put(137, 42){{$\mathfrak{Q}_{6}^{11}\big[\bfY^{1}(r;l,m)\big]$}}
\put(137, 22){{$\mathfrak{Q}_{6}^{11}\big[\bfY^{3}(r;l,m)\big]$}}
\end{picture}}
\caption{Partial radial quadratures using local radial nodes.}
\label{fig1:big_fig_6}
\end{subfigure}     
\begin{subfigure}[h]{0.45\textwidth}
\centering
\scalebox{0.73}{
\begin{picture}(235,170)
\thicklines
\multiput(55, 130)(120, 0 ){2}{\oval(100,70)}
\multiput(55, 40)(120, 0 ){2}{\oval(100,70)}

% Process numbers
\put(38, 150){\color{red}{Proc. 0}}
\put(38, 60){\color{red}{Proc. 1}}
\put(158, 150){\color{red}{Proc. 2}}
\put(158, 60){\color{red}{Proc. 3}}

% Proc 0
\put(24, 124){{$\bfY^0_{0,2,4}\;,\;\bfY^2_{0,2,4}$}}

% Proc 1
\put(24, 34){{$\bfY^0_{1,3,5}\;,\;\bfY^2_{1,3,5}$}}

% Proc 2
\put(144, 124){{$\bfY^1_{0,2,4}\;,\;\bfY^3_{0,2,4}$}}

% Proc 3
\put(144, 34){{$\bfY^1_{1,3,5}\;,\;\bfY^3_{1,3,5}$}}

% Vectors
\put(50,95){\color{blue}\vector(0,-1){20}}
\put(60,75){\color{blue}\vector(0,1){20}}

\put(170,95){\color{blue}\vector(0,-1){20}}
\put(180,75){\color{blue}\vector(0,1){20}}
\end{picture}}
\caption{Column-wise communication to evaluate radial quadratures (eq.~\ref{Arlm_to_coeff}) from partial results.}
\label{fig:big_fig_7}
\end{subfigure}
\caption{(Continued) Schematic of computation of the Hamiltonian times block of vectors product over $2\times2$ process grid. The product of the Hamiltonian with the block of vectors is denoted as $\bfY$ here. The symbol $\mathfrak{Q}_{i}^{j}\big[t(r)\big]$ is used to denote a partial radial quadrature (i.e., evaluation of eq.~\ref{Arlm_to_coeff}) over the radial nodes $r_i,\ldots,r_j$.}
%\label{fig:big_figure}
\end{figure}
%%%%%%%%%%%%%% End of Big Figure
\subsubsection{Scaling performance and process grid geometry choice}
\label{subsubsec:scaling_perf_and_PG}
We now discuss the scaling performance of the matrix vector product routine. In order to properly assess and interpret the scaling performance, we need to be mindful of the two dimensional nature of the underlying parallelization scheme. In particular, the choice of an optimal process grid geometry for a fixed basis set size, may be done as follows. First, with the given basis set size, we observe the computation time for the matrix vector product routine using only one electronic state. We increase the process grid height (keeping the process grid width fixed at 1) till the optimum performance is reached. Due to increasing communication costs during basis transforms, the performance saturates after a sufficiently large process grid height. In Figure~\ref{subfig:single_band_saturation}, we used one million basis functions for our study and for this basis set size, saturation occurs\footnote{Note that while the number of basis functions used was one million, the number of radial points used was only $200$. This helps us understand why the parallelization based on decomposition of the real space grid (which is based on the radial variable), has a relatively quick saturation at a grid height of 16.} after a grid height of 16.

Now, keeping the process grid height at 16, the process grid width may be varied in proportion to the number of Kohn-Sham states that are required for the calculation. The reason for this strategy is clear from Figure~\ref{subfig:grid_width_scaling} -- the two level parallelization scheme is able to make use of the embarrassingly parallel nature of the problem with the number of Kohn-Sham states in use and this is reflected from the nearly perfect weak scaling performance of the code.\footnote{A weak scaling parallel efficiency of over 98 \% is attained with a process grid width of 32, i.e., a total of 512 processes.  We changed the process grid width (from 1 to 32, in multiples of 2) in proportion to the number of states for the weak scaling study, thus varying the total number of MPI processes between 16 and 512.} For this same reason, we have also been able to verify that the strong scaling behavior of our code remains highly favorable\footnote{The strong parallel efficiency remains well above 97 \% at process grid width of 32 (i.e. 512 total MPI processes). {For testing the strong scalability, we kept the basis set size as well as the number of Kohn-Sham states constant (at $10^6$ and $256$ respectively). The process grid height was kept constant, while the process} {grid width was allowed to vary with increase in the number of MPI processes employed. More details may be found in \citep{My_PhD_Thesis}.}} at even 512 total MPI processes. We anticipate that the scaling performance is likely to remain at such favorable levels for even larger numbers of processors.
\begin{figure}[ht]
\centering
\begin{subfigure}[h]{\textwidth}
\centering
\includegraphics[scale=0.20]{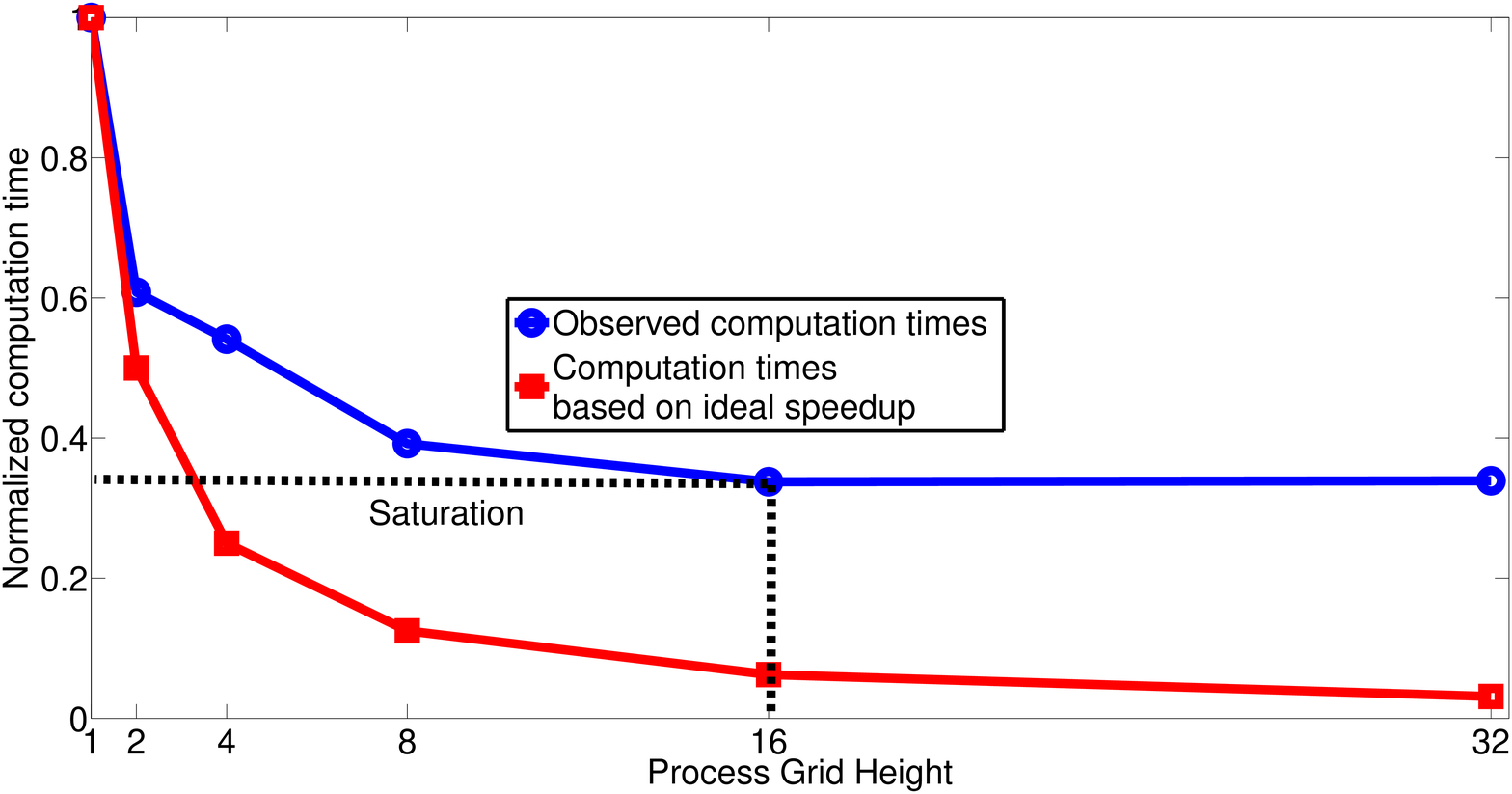}
\caption{Normalized matrix vector product computation time variation with increasing process grid height. A single Kohn-Sham state and $10^6$ basis functions was used for the computation. Saturation occurs at a grid height of 16.\\}
\label{subfig:single_band_saturation}
\end{subfigure}
\begin{subfigure}[h]{\textwidth}
\centering
\includegraphics[scale=0.19]{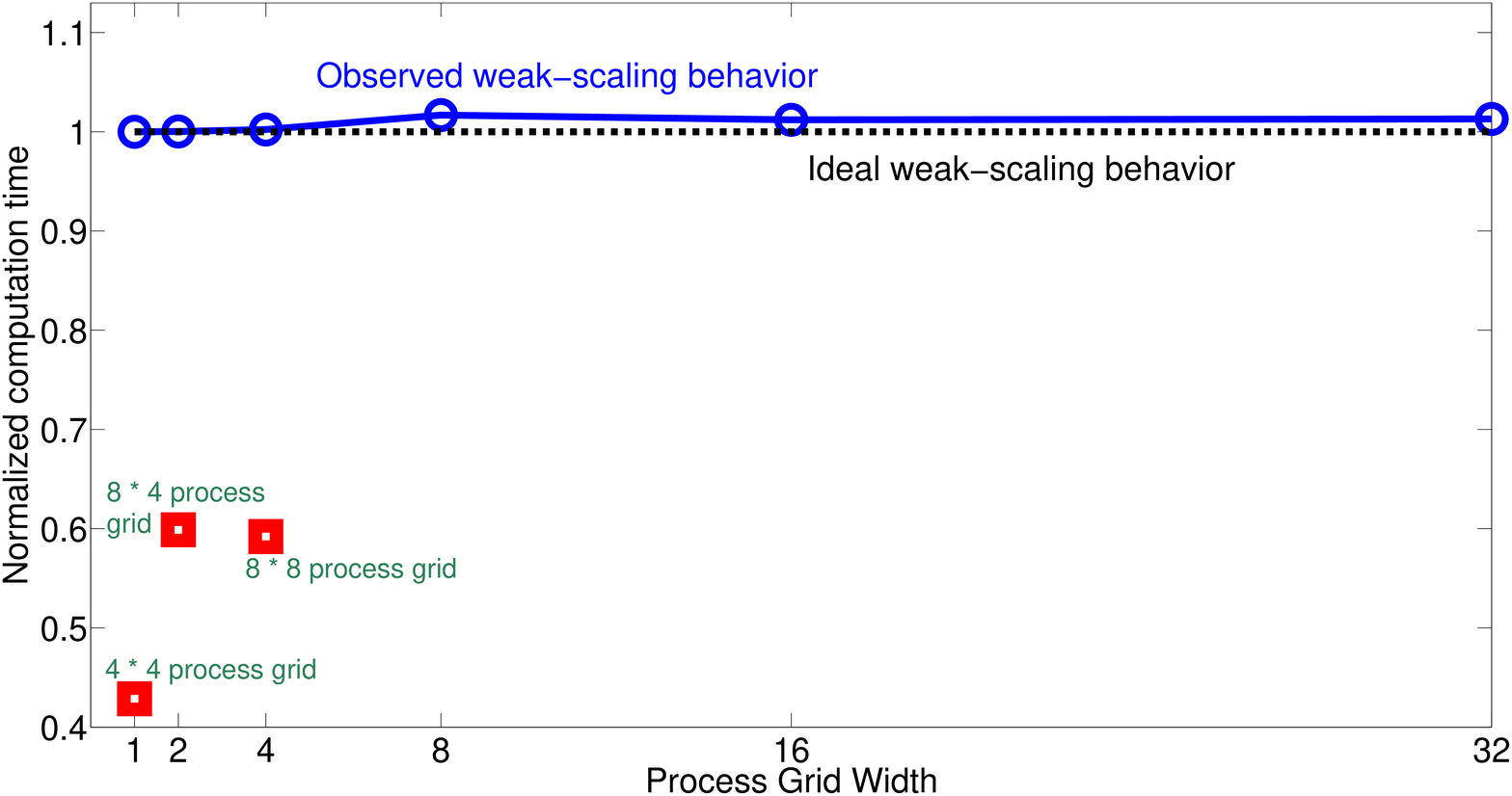}
\caption{Normalized matrix vector product computation time variation with increasing process grid width and proportionate increase in the number of Kohn-Sham states ($10^6$ basis functions for each state). The grid height was kept fixed at 16 for the blue curve and so, a total of 16 to 512 MPI processes were used.}
\label{subfig:grid_width_scaling}
\end{subfigure}
\caption{Parallel scaling efficiency of the matrix-vector product routine and its dependence on process grid geometry.}\label{fig:matvec_scaling}
\end{figure}

Although this straight-forward method for obtaining optimum process grid geometries (based on insights into the scaling performance) is useful, in practice it is sometimes possible to get even better performance from individual process grid configurations, depending on the specific number of basis functions and states in use.  Some examples of such cases are also shown\footnote{The problem sizes used for these individual examples was commensurate with the total number of processes in use, i.e., the $8\times8$ and $16\times4$ process grids were made to use the same number of processes, for example.} in Figure~\ref{subfig:grid_width_scaling}. The overall conclusion that we were able to draw from these studies is that out of the two levels of parallelism available in our implementation, state parallelism is more effective and scalable than the physical domain parallelism.

The scaling performance of the electron density computation routine can be understood on similar lines. We choose to skip further details of this since, unlike the time spent on matrix vector products, the time spent on computing the electron density is typically a relatively minor fraction of the total time spent in every SCF step.
\subsection{Miscellaneous details}
\label{subsec:misc_impl}
We briefly outline miscellaneous implementation related details in this section. 
\subsubsection{Mixing and smearing schemes}
\label{subsubsec:mixing_smearing}
As mentioned earlier, SCF iterations typically employ mixing schemes in order to accelerate convergence towards the fixed point of the Kohn-Sham map \citep{Martin_ES}. The importance of mixing schemes in SCF iterations has been recognized both empirically and theoretically \citep{dederichs_zeller}, leading to the development of various methods over the years \citep{anderson_mixing, broyden_mixing, pulay_mixing, johnson_modified_broyden, cances_mixing, secant_mixing_saad}. We employed the multiple stage Anderson mixing scheme \citep{anderson_mixing, Kohanoff} in this work. Our implementation allows for mixing of the total effective potentials or of the electron density. We have found that potential mixing tends to result in faster convergence of the total energies in most systems. A complete mixing history was used in all the examples and the associated linear mixing parameter used was between 0.1 and 0.3.

Regardless of the mixing procedure, materials systems which have small or no band gaps (metallic systems, for instance) tend to experience convergence issues in the SCF iterations. This occurs due to degenerate energy levels near the Fermi surface in these systems, and it usually manifests itself as \textit{charge sloshing} \citep{Kresse_abinitio_MD}. A common solution to this problem is to introduce \textit{smearing} of the Fermi surface by prescribing a distribution of occupation numbers for the Kohn-Sham states \citep{Kohanoff} . We implemented the widely used Fermi-Dirac distribution for this purpose. This scheme introduces electronic temperature dependent orbital occupations as:
\begin{align}
f_{i} &= \frac{1}{1+\exp\big(\frac{\lambda_i - \epsilon_F}{K_B\,\Theta}\big)}\,,
\intertext{in which, the Fermi level $\epsilon_F$ can be determined by solving the constraint:}
\label{eq:Fermi_constraint}
\int_{\rz^3}\!\rho &= N_e\quad \implies \sum_{i=1}^{N_e / 2}f_{i} = N_e / 2\,.
\end{align}
We solved eq.~\ref{eq:Fermi_constraint} using Brent's method \citep{Brent_Method_Book} and we set the electronic temperature $\Theta$ to 100 -- 200 Kelvin for all simulations where Fermi-Dirac smearing was used.
\subsubsection{The ClusterES package}
\label{subsubsec:ClusterES_package}
We have incorporated all the methods and algorithms discussed so far into an efficient and reliable package called ClusterES (Cluster Electronic Structure). Since our package makes heavy use of Spherical Harmonics Transforms, access to optimized and efficient routines for carrying out these transforms is essential for good performance of our code. We adopted the state of the art SHTns\footnote{We are grateful to Nathana{\"e}l Schaeffer (CNRS, France) for his help and support with the SHTns library.} library \citep{shtns} for this purpose. In spite of using a traditional cubic order algorithm for computation (as opposed to algorithms which are asymptotically faster, e.g. \cite{Mohlenkamp_SHT}) this library has been shown to far outperform other Spherical Harmonics Transform routines because of its use of various hardware level optimizations \citep{shtns}.

The spherical Bessel functions and the Associated Legendre polynomials required for various computations in our code were generated using routines from the GNU Scientific Library \citep{GSL_manual}. Evaluation of the Gauss quadrature weights and nodes were carried out using the algorithm presented in \citep{Golub_Gauss_Quadrature}. Computation of the roots of the Bessel functions was carried out by Halley's method \citep{CS_phase_encyclopedia}.
\subsubsection{Computational platform}
\label{subsubsec:platform_details}
All computations were carried out on the Itasca cluster of the Minnesota Supercomputing Institute. Itasca is an HP Linux cluster with 1,091 HP ProLiant BL280c G6 blade servers, each of which have two-socket, quad-core 2.8 GHz Intel Xeon X5560 ``Nehalem EP" processors sharing 24 GB of system memory, with a 40 gigabit QDR InfiniBand (IB) interconnect. The GNU g++ compiler (ver. 4.8.1) along with Open MPI (ver. 1.7.1) was used and all serial linear algebra and FFT operations were carried out using the hardware optimized Intel Math Kernel Library (ver. 11.0).
\section{Numerical results, example systems and applications}
\label{sec:examples}
We finally describe various numerical results obtained using our method in this section. For all the computations described here, the radius of the spherical domain was chosen by following the procedure suggested in \citep{Chelikowsky_Saad_1}: We first center the cluster~/~molecular system under study at the origin and then ensure that the atom(s) in the system that are furthest from the origin, are atleast 8--16 atomic units away from the boundary of the sphere.\footnote{The specific choice is dictated by computing single atom solutions and observing the decay rates of the electron density in such solutions. If multiple atom species are present, the atom with the slowest decay rate of the electron density is used to set the radius for the spherical domain in case of the cluster system.}
\subsection{Convergence properties}
\label{subsec:convergence}
We begin by studying the convergence properties of our method using numerical examples. First, we computed the ground state of the Hydrogen atom based on the Schr\"odinger equation (i.e., Kohn-Sham self-consistent iterations were not used). This system has the particular advantage that the ground state energy is known analytically to be $-0.5$ Hartrees and so, it serves as an accurate reference for convergence studies. Since the ground state wave function is radially symmetric, we used only the $l =0, m = 0$ spherical harmonic in the angular direction. Figure~\ref{fig:spectral_conv} shows the convergence of the numerical solution to the analytical one with increasing number of basis functions. Due to the Coulombic singularity in the nuclear potential, the plot is a straight line indicating that the convergence rate is polynomial.

Next, we replaced the Coulombic potential for Hydrogen with a smooth pseudopotential as parametrized in \cite{GTH_pseudoptential}. We computed the Kohn-Sham ground state of the Hydrogen (pseudo) atom with this pseudopotential for increasing values of $\calN$, while using only the $l =0, m = 0$ spherical harmonic in the angular direction in every case. We used the $\calN = 50$ case as a reference\footnote{{We verified that this reference case agrees with results from a standard plane-wave code upto atleast $10^{-5}$ Hartrees.}} and plotted the (logarithmic) relative errors with increasing basis set size (with respect to the reference) in Figure~\ref{fig:spectral_conv}. In this case, due to the smoothness of the potential used, the plot has an overall curvature indicating a faster than polynomial rate of convergence (i.e., spectral convergence). 
\begin{figure}[ht]
\centering
\includegraphics[scale=0.28]{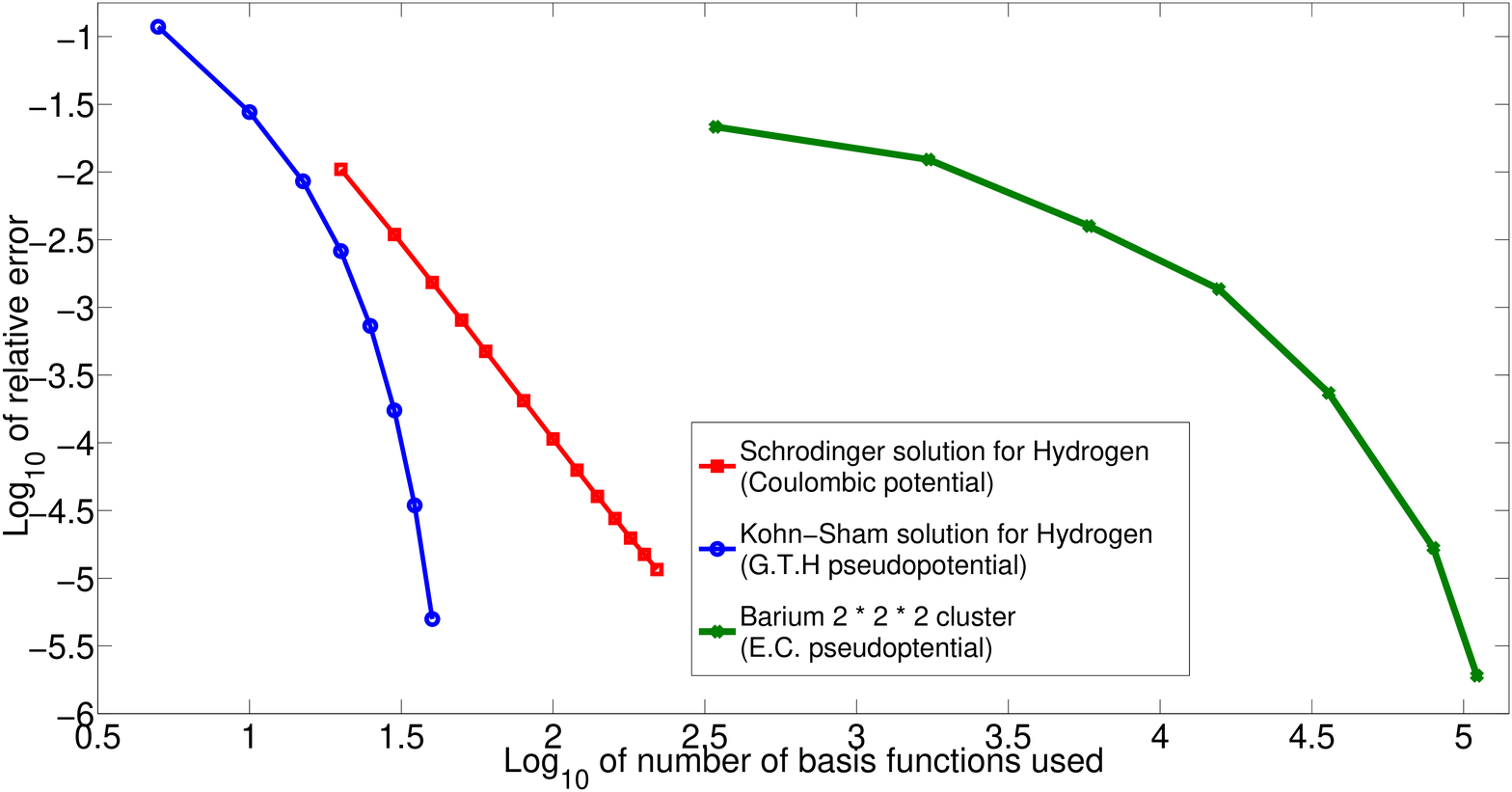}
\caption{Convergence of numerical solutions with increasing number of basis functions. The Coulombic nuclear potential shows a polynomial rate of convergence while smooth pseudopotential solutions show a faster than polynomial rate of convergence.}
\label{fig:spectral_conv}
\end{figure}

Finally, in order to assess the convergence properties for a full scale problem, we computed the ground state of a $2\times2\times2$ body centered cubic (BCC) cluster of Barium. This cluster system has 35 atoms. We employed the smooth `Evanescent Core' local pseudopotential \citep{EC_Fiolhais} for simulation of the Barium atoms. In order to be able to obtain results which can be compared to the literature, we followed \citep{Gavini_higher_order} to use a lattice constant of 9.5 a.u. and an electronic temperature of 200 K for our calculations. To make apparent the convergence rate of our method, we computed the ground state energy of this system using $\calL = 100, \calN = 100$ (i.e., one million basis functions) and used it as a reference value. Thereafter, starting from $\calL = 7, \calN = 7$ we computed the ground state energy for increasing values of $\calL$ and $\calN$ and plotted the (logarithmic) relative errors (compared to the reference value) as a function of the (logarithmic) basis set size as shown in Figure~\ref{fig:spectral_conv}. We see from this figure that because of our use of smooth pseudopotentials, once again, the convergence is rapid: even on a logarithmic scale, there is an overall curvature in the plot, thus indicating faster than polynomial rate of convergence (i.e., spectral convergence). Figure~\ref{fig:Ba2x2x2} shows contour plots of the electron density for the barium cluster obtained using our code. 

For the Barium cluster, the ground state energy per atom obtained using our code  comes out to be $-0.6386253$ Ha which compares well with the value of $-0.6386277$ Ha obtained using a plane-wave code\footnote{The relative difference is of the order of $10^{-6}$.} in \citep{Gavini_higher_order}. This indicates not only rapid convergence of our code but also convergence to the correct value. As a matter of further comparison, we mention that in order to reach these aforementioned numbers, the plane-wave code needed to use over two million plane-waves (mainly arising due to a large vacuum region that had to be used around the cluster) whereas, our code used only $216,000$ basis functions. Even with approximately $55,000$ basis functions (a calculation that took only about 15 c.p.u. minutes on a laptop), we were able to reach convergence levels of about $2\times 10^{-4}$ eV/atom, which is an order of magnitude smaller than the usual levels of convergence demanded in accurate ground state energy calculations.

Aside from the numerical observations presented here, we should point out that an application of the analysis presented in \citep{Zhou_finite_dimensional_numerical_analysis} rigorously establishes that our basis set correctly approximates the Kohn-Sham ground states. A full scale mathematical investigation of the convergence rates of our basis set (similar to the results presented in \citep{Cances_planewave_numerical_analysis} in the context of the plane-wave method) is the scope of future work.\footnote{We are grateful to Eric Canc{\`e}s  (Ecole des Ponts ParisTech, France) for providing us with useful suggestions in this direction.}
\begin{figure}[ht]
\centering
\includegraphics[scale=0.32]{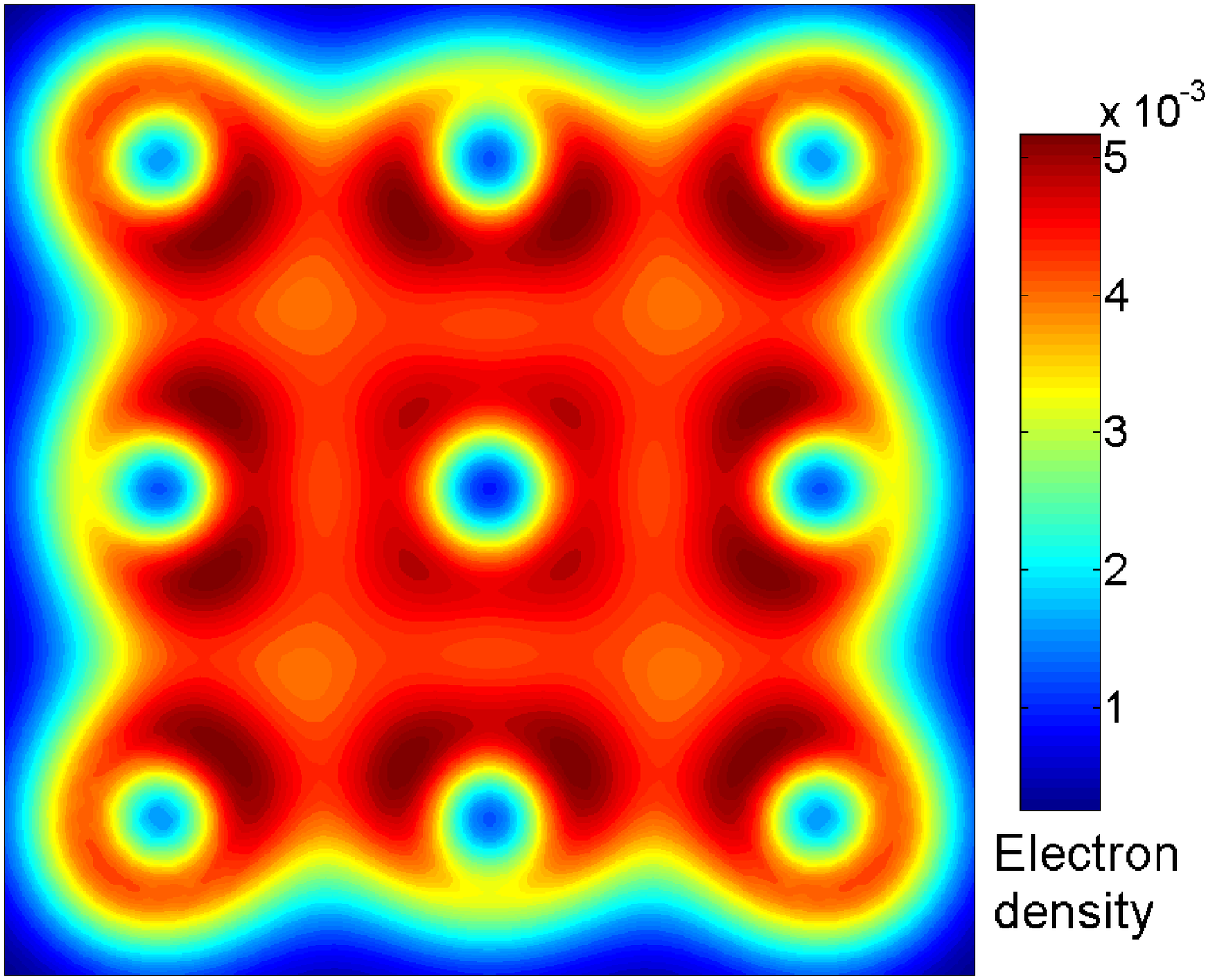}
\caption{Electron density contour plot of the 
$2\times2\times2$ BCC Barium cluster computed using ClusterES.}
\label{fig:Ba2x2x2}
\end{figure}
\subsection{Example calculations of various materials systems}
\label{subsec:example_calculations}
Having ascertained the convergence properties of our method, we now compute the ground state properties of various metallic and non-metallic materials systems using our code and compare our results with the literature. Many of our examples related to metallic clusters (computed using local pseudopotentials) are motivated by recent work in finite element methods for the Kohn-Sham equations \citep{Gavini_Kohn_Sham, Gavini_higher_order}. These examples gave us a way of verifying and benchmarking our calculations, as well as helping establish the relative ease and effectiveness by which such computations can be done routinely using our method. 
\subsubsection{All-electron calculations of light atoms}
\label{subsubsec:light_atom_AE}
We begin by computing the ground state electronic structures of the first few elements of the periodic table. This serves as a simple test of our implementation. No pseudopotential was used. That is, these are all electron calculations. We used the parametrization of the Local Density Approximation as presented in \cite{perdew_zunger, ceperley_alder}. The results of our computations are shown in Table~\ref{table:light_atoms} and compared with values from the literature. 
\begin{table}[ht]
\begin{center}
\begin{tabular}{ | c || c | c | c |}
\hline
Element  & ClusterES & Ref. \cite{Kotochigova_NIST} & 
Ref. \cite{Phanish_max_ent}\\\hline\hline
Hydrogen & -0.445 & -0.445 & -0.445 \\\hline
Helium   & -2.833 & -2.834 & -2.830 \\\hline
Lithium &  -7.327* & -7.335 & -7.338 \\\hline
Beryllium & -14.265* & -14.447 & -14.434 \\\hline
\end{tabular}
\caption{Ground state energies of a few light atoms (Hartree units used). Items marked with * indicate results where basis set convergence was not pursued due to the requirement of a large number of radial basis functions.}
\label{table:light_atoms}
\end{center}
\end{table}
While the results are largely positive, they also illustrate the difficulty that our code faces when dealing with all-electron calculations. In spite of the spherical symmetry of the ground state, the Coulombic singularity at the origin makes it necessary to use a large number of radial basis functions to converge towards expected results. As the atomic numbers increase, so does the strength of the singularity and hence the increased difficulty of the computation. Therefore, we did not pursue the Lithium and Beryllium atom calculations after we ascertained that our results were within about 1\% of the values from the literature. All subsequent calculations reported here employ pseudopotentials to mitigate this issue.
\subsubsection{Local pseudopotential calculations}
\label{subsubsec:local_pseudo}
Having validated the basic correctness of our methodology and implementation using the all electron atomic calculations, we now move to pseudopotential calculations. We first work with the smooth local `Evanescent Core' pseudopotential \citep{EC_Fiolhais}. This bulk-fitted  pseudopotential has been designed to deal with various simple metallic systems and because of the lack of non-local projectors, it is relatively computationally inexpensive. Due to the smoothness of the pseudopotential, we witnessed rapid convergence of our code with increasing basis set size in all the examples that follow.

We first compute the ground state energies of various pseudo-atoms using the pseudopotential and compare with the values from the literature. The results displayed in Table~\ref{Table:ec_atoms} show perfect agreement.
\begin{table}[ht]
\begin{center}
\begin{tabular}{ | c || c | c | c |}
\hline
Element  & ClusterES & Ref. \citep{Nogueira_EC_transferability} & Ref. \cite{Gavini_Kohn_Sham}\\\hline\hline
Lithium & -5.97 & -5.97 & -5.97 \\\hline
Sodium   & -5.21 & -5.21 & -5.21 \\\hline
Magnesium &  -23.06 & -23.06 & -23.05 \\\hline
\end{tabular}
\caption{Ground state energies of a few light atoms (electron volt units used).}
\label{Table:ec_atoms}
\end{center}
\end{table}
Next, we computed the ground state properties of lithium and sodium dimers and octahedral clusters. We computed the binding energy (in electron volts per atom units) and the bond length (in atomic units)  of these systems. For the octahedral clusters, as in \citep{Nogueira_EC_transferability, Phanish_max_ent}, we did not perform any geometry optimization but only sought minima in terms of the nearest neighbour bond length. Also, following these authors, the cluster system ground states were computed without spin polarization while the individual atomic data used spin polarization. {For reference purposes, we also carried out well converged plane-wave calculations\footnote{{We employed an energy cutoff of $30$ Hartrees and a cell length of $30$ atomic units or more in ABINIT.}} on these cluster systems using the ABINIT \citep{Gonze_ABINIT_1} code .} The results are shown in Table~\ref{Table:Li_Na_clusters}.
\begin{table}[ht]
\begin{center}
\begin{tabular}{ |c|c||c|c|c|c| }
\hline
{Cluster} & {Parameters} & {ClusterES} & {{Plane-wave}} &
{Ref. \citep{Phanish_max_ent}} & {Ref.\citep{Nogueira_EC_transferability}}
\\\hline\hline
\multirow{2}{*}{Li$_2$} & Binding Energy  & {-0.48} & {-0.48} & -0.49 & -0.52\\
& Bond Length  & {4.75} & {4.75} & 4.86 & 4.92
\\\hline
\multirow{2}{*}{Na$_2$} & Binding Energy & {-0.35} & {-0.35} & -0.36 & -0.46\\
 &  Bond Length & {5.71} & {5.72} & 5.72 & 5.77
 \\\hline
\multirow{2}{*}{Li$_6$} & Binding Energy & {-0.54} & {-0.54} & -0.50 & -0.72\\
 & Bond Length  & {5.72} & {5.72} & 5.69 & 5.79
\\\hline 
\multirow{2}{*}{Na$_6$} & Binding Energy & {-0.43} & {-0.43} & -0.42 & -0.53\\
 & Bond Length & {6.79} & {6.79} & 6.80 & 6.87
\\\hline
\end{tabular}
\caption{Binding energy in electron volts per atom and bond length in atomic units for sodium and lithium dimers and octahedral clusters.}
\label{Table:Li_Na_clusters}
\end{center}
\end{table}
{We see that our results match with the plane-wave results almost exactly. Additionally, the overall agreement with the values in the literature is also good.} The observable discrepancies with the results of \citep{Nogueira_EC_transferability} is probably because of the use of the LCAO method by those authors. {We are not completely sure of the reasons behind the minor discrepancies with the results of \citep{Phanish_max_ent}. However, there seems to be some confusion in the literature about the correct values of the parameters used by the evanescent core pseudopotential \citep{EC_Fiolhais_erratum} and this might have caused a slightly different set of parameters to have been used in \citep{Phanish_max_ent}. Also, as noted in more recent work \citep{Gavini_higher_order}, higher order finite elements are often necessary for well converged reliable calculations and these were not employed in \citep{Phanish_max_ent}. We believe however, that the precise agreement between our results and the plane-wave code lend support to the credibility of our results.}

Next, we study the properties of a few larger clusters of sodium consisting of $2\times2\times 2$, and $3\times3\times3$ body centered cubic unit cells. We calculated the binding energy per atom and lattice constant for these clusters by computing the total energy for various values of the lattice parameter and then fitting this data to a cubic polynomial. {Our results, compare essentially exactly to the results from well converged plane-wave calculations as Table~\ref{Table:BCC_sodium} shows, assuring us of the efficacy of our method.} {{As a matter of further illustration, let us mention that for the $3\times3\times3$ sodium cluster, at the minimum energy bond length, the total ground state free energies from the plane-wave code and our code are $-20.010982$ Hartrees and $-20.011008$ Hartrees respectively. This corresponds to a difference of less than $0.5$ micro-Hartrees per atom, demonstrating the extremely high accuracies that are easily accessible with our code.}} 

{Other values from the literature are also shown in that table. The overall agreement with these values is also very good (the bond lengths agree to within 1\%, for example). The minor discrepancies from \citep{Gavini_Kohn_Sham} are likely to be explained by the factors mentioned above.}\footnote{{Also, it was not completely clear to us if the authors of \citep{Gavini_Kohn_Sham} used spin-polarization for these particular set of calculations. As before, we computed the cluster system ground states without spin polarization while the individual atomic data used spin polarization.}}
\begin{table}[ht]
\begin{center}
\begin{tabular}{ |c|c||c|c|c| }
\hline
{Sodium cluster} & {Properties} & {ClusterES} & {{Plane-wave}} & Ref. \citep{Gavini_Kohn_Sham}
\\\hline\hline
\multirow{2}{*}{$2\times 2 \times 2$} & Binding Energy (eV/atom) & {-0.71} & {-0.71} & -0.71\\
 & Bond Length (a.u.) & {7.61} & {7.61} & 7.55
\\\hline
\multirow{2}{*}{$3 \times 3 \times 3$} & Binding Energy (eV/atom) & {-0.78} & {-0.78} &-0.80\\
 & Bond Length (a.u.) & {7.78} & {7.78} & 7.75
\\\hline 
\end{tabular}
\caption{Binding energy per atom and lattice constant of sodium BCC unit cells.}
\label{Table:BCC_sodium}
\end{center}
\end{table}
\subsubsection{Non-local pseudopotential calculations}
\label{subsubsec:non_local_pseudo}
In order to deal with a wider variety of materials systems, we now turn to calculations involving {ab initio} norm-conserving non-local pseudopotentials. This class of  pseudopotentials is attractive because the pseudopotentials are accurate and transferable and at the same time, they are available for all elements in the periodic table (including ones which require relativistic treatment of the core electrons). Here, we look at the results obtained using the separable dual space Gaussian pseudopotentials introduced in \cite{GTH_pseudoptential, GTH_relativistic}. This pseudopotential is available in analytical form with a small set of parameters for every element (thus allowing for easy implementation) and it satisfies an optimality criterion for the real-space integration of the nonlocal part. While this pseudopotential is known to be harder than other norm conserving pseudopotentials (i.e., it requires many more basis functions per atom for converged results), it is also known to be more accurate and transferable than other pseudopotentials \citep{GTH_pseudoptential}.

We computed the bond lengths of a few small molecules using our spectral code and compared our results with values from literature, as presented in  Table~\ref{Table:GTH_molecules}. Our results all agree to within 0.2\% of values obtained by the authors of \cite{GTH_pseudoptential}.
\begin{table}[ht]
\begin{center}
\begin{tabular}{| c || c | c |}
\hline
{Molecule} & {Bond length: ClusterES} & {Bond length: Ref. \citep{GTH_pseudoptential}}
\\\hline\hline
CO & {2.128 a.u.} & 2.127 a.u.\\\hline
CH$_4$ & {2.074 a.u.} & 2.072 a.u.\\\hline
SiH$_4$ & {2.810 a.u.} & 2.810 a.u.\\\hline
NH$_3$ & {1.928 a.u.} & 1.931 a.u.\\\hline
H$_2$O & {1.833 a.u.} & 1.835 a.u. 
\\\hline
\end{tabular}
\caption{Bond lengths of a few small molecules computed using the Goedecker-Teter-Hutter pseudopotentials.}
\label{Table:GTH_molecules}
\end{center}
\end{table}

Next, we computed the ground state properties of a few larger systems consisting of organic molecules and fullerenes.\footnote{We are grateful to Qing-Bo Yan (UCAS, China) for making the coordinates of the Boron fullerene available to us.} {We compared our results with the literature as well as with plane-wave code\footnote{{The hardness of the pseudopotentials used often required energy cutoffs as large as $200$ Hartrees to be employed for the plane-wave code.}} calculations (using ABINIT \citep{Gonze_ABINIT_1}) and finite difference method calculations (using the Octopus code \citep{Octopus_1}). The results are presented in Table~\ref{table:buckyball_etc}. The agreement with the plane-wave and finite difference method results is excellent, thereby confirming the efficacy of our method. The overall agreement with other independent sources from the literature is also very good.} The relatively minor differences with the results presented in \citep{zhou_hexahedral_fem} is most likely because of the use of a different pseudopotential by the authors of that work, while the difference from \citep{Boron_fullerenes} occurs probably because of the use of an LCAO basis with a gradient corrected functional by those authors. Figure~\ref{fig:buckyball} shows the electron density iso-surfaces of the Buckyball cluster while figure ~\ref{fig:azobenzene_contours} shows the electron density contour plots for the Azobenzene molecule.
\begin{table}[ht]
\begin{center}
\resizebox{15.5cm}{!}{
\begin{tabular}{ |c|c||c|c|c|c| }
\hline
{System} & {Properties} & {ClusterES} & {Plane-wave} & F.D.M & Other sources
\\\hline\hline
\multirow{2}{*}{} \footnotesize{Benzene} & \footnotesize{Ground State Energy} & {-85.47} & {-85.47} & {-85.48} & {-85.65} \small{(Ref. \citep{zhou_hexahedral_fem})}\\
$\text{C}_{6}\text{H}_{6}$ & \footnotesize{HOMO-LUMO gap} & {5.15} & {5.15} & {5.15} & {5.22}  \small{(Ref. \citep{zhou_hexahedral_fem})}
\\\hline 
\multirow{2}{*}{} \footnotesize{Buckyball} & \footnotesize{Ground State Energy} & {-155.09} & {-155.09} & {-155.09} &-155.02  \small{(Ref. \citep{zhou_hexahedral_fem})}\\
$\text{C}_{60}$ & \footnotesize{HOMO-LUMO gap} & {1.64} & {1.64} & 1.64 & 1.64 \small{(Ref. \citep{C60_bandgap})}
\\\hline 
\multirow{2}{*}{} \footnotesize{Azobenzene} & \footnotesize{Ground State Energy} & {-106.68} & {-106.68} & {-106.68} & {--}\\
$\text{C}_{12}\text{H}_{10}\text{N}_{2}$ & \footnotesize{HOMO-LUMO gap} & {1.39} & {1.39} & {1.39} & --
\\\hline 
\multirow{2}{*}{} \footnotesize{Boron fullerene} & \footnotesize{Ground State Energy} & {-76.94} & {-76.94} & {-76.95} & --\\
$\text{B}_{96}$ & \footnotesize{HOMO-LUMO gap} & {0.79} & {0.79} & 0.79 & 0.78 \small{(Ref. \cite{Boron_fullerenes})}
\\\hline 
\end{tabular}}
\caption{Ground State Energy (eV/atom) and HOMO-LUMO gap (eV) of some organic molecules and fullerenes computed using our code and compared with  results obtained from other sources. F.D.M denotes finite difference method  calculations done using the Octopus \citep{Octopus_1} code. Plane-wave calculations were carried out using the ABINIT code \citep{Gonze_ABINIT_1}.}
\label{table:buckyball_etc}
\end{center}
\end{table}
\begin{figure}[ht]
\centering
\begin{subfigure}[h]{0.45\textwidth}
\centering
\includegraphics[scale=0.30]{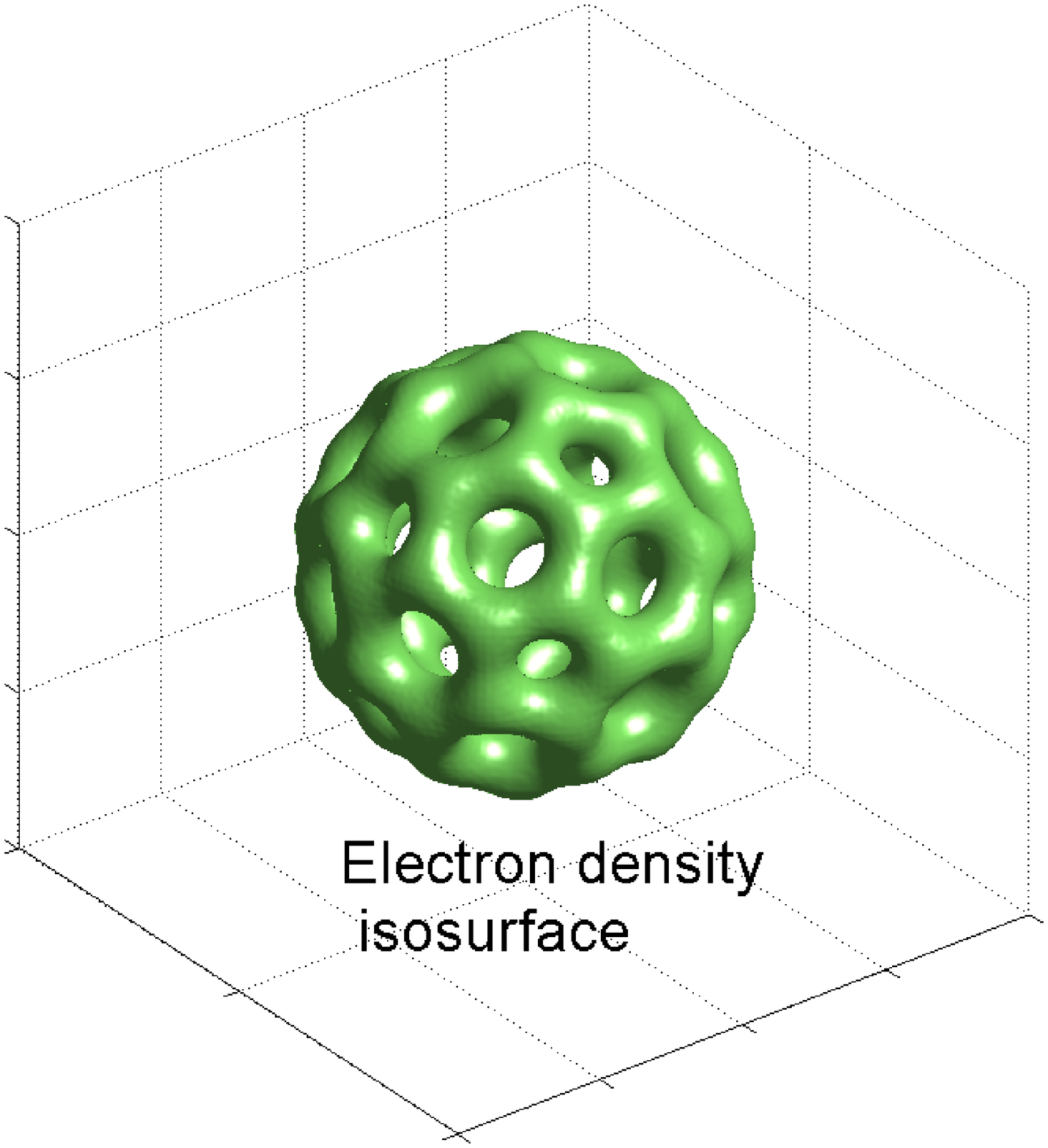}
\caption{Electron density isosurface of $\text{C}_{60}$.}
\label{fig:buckyball}
\end{subfigure}
\begin{subfigure}[h]{0.45\textwidth}
\centering
\includegraphics[scale=0.30]{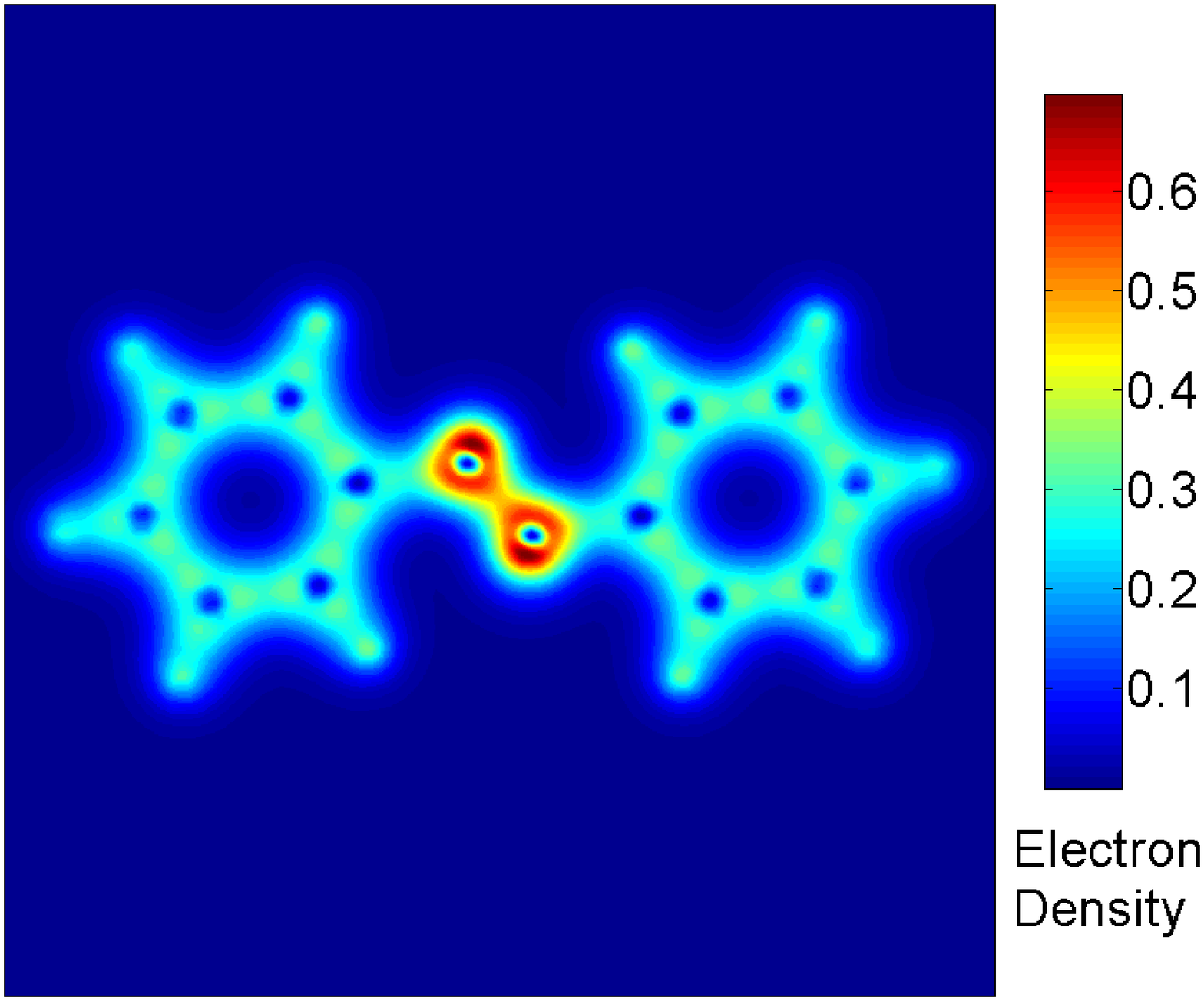}
\caption{Electron density contours of azobenzene.}
\label{fig:azobenzene_contours}
\end{subfigure}
\caption{Ground state electron density plots for the $\text{C}_{60}$ Buckyball and the azobenzene molecule.}\label{fig:C60_azobenzene}
\end{figure}

Some of the examples presented above (for both local and non-local pseudopotentials) highlight the fact that our code is easily able to handle arbitrary cluster / molecule shapes and geometries. Indeed, thanks to the convergence properties of our basis set and our efficient implementation, even linear or planar molecules, which are quite far from spherical shapes, present no issues.\footnote{{In order to further verify that our code does not face any difficulties in dealing with  asymmetric or non-spherical systems, we carried out the following test (suggested to us by an anonymous reviewer): We computed the ground state energy of single silicon atom placed at the origin of a domain of radius $20$ a.u.} {and then observed the change in energy of the system as this atom was moved outward radially. We observed that even at a radial distance of 8 a.u. from the origin, the energy change in the system remained less than $0.1$ milli eV. The basis set size used in these examples was not particularly large - in fact even when we used a basis set which was $40$ \% smaller in size, the change in energy of the system changed only by $0.25$ milli eV when placed at a distance of $8$ a.u. from the origin. Thus, the origin does not have a special status in our method and our code does not seem to require the use of very large basis sets in dealing with highly asymmetric systems. Of course, in most practical situations, the system under study is placed such that it's center of mass coincides with the origin, atleast approximately.}}

The use of spherical harmonics in our basis set makes it very convenient to systematically compute several other quantities of interest, such as electrostatic multipole moments. Indeed, following the expressions presented in \citep[page 108]{dipole_electrostatic_book} we see that quadruple or dipole moments can be easily obtained in our method by carrying out computations similar to the computation of forward basis transforms. We carried out this exercise for obtaining the dipole moment of the carbon monoxide molecule. We chose this particular example since it appears to us that various authors seem to have obtained a wide range of values of this quantity due to systematic errors in their computations -- probably, either through incomplete basis sets (see \citep{gunnarsson_diatomic} and \citep{dipole_electrostatic_book} where values of -0.01 D and -0.60 D are respectively mentioned) or possibly through the use of unconverged grids or inaccurate pseudopotentials (as apparently obtained in \citep{Chelikowsky_Saad_1}). The currently accepted value of this quantity at equilibrium bond length (using LDA calculations)\footnote{This differs from the experimental value by about a factor of two \citep[see e.g.][]{dipole_moments_1}. This discrepancy is usually ascribed to correlation effects being insufficiently modelled in Kohn-Sham LDA calculations \citep{dipole_electrostatic_book}.} appears to be about -0.22 D \citep{dipole_electrostatic_book, dipole_moments_1, dipole_moments_2} which agrees well with our results as Table~\ref{table:CO_Dipole} shows.
\begin{table}[ht]
\begin{center}

\begin{tabular}{ |c|c||c|c|c| }
\hline
{System} & {Property} & {ClusterES} & Ref. \citep{dipole_moments_1} & F.D.M
\\\hline\hline
\multirow{2}{*}{} CO  & Dipole Moment (Debye) & {-0.23} & -0.22 & {-0.23}
\\\hline 
\end{tabular}
\caption{Dipole moment of the carbon monoxide molecule at equilibrium bond length. F.D.M stands for a calculation using the finite difference method carried out using the Octopus code \citep{Octopus_1}.}
\label{table:CO_Dipole}
\end{center}
\end{table}
\subsection{Benchmark calculations on large systems}
\label{subsec:benchmark}
Finally, in order to demonstrate the capabilities of our method in dealing with large systems efficiently, we carry out computations of the ground states of large aluminum clusters. We looked at $3\times3\times3$, $5\times5\times5$ and $7\times7\times7$ face centered cubic (FCC) clusters for this study. The lattice spacing was fixed at 7.45 a.u. for all the clusters and we used the `Evanescent Core' pseudopotential \citep{EC_Fiolhais} for these calculations. A thermalization temperature of 100 Kelvin was used. For the $3\times3\times3$ and $5\times5\times5$ clusters, in order to assess the efficacy of our method, we aimed to converge our ground state energies (per atom) to within one--two milli electron volts of the plane-wave and higher order finite element method (FEM) results\footnote{This corresponds to relative errors of the order of $10^{-5}$.}\footnote{{It was pointed out to us by an anonymous reviewer that these levels are convergence are more demanding than the standards typically adhered to in many electronic structure calculations. Often, it is sufficient for the ground state energy to reach convergence levels of about $1$ milli-Hartree per atom. This does not change the conclusions about the benchmark calculations laid out in this section, in the sense that the performance of our code remains very favourable when compared with a standard plane-wave code (ABINIT, \citep{Gonze_ABINIT_1}), if both codes are made to use the minimum number of basis functions that would allow them to reach convergence levels of $1$ milli-Hartree per atom in the ground state energies. In a cluster system involving 62 aluminum atoms (FCC structure with a single mono-vacancy), our code had a wall clock time which was about $1.8$} {times smaller than the wall clock time registered by the plane-wave code when both codes were made to attain this level of convergence. {Both codes were executed with the same computational resources -- specifically, each code was run on a single node of the Itasca cluster (hardware described at the beginning of Section \ref{sec:examples}) and a single computational thread on a single core of this node was used}. For reaching this convergence level, the plane-wave code used over $370,000$ basis functions, while our code used only about $43,000$. We systematically ensured that the minimum energy cutoff and cell size that would be required by the plane-wave code to reach the desired accuracy was used. We intend to present more details of these kinds of benchmarks in future work.}} presented in \cite{Gavini_higher_order}. For the 7$\times$7$\times$7 cluster, due to computational resource constraints, we used a somewhat smaller basis set than what would be required to achieve this same level of convergence. So we present here results in which 
the total energy was within 0.01 electron volts per atom of the higher order finite element method (FEM) results. The results are shown in Table~\ref{table:Al_clusters}.
\begin{table}[ht]
\small
\begin{center}
\begin{tabular}{ | c | c | c || c | c | c |}
\hline
System  & No. atoms & No. electrons & ClusterES & Plane-wave & FEM\\\hline\hline
$3\times3\times3$ & 172 & 516 & -56.01809  & -56.01814
 & -56.01776\\\hline
$5\times5\times5$ & 666 & 1998 & -56.05057 &  -56.05068 & -56.04906\\\hline
$7\times7\times7$ & 1688 & 5064 & -56.05812  & -- &  -56.06826\\\hline
\end{tabular}
\caption{Ground state energy per atom of large aluminum clusters. Electron-volt units used. The plane wave and FEM results were obtained from reference \cite{Gavini_higher_order}.}
\label{table:Al_clusters}
\end{center}
\end{table}
To show that our methodology and its implementation is highly competitive with existing methods, we display in Table~\ref{table:run_times} timing results\footnote{{In order to enable comparison with the results in \citep{Gavini_higher_order}, we report here timings in total c.p.u hours. This was estimated by multiplying the wall clock timings with the number of MPI processes used. The $3\times3\times3$} {and $5\times5\times5$ aluminum clusters used 16 MPI and 256 MPI processes respectively.}} of the $3\times3\times3$ and $5\times5\times5$ systems and compare it with the results presented in \cite{Gavini_higher_order}. The computational platform used by the authors in \cite{Gavini_higher_order} was quite similar to our own, if not by some measures, superior to ours.  Nevertheless, due to the fast convergence of our spectral basis set, the efficient basis transforms,  and various other algorithmic methodologies adopted here, our code was able to well outperform the plane-wave and finite element codes. In particular, in spite of having access to highly efficient FFTs, the plane-wave code performance seems to have suffered due to the requirement of having large supercells (with large vacuum regions) for obtaining converged results with these clusters.
\begin{table}[ht]
%\small
\begin{center}
\begin{tabular}{ | c || c | c | c |}
\hline
System  & ClusterES & Plane-wave & FEM\\\hline\hline
$3\times3\times3$ FCC Aluminum cluster &  18  & 646 & 371\\\hline
$5\times5\times5$ FCC Aluminum cluster & 1948 & 7307 & 6619\\\hline
\end{tabular}
\caption{Computational run times of ClusterES compared against existing plane-wave and FEM codes. All run times are presented in c.p.u. hours. The plane wave and FEM results were obtained from reference \cite{Gavini_higher_order}.}
\label{table:run_times}
\end{center}
\end{table}
\subsection{Brief comments on symmetry adaptation}
\label{subsec:symmetry_adaptation}
Most plane-wave codes allow for some method of symmetry adaptation, usually in the form of special point sampling methods for the Brillouin zone \citep{chadi_cohen_special_points, evarestov_special_points}. Due to the formal similarities of our method with the plane-wave method, it is natural to investigate if symmetry adaptation can be carried out in a straight-forward way in our setting. 

There indeed seems to be a relatively simple way of carrying out this enterprise. The key point is that (like in the case of plane-waves), the basis functions in use arise as eigenfunctions of the Laplacian operator. This operator commutes with all relevant symmetry operations -- it commutes with translational symmetry in a periodic setting and similarly it commutes with all point group operations in our setting. In our case, this results in the fact that point group actions on the basis set can be computed easily : the spherical harmonics ensure that symmetry group action on the basis set can be written down analytically in terms of Wigner D-matrices \citep{Edmonds_angular_quantum}. Therefore, this gives us an efficient method of constructing Peter-Weyl projectors \citep{Folland_Harmonic, Barut_Reps} onto the symmetry invariant irreducible subspaces of the problem at hand. These projected subspaces can then be employed, in conjunction with subspace iteration methods (such as Chebyshev-Filtered SCF iterations) to obtain a symmetry adapted reduction of a given problem. 

A full scale report on symmetry adaptation within our basis set highlighting these points is currently  under preparation.
\section{Conclusions and future directions}
\label{sec:conclusions}
In summary, we have proposed and implemented a method for efficient and accurate solution of the Kohn-Sham equations for clusters. This method serves as an analog of the plane-wave method for periodic systems and similar to that method, it shows  rapid and systematic convergence properties. We have demonstrated that with the adoption of various algorithmic strategies, our method produces reliable results for a vast array of materials systems. In terms of performance metrics, benchmark calculations on various cluster systems show that our method is highly competitive when compared with other established basis sets and methods, both in terms of accuracy and speed. The formal analogies of our method with the plane-wave method allow us to adopt, mutatis mutandis, a multitude of numerical and algorithmic strategies commonly employed by the plane-wave method, which eventually lead to the efficient and reliable performance of our implementation.

An additional outcome is that our method forms a systematic generalization of approximate spherical basis function based methods introduced earlier in the literature in a variety of contexts (most commonly for the purpose of jellium calculations). Our method retains the basic simplicity of those methods (since the basis functions employed are of a similar nature), but it has far superior performance and applicability than any of those approximate methods. Our basis functions allow arbitrary point group symmetries to be exploited systematically, and obtaining leverage out of this fact constitutes the subject of on going and future work. A promising area of research is to use the ClusterES package for first principles materials discovery based on the ideas broadly outlined in \citep{James_OS}.

{In a separate contribution, we are currently following up on this work by demonstrating the use of our method in the accurate computation of quantum mechanical forces. It appears to us that the application of the Hellman-Feynman force formula is straight forward in our method: the global nature of the basis results in the absence of Pulay forces and further, the spectral convergence properties that are inherent to our basis also carry over to the forces. In the near future, we aim to carry out abinitio molecular dynamics simulations of various cluster systems of interest using our method.}

{Currently, one of the main computational bottlenecks in carrying out basis transforms is due to the transform in the radial direction, which scales quadratically in the number of radial basis functions (see the discussion following eq.~\ref{eq:G_lmr}). So far, our use of Gauss quadrature and that of machine optimized libraries has enabled us to keep the constant in front of this asymptotic expression small, thus leading to the competitive run times of our code in practice. In the long run, however, an asymptotically faster algorithm should be employed and we intend to explore various possibilities in this direction since it has a direct bearing on our ability to successfully tackle even larger systems of interest.}

{Finally, due to its use of Dirichlet boundary conditions, our proposed method allows for charged systems to be easily studied without the need for introducing an artificial background charge (as currently used in plane-wave codes). Thus, the study of charged cluster systems\footnote{{This possible avenue of research was suggested to us by an anonymous reviewer.}} using our approach is likely to be a fruitful avenue of research in the near future.}
\section*{Acknowledgement}
This work was primarily supported by Russell Penrose.  It also benefited from the support of NSF-PIRE Grant No. OISE-0967140, ONR
N00014-14-1-0714 and the MURI project FA9550-12-1-0458 (administered by AFOSR). We would like to thank the Minnesota Supercomputing Institute for making the parallel computing resources used in this work available. We would like to thank Phanish Suryanarayana (Georgia Tech.) for his many insightful comments and suggestions at various stages of this work. We would like to thank Vikram Gavini and Phani Motamarri (U. Michigan) for stimulating discussions as well as for making available some of their Finite Element Method results which helped us in carrying out validation studies. We would also like to thank Gero Friesecke (TU Munich, Germany) and Michael Ortiz (Caltech) for informative discussions. {We gratefully acknowledge comments from the anonymous reviewers which helped us in improving the presentation of our work.} ASB and RDJ would like to acknowledge the hospitality of the Hausdorff Research Institute for Mathematics, Bonn, Germany where this work was partially carried out.
\bibliographystyle{elsarticle-num}
\bibliography{main}

\begin{thebibliography}{100}
\expandafter\ifx\csname url\endcsname\relax
  \def\url#1{\texttt{#1}}\fi
\expandafter\ifx\csname urlprefix\endcsname\relax\def\urlprefix{URL }\fi
\expandafter\ifx\csname href\endcsname\relax
  \def\href#1#2{#2} \def\path#1{#1}\fi

\bibitem{Martin_ES}
R.~M. Martin, Electronic Structure: Basic Theory and Practical Methods, 1st
  Edition, Cambridge University Press, 2004.

\bibitem{LeBris_ReviewBook}
C.~{Le Bris} (Ed.), Computational Chemistry, Vol.~X of Handbook of Numerical
  Analysis, North-Holland, 2003.

\bibitem{Saad_Chelikowsky_Shontz_review}
Y.~Saad, J.~R. Chelikowsky, S.~M. Shontz, Numerical methods for electronic
  structure calculations of materials, SIAM Review 52~(1) (2010) 3--54.

\bibitem{Kresse_abinitio_iterative}
G.~Kresse, J.~Furthmuller, Efficient iterative schemes for ab initio
  total-energy calculations using a plane-wave basis set, Physical Review B 54
  (1996) 11169--11186.

\bibitem{Kresse_metal_semiconductor}
G.~Kresse, J.~Furthmuller, Efficiency of ab-initio total energy calculations
  for metals and semiconductors using a plane-wave basis set., Computational
  Materials Science 6~(1) (1996) 15 -- 50.

\bibitem{Hutter_abinitio_MD}
D.~Marx, J.~Hutter, Ab initio molecular dynamics: basic theory and advanced
  methods, 1st Edition, Cambridge University Press, 2009.

\bibitem{Teter_Payne_Allan_2}
M.~C. Payne, M.~P. Teter, D.~C. Allan, T.~A. Arias, J.~D. Joannopoulos,
  Iterative minimization techniques for \textit{ab initio} total-energy
  calculations: molecular dynamics and conjugate gradients, Rev. Mod. Phys. 64
  (1992) 1045--1097.

\bibitem{Barnett_Landman}
R.~N. Barnett, U.~Landman, Born-oppenheimer molecular-dynamics simulations of
  finite systems: Structure and dynamics of $(\text{H}_2\text{O})_2$, Physical
  review B 48~(4) (1993) 2081.

\bibitem{Gonze_ABINIT_1}
X.~Gonze, J.-M. Beuken, R.~Caracas, F.~Detraux, M.~Fuchs, G.-M. Rignanese,
  L.~Sindic, M.~Verstraete, G.~Zerah, F.~Jollet, M.~Torrent, A.~Roy, M.~Mikami,
  P.~Ghosez, J.-Y. Raty, D.~Allan, First-principles computation of material
  properties: the {ABINIT} software project, Computational Materials Science
  25~(3) (2002) 478 -- 492.

\bibitem{CASTEP_1}
M.~D. Segall, P.~J.~D. Lindan, M.~J. Probert, C.~J. Pickard, P.~J. Hasnip,
  S.~J. Clark, M.~C. Payne, First-principles simulation: ideas, illustrations
  and the {CASTEP} code, Journal of Physics: Condensed Matter 14~(11) (2002)
  2717.

\bibitem{Quantum_Espresso_1}
P.~Giannozzi, S.~Baroni, N.~Bonini, M.~Calandra, R.~Car, C.~Cavazzoni,
  D.~Ceresoli, G.~L. Chiarotti, M.~Cococcioni, I.~Dabo, A.~D. Corso,
  S.~de~Gironcoli, S.~Fabris, G.~Fratesi, R.~Gebauer, U.~Gerstmann,
  C.~Gougoussis, A.~Kokalj, M.~Lazzeri, L.~Martin-Samos, N.~Marzari, F.~Mauri,
  R.~Mazzarello, S.~Paolini, A.~Pasquarello, L.~Paulatto, C.~Sbraccia,
  S.~Scandolo, G.~Sclauzero, A.~P. Seitsonen, A.~Smogunov, P.~Umari, R.~M.
  Wentzcovitch, {QUANTUM ESPRESSO}: a modular and open-source software project
  for quantum simulations of materials, Journal of Physics: Condensed Matter
  21~(39).

\bibitem{Cances_planewave_numerical_analysis}
E.~Canc{\`e}s, R.~Chakir, Y.~Maday, Numerical analysis of the planewave
  discretization of some orbital-free and {K}ohn-{S}ham models, ESAIM:
  Mathematical Modelling and Numerical Analysis 46 (2012) 341--388.

\bibitem{Rappe_planewaves_molecules}
A.~M. Rappe, J.~Joannopoulos, P.~Bash, A test of the utility of plane-waves for
  the study of molecules from first principles, Journal of the American
  Chemical Society 114~(16) (1992) 6466--6469.

\bibitem{LCAO_3}
W.~Hehre, R.~Stewart, J.~Pople, Self-consistent molecular-orbital methods. {I}.
  use of {G}aussian expansions of slater-type atomic orbitals, The Journal of
  Chemical Physics 51~(6) (1969) 2657--2664.

\bibitem{LCAO_famous}
J.~C. Slater, G.~F. Koster, Simplified {LCAO} method for the periodic potential
  problem, Phys. Rev. 94 (1954) 1498--1524.

\bibitem{SIESTA_1}
J.~Soler, E.~Artacho, J.~Gale, A.~García, J.~Junquera, P.~Ordejón,
  D.~Sánchez-Portal, The {SIESTA} method for ab initio order-n materials
  simulation, Journal of Physics Condensed Matter 14~(11) (2002) 2745--2779.

\bibitem{Chelikowsky_Saad_1}
J.~R. Chelikowsky, N.~Troullier, K.~Wu, Y.~Saad, {Higher order} {finite
  difference} pseudopotential method: An application to diatomic molecules,
  Phys. Rev. B 50 (1994) 11355--11364.

\bibitem{Octopus_1}
A.~Castro, H.~Appel, M.~Oliveira, C.~Rozzi, X.~Andrade, F.~Lorenzen,
  M.~Marques, E.~Gross, A.~Rubio, Octopus: A tool for the application of
  time-dependent density functional theory, Physica Status Solidi (B) Basic
  Research 243~(11) (2006) 2465--2488.

\bibitem{Pask_FEM_review}
J.~Pask, P.~Sterne, Finite element methods in ab initio electronic structure
  calculations, Modelling and Simulation in Materials Science and Engineering
  13~(3) (2005) R71.

\bibitem{Gavini_Kohn_Sham}
P.~Suryanarayana, V.~Gavini, T.~Blesgen, K.~Bhattacharya, M.~Ortiz,
  Non-periodic finite-element formulation of {K}ohn-{S}ham density functional
  theory, Journal of the Mechanics and Physics of Solids 58~(2) (2010)
  256--280.

\bibitem{Gavini_higher_order}
P.~Motamarri, M.~Nowak, K.~Leiter, J.~Knap, V.~Gavini, Higher-order adaptive
  finite-element methods for {K}ohn-{S}ham density functional theory, Journal
  of Computational Physics 253 (2013) 308--343.

\bibitem{Optical_magnetic_Boron_fullerene}
S.~Botti, A.~Castro, N.~N. Lathiotakis, X.~Andrade, M.~A.~L. Marques, Optical
  and magnetic properties of boron fullerenes, Phys. Chem. Chem. Phys. 11
  (2009) 4523--4527.

\bibitem{Chelikowsky_silicon_nanostructures}
X.~Jing, N.~Troullier, J.~R. Chelikowsky, K.~Wu, Y.~Saad, Vibrational modes of
  silicon nanostructures, Solid State Communications 96~(4) (1995) 231 -- 235.

\bibitem{PARSEC}
L.~Kronik, A.~Makmal, M.~L. Tiago, M.~M.~G. Alemany, M.~Jain, X.~Huang,
  Y.~Saad, J.~R. Chelikowsky, Parsec – the pseudopotential algorithm for
  real-space electronic structure calculations: recent advances and novel
  applications to nano-structures, Physica Status Solidi (b) 243~(5) (2006)
  1063--1079.

\bibitem{Parallel_Chebyshev}
Y.~Zhou, Y.~Saad, M.~L. Tiago, J.~R. Chelikowsky, Parallel
  self-consistent-field calculations via {Chebyshev}-filtered subspace
  acceleration, Phys. Rev. E 74 (2006) 066704.

\bibitem{Abinito_Fullerenes_Science}
G.~E. Scuseria, Ab initio calculations of fullerenes, Science 271~(5251) (1996)
  942--945.

\bibitem{Small_Metal_Clusters}
V.~Gurin, Small metal clusters: Ab initio calculated bare clusters and models
  within fullerene cages, in: E.~Buzaneva, P.~Scharff (Eds.), Frontiers of
  Multifunctional Integrated Nanosystems, Vol. 152 of NATO Science Series II:
  Mathematics, Physics and Chemistry, Springer Netherlands, 2005, pp. 31--38.

\bibitem{B80_abinitio}
N.~Gonzalez~Szwacki, A.~Sadrzadeh, B.~I. Yakobson, {${\mathrm{B}}_{80}$}
  fullerene: An \textit{Ab~Initio} prediction of geometry, stability, and
  electronic structure, Phys. Rev. Lett. 98 (2007) 166804.

\bibitem{Gold_atomic_electronic_structure}
H.~Hakkinen, Atomic and electronic structure of gold clusters: understanding
  flakes, cages and superatoms from simple concepts, Chem. Soc. Rev. 37 (2008)
  1847--1859.

\bibitem{Super_Atoms_1}
A.~W. Castleman, S.~N. Khanna, Clusters, superatoms, and building blocks of new
  materials†, The Journal of Physical Chemistry C 113~(7) (2009) 2664--2675.

\bibitem{Super_Atoms_2}
A.~W. Castleman, From elements to clusters: The periodic table revisited, The
  Journal of Physical Chemistry Letters 2~(9) (2011) 1062--1069.

\bibitem{electronic_sodium_magnesium_clusters}
M.~I{\~n}iguez, M.~Lopez, J.~Alonso, J.~Soler, Electronic and atomic structure
  of {Na}, {Mg}, {Al} and {Pb} clusters, Zeitschrift f{\"u}r Physik D Atoms,
  Molecules and Clusters 11~(2) (1988) 163--174.

\bibitem{spherical_averaged_jellium}
G.~Mattei, F.~Toigo, Spherical averaged jellium model with norm-conserving
  pseudopotentials, The European Physical Journal D $-$ Atomic, Molecular,
  Optical and Plasma Physics 3~(3) (1998) 245--256.

\bibitem{C60_in_spherical_basis}
K.~Yabana, G.~F. Bertsch, Electronic structure of {$\textrm{C}_{60}$} in a
  spherical basis, Physica Scripta 48~(5) (1993) 633.

\bibitem{Broglia_original_paper}
F.~Alasia, R.~A. Broglia, H.~E. Roman, L.~Serra, G.~Colo, J.~M. Pacheco,
  Single-particle and collective degrees of freedom in {$\textrm{C}_{60}$},
  Journal of Physics B: Atomic, Molecular and Optical Physics 27~(18) (1994)
  L643.

\bibitem{solid_state_finite}
R.~Broglia, G.~Col{\'o}, G.~Onida, H.~Roman, Solid State Physiscs of Finite
  Systems: Metallic Clusters, Fullerenes, Atomic Wires, 1st Edition, Advanced
  Texts in Physics, Springer, 2004.

\bibitem{Review_metal_clusters}
M.~Brack, The physics of simple metal clusters: self-consistent jellium model
  and semiclassical approaches, Rev. Mod. Phys. 65 (1993) 677--732.

\bibitem{Gauss_Jacobi_Quad}
P.~P. Teodorescu, N.-D. Stanescu, N.~Pandrea, Numerical Analysis with
  Applications in Mechanics and Engineering, 1st Edition, John Wiley \& Sons,
  2013.

\bibitem{shtns}
N.~Schaeffer, Efficient spherical harmonic transforms aimed at pseudospectral
  numerical simulations, Geochemistry, Geophysics, Geosystems 14~(3) (2013)
  751--758.

\bibitem{Hockney_1970}
R.~Hockney, Potential calculation and some applications., Methods Comput. Phys.
  9: 135-211(1970).

\bibitem{Eastwood_Brownrig}
J.~Eastwood, D.~Brownrigg, Remarks on the solution of {P}oisson's equation for
  isolated systems, Journal of Computational Physics 32~(1) (1979) 24 -- 38.

\bibitem{Reciprocal_Hockney}
G.~J. Martyna, M.~E. Tuckerman, A reciprocal space based method for treating
  long range interactions in ab initio and force-field-based calculations in
  clusters, The Journal of Chemical Physics 110~(6) (1999) 2810--2821.

\bibitem{genovese2006efficient}
L.~Genovese, T.~Deutsch, A.~Neelov, S.~Goedecker, G.~Beylkin, Efficient
  solution of poisson's equation with free boundary conditions, The Journal of
  chemical physics 125~(7) (2006) 074105.

\bibitem{various_eigensolvers}
C.~V{\"o}mel, S.~Z. Tomov, O.~A. Marques, A.~Canning, L.-W. Wang, J.~J.
  Dongarra, State-of-the-art eigensolvers for electronic structure calculations
  of large scale nano-systems, Journal of Computational Physics 227~(15) (2008)
  7113 -- 7124.

\bibitem{LOBPCG_1}
A.~Knyazev, Toward the optimal preconditioned eigensolver: Locally optimal
  block preconditioned conjugate gradient method, SIAM Journal on Scientific
  Computing 23~(2) (2001) 517--541.

\bibitem{ABINIT_LOBPCG}
F.~Bottin, S.~Leroux, A.~Knyazev, G.~Z{\'e}rah, Large-scale ab initio
  calculations based on three levels of parallelization, Computational
  Materials Science 42~(2) (2008) 329 -- 336.

\bibitem{Octopus_LOBPCG}
F.~Lorenzen, Massively-parallel eigensolver for the {Octopus} code, (Online
  article) (2006).

\bibitem{Serial_Chebyshev}
Y.~Zhou, Y.~Saad, M.~L. Tiago, J.~R. Chelikowsky, Self-consistent-field
  calculations using {Chebyshev}-filtered subspace iteration, Journal of
  Computational Physics 219 (2006) 172--184.

\bibitem{Gygi_2D_parallel}
F.~Gygi, E.~W. Draeger, M.~Schulz, B.~R. de~Supinski, J.~A. Gunnels, V.~Austel,
  J.~C. Sexton, F.~Franchetti, S.~Kral, C.~W. Ueberhuber, J.~Lorenz,
  Large-scale electronic structure calculations of high-z metals on the
  bluegene/l platform, in: Proceedings of the 2006 ACM/IEEE conference on
  Supercomputing, SC '06, ACM, New York, NY, USA, 2006.

\bibitem{KohnSham_DFT}
W.~Kohn, L.~J. Sham, Self-consistent equations including exchange and
  correlation effects, Physical Review 140~(4A) (1965) 1133--1138.

\bibitem{Folland_Real}
G.~B. Folland, Real Analysis: Modern Techniques and Their Applications, 2nd
  Edition, Wiley, 1999.

\bibitem{Troullier_Martins_pseudo}
N.~Troullier, J.~L. Martins, Efficient pseudopotentials for plane-wave
  calculations, Physical Review B 43~(3) (1991) 1993.

\bibitem{Parr_Yang}
R.~G. Parr, W.~Yang, Density-Functional Theory of Atoms and Molecules, Vol.~16
  of International Series of Monographs on Chemistry, Oxford University Press,
  USA, 1994.

\bibitem{perdew_zunger}
J.~P. Perdew, A.~Zunger, Self interaction correction to density functional
  approximations for many electron systems, Physical Review B 23~(10) (1981)
  5048--5079.

\bibitem{ceperley_alder}
D.~M. Ceperley, B.~J. Alder, Ground state of the electron gas by a stochastic
  method, Physical. Review Letters 45~(7) (1980) 566--569.

\bibitem{GGA_made_simple_perdew}
J.~P. Perdew, K.~Burke, M.~Ernzerhof, Generalized gradient approximation made
  simple, Phys. Rev. Lett. 77 (1996) 3865--3868.

\bibitem{Dederichs_Zeller_SCF}
P.~H. Dederichs, R.~Zeller, Self-consistency iterations in electronic-structure
  calculations, Phys. Rev. B 28 (1983) 5462--5472.

\bibitem{wavefunc_decay1}
M.~Hoffmann-Ostenhof, T.~Hoffmann-Ostenhof, R.~Ahlrichs, J.~Morgan, On the
  exponential fall off of wavefunctions and electron densities, in:
  K.~Osterwalder (Ed.), Mathematical Problems in Theoretical Physics, Vol. 116
  of Lecture Notes in Physics, Springer Berlin / Heidelberg, 1980, pp. 62--67.

\bibitem{wavefunc_decay2}
R.~Ahlrichs, M.~Hoffmann-Ostenhof, T.~Hoffmann-Ostenhof, J.~D. Morgan, Bounds
  on the decay of electron densities with screening, Physical Review A 23~(5)
  (1981) 2106--2117.

\bibitem{Chelikowsky_Saad_2}
X.~Jing, N.~Troullier, D.~Dean, N.~Binggeli, J.~R. Chelikowsky, K.~Wu, Y.~Saad,
  Ab initio molecular-dynamics simulations of si clusters using the
  higher-order finite-difference-pseudopotential method, Physical Review B
  50~(16) (1994) 12234--12237.

\bibitem{My_MS_Thesis}
A.~S. Banerjee, Harmonic analysis on isometry groups of {O}bjective
  {S}tructures and its applications to {O}bjective {D}ensity {F}unctional
  {T}heory, Master's thesis, University of Minnesota, Minneapolis (2011).

\bibitem{Evans_PDE}
L.~C. Evans, Partial Differential Equations, Vol.~19 of Graduate Studies in
  Mathematics, American Mathematical Society, 1998.

\bibitem{Kato}
T.~Kato, Perturbation Theory for Linear Operators, Classics in Mathematics,
  Springer, 1995.

\bibitem{CS_phase_encyclopedia}
E.~W. Weisstein, CRC Concise Encyclopedia of Mathematics, 2nd Edition, Chapman
  \& Hall / CRC, 2010.

\bibitem{FFT_Cooley_Tukey}
J.~W. Cooley, J.~W. Tukey, An algorithm for the machine calculation of complex
  fourier series, Mathematics of Computation 19~(90) (1965) 297--301.

\bibitem{Mohlenkamp_SHT}
M.~Mohlenkamp, A fast transform for spherical harmonics, Journal of Fourier
  Analysis and Applications 5~(2-3) (1999) 159--184.

\bibitem{Driscoll_Healey_SHT}
J.~R. Driscoll, D.~Healy, Computing fourier transforms and convolutions on the
  2-sphere, Advances in Applied Mathematics 15~(2) (1994) 202 -- 250.

\bibitem{Messiah_3j_symbol}
A.~Messiah, {Appendix C - Clebsch-Gordan} coefficients and 3j symbols, in:
  Quantum Mechanics, Volume 2, North-Holland, 1962, pp. 1054--1060.

\bibitem{MRRR}
I.~S. Dhillon, B.~N. Parlett, C.~{V\"{o}}mel, The design and implementation of
  the {MRRR} algorithm, ACM Trans. Math. Softw. 32~(4) (2006) 533--560.

\bibitem{Saad_large_eigenvalue_book}
Y.~Saad, Numerical Methods for Large Eigenvalue Problems, {R}evised Edition,
  SIAM, 2011.

\bibitem{Large_scale_plane_wave}
E.~Bylaska, K.~Tsemekhman, N.~Govind, M.~Valiev, {Large-Scale}
  {Plane-Wave-Based} {D}ensity {F}unctional {T}heory: Fomalism,
  {P}arallelization and {A}pplications, in: J.~R. Reimers (Ed.), Computational
  Methods for Large Systems: Electronic Structure Approaches for Biotechnology
  and Nanotechnology, John Wiley \& Sons, Inc., 2011, pp. 77--116.

\bibitem{Spectral_Methods_book}
C.~Canuto, M.~Y. Hussaini, A.~Quarteroni, T.~A. Zang, Spectral Methods:
  Fundamentals in Single Domains, 1st Edition, Scientific Computation,
  Springer, 2006.

\bibitem{orszag_1}
S.~A. Orszag, Transform method for the calculation of vector-coupled sums:
  Application to the spectral form of the vorticity equation, Journal of the
  Atmospheric Sciences 27~(6) (1970) 890--895.

\bibitem{orszag_2}
S.~A. Orszag, Numerical methods for the simulation of turbulence, Physics of
  Fluids 12~(12) (1969) II--250.

\bibitem{rasch_wu_hash}
J.~Rasch, A.~Yu, Efficient storage scheme for precalculated {W}igner 3 j, 6 j
  and {G}aunt coefficients, SIAM Journal on Scientific Computing 25~(4) (2004)
  1416--1428.

\bibitem{BigDFT}
L.~Genovese, A.~Neelov, S.~Goedecker, T.~Deutsch, S.~A. Ghasemi, A.~Willand,
  D.~Caliste, O.~Zilberberg, M.~Rayson, A.~Bergman, et~al., Daubechies wavelets
  as a basis set for density functional pseudopotential calculations, The
  Journal of Chemical Physics 129 (2008) 014109.

\bibitem{Jackson_ElectroDyn}
J.~D. Jackson, Classical Electrodynamics, John Wiley {\&} Sons, New York, 1975.

\bibitem{Lowdin_1956_quantum}
P.~O. L{\"o}wdin, Quantum theory of cohesive properties of solids, Advances in
  Physics 5~(17) (1956) 1--171.

\bibitem{king_smith_non_local_pseudo}
R.~King-Smith, M.~Payne, J.~Lin, Real-space implementation of nonlocal
  pseudopotentials for first-principles total-energy calculations, Physical
  Review B 44~(23) (1991) 13063.

\bibitem{Teter_Payne_Allan_1}
M.~P. Teter, M.~C. Payne, D.~C. Allan, Solution of {S}chr{\"o}dinger's equation
  for large systems, Physical Review B 40 (1989) 12255--12263.

\bibitem{Kleinman_Bylander_band_by_band_CG}
D.~Bylander, L.~Kleinman, S.~Lee, Self-consistent calculations of the energy
  bands and bonding properties of {$\text{B}_{12}\text{C}_3$}, Physical Review
  B 42~(2) (1990) 1394.

\bibitem{LOBPCG_support}
A.~V. Knyazev, K.~Neymeyr, A geometric theory for preconditioned inverse
  iteration {III}: A short and sharp convergence estimate for generalized
  eigenvalue problems, Linear Algebra and its Applications 358~(1) (2003)
  95--114.

\bibitem{Meza_Yang_DCM}
C.~Yang, J.~C. Meza, L.-W. Wang, A constrained optimization algorithm for total
  energy minimization in electronic structure calculations, Journal of
  Computational Physics 217~(2) (2006) 709--721.

\bibitem{E_Lin_LOBPCG_F}
L.~Lin, S.~Shao, W.~E, Efficient iterative method for solving the
  {D}irac-{K}ohn-{S}ham density functional theory, Journal of Computational
  Physics 245~(15) (2013) 205--217.

\bibitem{LOBPCG_3}
A.~V. Knyazev, M.~E. Argentati, I.~Lashuk, E.~Ovtchinnikov, Block locally
  optimal preconditioned eigenvalue xolvers ({BLOPEX}) in {HYPRE} and {PETSc},
  SIAM Journal on Scientific Computing 29~(5) (2007) 2224--2239.

\bibitem{Knyazev_email}
A.~Knyazev, private communication (2013).

\bibitem{Elemental_Poulson}
J.~Poulson, B.~Marker, R.~A. van~de Geijn, J.~R. Hammond, N.~A. Romero.,
  Elemental: A new framework for distributed memory dense matrix computations,
  ACM Trans. Math. Softw. 39~(2) (2013) 13:1--13:24.

\bibitem{ScaLAPACK_1}
L.~S. Blackford, J.~Choi, A.~Cleary, E.~D'Azevedo, J.~Demmel, I.~Dhillon,
  J.~Dongarra, S.~Hammarling, G.~Henry, A.~Petitet, K.~Stanley, D.~Walker,
  R.~C. Whaley, {ScaLAPACK} Users' Guide, Society for Industrial and Applied
  Mathematics, Philadelphia, PA, 1997.

\bibitem{ScaLAPACK_2}
J.~Choi, J.~J. Dongarra, R.~Pozo, D.~W. Walker, Scalapack: A scalable linear
  algebra library for distributed memory concurrent computers, in: Frontiers of
  Massively Parallel Computation, 1992., Fourth Symposium on the, IEEE, 1992,
  pp. 120--127.

\bibitem{PLAPACK_1}
R.~A. van~de Geijn, Using PLAPACK: parallel linear algebra package, The MIT
  Press, 1997.

\bibitem{PLAPACK_2}
P.~Alpatov, G.~Baker, C.~Edwards, J.~Gunnels, G.~Morrow, J.~Overfelt, R.~{van
  de Geijn}, Y.-J.~J. Wu, {PLAPACK}: parallel linear algebra package design
  overview, in: Proceedings of the 1997 ACM/IEEE conference on Supercomputing
  (CDROM), ACM, 1997, pp. 1--16.

\bibitem{poulson_parallel_matmul}
M.~D. Schatz, J.~Poulson, R.~A. van~de Geijn, Scalable universal matrix
  multiplication algorithms: 2d and 3d variations on a theme, submitted to ACM
  Transactions on Mathematical Software.

\bibitem{My_PhD_Thesis}
A.~S. Banerjee, Density functional methods for {O}bjective {S}tructures: Theory
  and simulation schemes, Ph.D. thesis, University of Minnesota, Minneapolis
  (2013).

\bibitem{Online_Template_Eigenvalue_Problem_Book}
M.~Gu, Single- and multiple-vector iterations, in: Z.~Bai, J.~Demmelc,
  J.~Dongarra, A.~Ruhe, H.~van~der Vorst (Eds.), Templates for the Solution of
  Algebraic Eigenvalue Problems: A Practical Guide, SIAM, Philadelphia, 2000.

\bibitem{stephan1998improved}
U.~Stephan, D.~A. Drabold, R.~M. Martin, Improved accuracy and acceleration of
  variational order-n electronic-structure computations by projection
  techniques, Physical Review B 58~(20) (1998) 13472.

\bibitem{bekas2005computing}
C.~Bekas, Y.~Saad, M.~L. Tiago, J.~R. Chelikowsky, Computing charge densities
  with partially reorthogonalized {L}anczos, Computer Physics Communications
  171~(3) (2005) 175--186.

\bibitem{baroni1992towards}
S.~Baroni, P.~Giannozzi, Towards very large-scale electronic-structure
  calculations, EPL (Europhysics Letters) 17~(6) (1992) 547.

\bibitem{zhou_2014_chebyshev}
Y.~Zhou, J.~R. Chelikowsky, Y.~Saad, Chebyshev-filtered subspace iteration
  method free of sparse diagonalization for solving the {K}ohn--{S}ham
  equation, Journal of Computational Physics 274 (2014) 770--782.

\bibitem{DGKS}
J.~W. Daniel, W.~B. Gragg, L.~Kaufman, G.~Stewart, Reorthogonalization and
  stable algorithms for updating the {G}ram-{S}chmidt {QR} factorization,
  Mathematics of {C}omputation 30~(136) (1976) 772--795.

\bibitem{dederichs_zeller}
P.~Dederichs, R.~Zeller, Self consistency iterations in electronic structure
  calculations, Physical Review B 28~(10) (1983) 5462.

\bibitem{anderson_mixing}
D.~G. Anderson, Iterative procedures for nonlinear integral equations, Journal
  of the ACM (JACM) 12~(4) (1965) 547--560.

\bibitem{broyden_mixing}
C.~G. Broyden, A class of methods for solving nonlinear simultaneous equations,
  Mathematics of {C}omputation 19~(92) (1965) 577--593.

\bibitem{pulay_mixing}
P.~Pulay, Convergence acceleration of iterative sequences. the case of scf
  iteration, Chemical Physics Letters 73~(2) (1980) 393--398.

\bibitem{johnson_modified_broyden}
D.~Johnson, Modified {B}royden's method for accelerating convergence in
  self-consistent calculations, Physical Review B 38~(18) (1988) 12807.

\bibitem{cances_mixing}
K.~N. Kudin, G.~E. Scuseria, E.~Canc{\`e}s, A black box self consistent field
  convergence algorithm: One step closer, The Journal of Chemical Physics 116
  (2002) 8255.

\bibitem{secant_mixing_saad}
H.~Fang, Y.~Saad, Two classes of multisecant methods for nonlinear
  acceleration, Numerical Linear Algebra with Applications 16~(3) (2009)
  197--221.

\bibitem{Kohanoff}
J.~Kohanoff, Electronic structure calculations for solids and molecules: Theory
  and computational methods, 1st Edition, Cambridge University Press, 2006.

\bibitem{Kresse_abinitio_MD}
G.~Kresse, J.~Hafner, Ab initio molecular dynamics for liquid metals, Physical
  Review B 47 (1993) 558--561.

\bibitem{Brent_Method_Book}
R.~P. Brent, Algorithms for Minimization Without Derivatives, Courier Dover
  Publications, 1973.

\bibitem{GSL_manual}
M.~Galassi, J.~Davies, J.~Theiler, B.~Gough, G.~Jungman, P.~Alken, M.~Booth,
  F.~Rossi, GNU Scientific Library Reference Manual, 3rd Edition, Network
  Theory Ltd., 2009.

\bibitem{Golub_Gauss_Quadrature}
G.~H. Golub, J.~H. Welsch, Calculation of {G}auss quadrature rules, Mathematics
  of Computation 23~(106) (1969) 221--230.

\bibitem{GTH_pseudoptential}
S.~Goedecker, M.~Teter, J.~Hutter, Separable dual-space {G}aussian
  pseudopotentials, Physical Review B 54 (1996) 1703--1710.

\bibitem{EC_Fiolhais}
C.~Fiolhais, J.~P. Perdew, S.~Q. Armster, J.~M. MacLaren, M.~Brajczewska,
  Dominant density parameters and local pseudopotentials for simple metals,
  Physical Review B 51~(20) (1995) 14001.

\bibitem{Zhou_finite_dimensional_numerical_analysis}
H.~Chen, X.~Gong, L.~He, Z.~Yang, A.~Zhou, Numerical analysis of finite
  dimensional approximations of {K}ohn-{S}ham models, Advances in Computational
  Mathematics 38~(2) (2013) 225--256.

\bibitem{Kotochigova_NIST}
S.~Kotochigova, Z.~H. Levine, E.~L. Shirley, M.~Stiles, C.~W. Clark,
  Local-density-functional calculations of the energy of atoms, Physical Review
  A 55~(1) (1997) 191.

\bibitem{Phanish_max_ent}
P.~Suryanarayana, K.~Bhattacharya, M.~Ortiz, A mesh-free convex approximation
  scheme for {K}ohn-{S}ham density functional theory, Journal of Computational
  Physics 230~(13) (2011) 5226--5238.

\bibitem{Nogueira_EC_transferability}
F.~Nogueira, C.~Fiolhais, J.~He, J.~P. Perdew, A.~Rubio, Transferability of a
  local pseudopotential based on solid-state electron density, Journal of
  Physics: Condensed Matter 8~(3) (1996) 287.

\bibitem{EC_Fiolhais_erratum}
C.~Fiolhais, J.~P. Perdew, S.~Q. Armster, J.~M. MacLaren, M.~Brajczewska,
  Erratum: Dominant density parameters and local pseudopotentials for simple
  metals, Physical Review B 53~(19) (1996) 13193.

\bibitem{GTH_relativistic}
C.~Hartwigsen, S.~Goedecker, J.~Hutter, Relativistic separable dual-space
  {G}aussian pseudopotentials from {H} to {Rn}, Physical Review B 58 (1998)
  3641--3662.

\bibitem{zhou_hexahedral_fem}
J.~Fang, X.~Gao, A.~Zhou, A {K}ohn-{S}ham equation solver based on hexahedral
  finite elements, Journal of Computational Physics 231~(8) (2012) 3166--3180.

\bibitem{Boron_fullerenes}
Q.-B. Yan, X.-L. Sheng, Q.-R. Zheng, L.-Z. Zhang, G.~Su, Family of boron
  fullerenes: General constructing schemes, electron counting rule, and ab
  initio calculations, Physical Review B 78~(20) (2008) 201401.

\bibitem{C60_bandgap}
W.~H. Green, S.~M. Gorun, G.~Fitzgerald, P.~W. Fowler, A.~Ceulemans, B.~C.
  Titeca, Electronic structures and geometries of {$\text{C}_{60}$} anions via
  density functional calculations, The Journal of Physical Chemistry 100~(36)
  (1996) 14892--14898.

\bibitem{dipole_electrostatic_book}
J.~Murray, K.~Sen, Molecular electrostatic potentials: concepts and
  applications, 1st Edition, Theoretical and computational chemistry, Elsevier,
  1996.

\bibitem{gunnarsson_diatomic}
O.~Gunnarsson, J.~Harris, R.~Jones, Density functional theory and molecular
  bonding. {I}. first-row diatomic molecules, The Journal of Chemical Physics
  67~(9) (1977) 3970--3979.

\bibitem{dipole_moments_1}
R.~M. Dickson, A.~D. Becke, Local density-functional polarizabilities and
  hyperpolarizabilities at the basis-set limit, The Journal of Physical
  Chemistry 100~(40) (1996) 16105--16108.

\bibitem{dipole_moments_2}
J.~Guan, P.~Duffy, J.~T. Carter, D.~P. Chong, K.~C. Casida, M.~E. Casida,
  M.~Wrinn, Comparison of local-density and {H}artree-{F}ock calculations of
  molecular polarizabilities and hyperpolarizabilities, The Journal of Chemical
  Physics 98~(6) (1993) 4753--4765.

\bibitem{chadi_cohen_special_points}
D.~Chadi, M.~L. Cohen, Special points in the {B}rillouin zone, Physical Review
  B 8~(12) (1973) 5747.

\bibitem{evarestov_special_points}
R.~A. Evarestov, V.~P. Smirnov, Special points of the brillouin zone and their
  use in the solid state theory, Physica Status Solidi (b) 119~(1) (1983)
  9--40.

\bibitem{Edmonds_angular_quantum}
A.~R. Edmonds, Angular Momentum in Quantum Mechanics, fourth printing Edition,
  Princeton University Press, 1996.

\bibitem{Folland_Harmonic}
G.~B. Folland, A Course in Abstract Harmonic Analysis, 1st Edition, Studies in
  Advanced Mathematics, Taylor {\&} Francis, 1994.

\bibitem{Barut_Reps}
A.~O. Barut, R.~Raczka, Theory of Group Representations and Applications,
  second revised Edition, World Scientific Publishing Company, 1986.

\bibitem{James_OS}
R.~D. James, Objective structures, Journal of the Mechanics and Physics of
  Solids 54~(11) (2006) 2354--2390.

\end{thebibliography}

\end{document}